\documentclass[nofootinbib,aps,11pt]{revtex4-1}
\usepackage{graphicx}
\usepackage{subfigure} 
\usepackage{hyperref}
\usepackage{cancel}
\usepackage{amssymb}
\usepackage{textcomp}
\usepackage{amsmath}
\usepackage{bm}
\usepackage{times}
\usepackage{epsfig}
\usepackage{color}
\usepackage{mathrsfs}
\newcommand{\GeV}{{\rm GeV}}

\newcommand{\eq}{{\rm eq}}
\newcommand{\cm}{{\rm cm}}
\begin{document}
	\title{\Large Reviving $Z^\prime$ Portal Dark Matter with Conversion Mechanism }
	\bigskip
	\author{Zhen-Wei Wang$^1$}
	\email{zwwang09@163.com}
	\author{Zhi-Long Han$^1$}
	\email{sps\_hanzl@ujn.edu.cn}
	\author{Fei Huang$^{1,2}$}
	\email{sps\_huangf@ujn.edu.cn}
	\author{Honglei Li$^1$}
	\email{sps\_lihl@ujn.edu.cn}
	\author{Ang Liu$^3$}
	\email{AL@jnxy.edu.cn}
	\affiliation{$^1$School of Physics and Technology, University of Jinan, Jinan 250022, China }
	\affiliation{$^2$State Key Laboratory of Dark Matter Physics, School of Physics and Astronomy, Shanghai Jiao Tong University, Shanghai 200240, China}	
	\affiliation{$^3$School of Physical Science and Intelligent Engineering, Jining University, Jining 273155, China}
	\date{\today}
	\begin{abstract} 
		In many new physics models with extended gauge symmetry, the new gauge boson $Z'$ could mediate the interactions between the dark matter and standard model particles. For the conventional $Z^\prime$ portal dark matter, the collider and the direct detection constraints typically pose a significant challenge. To address this pressing issue, we present in this paper a new benchmark model based on the gauged $U(1)_{B-L}$ symmetry, which introduces a Dirac dark fermion $\tilde{\chi}_1$  and a heavier partner $\tilde{\chi}_2$ with zero and nonzero $U(1)_{B-L}$ charge, respectively. Including the mass term $\delta m \bar{\tilde{\chi}}_1\tilde{\chi}_2$ results in the dark fermions $\chi_1$ and $\chi_2$ in the mass eigenstate, where the lighter one $\chi_1$ is regarded as the dark matter candidate. Various intriguing processes for the relic density arise with the compressed mass spectrum $m_{\chi_1}\simeq m_{\chi_2}$, such as the coscattering $\chi_2f\to\chi_1f$, the conversion $\chi_2\chi_i\to\chi_1\chi_j$, and the coannihilation $\chi_1\chi_2\to f\bar{f}$ processes. Suppressed by the small mixing angle $\theta$ between the dark fermions, the small effective gauge coupling of dark matter $\chi_1$ to the gauge boson $Z'$ is one distinct feature of this model, rendering phenomenology in many aspects more promising. In this paper, we investigate the production of dark matter through new mechanisms within the frameworks of resonance and secluded scenarios. The impacts of phenomenological constraints from collider, dark matter, and cosmology are also taken into account. We report that the conversion mechanism is both favored by the resonance and secluded scenarios under current constraints.
	\end{abstract}
	\maketitle

\section{Introduction}

The cosmological and astrophysical observations  provide evidence for the feasibility of particle dark matter (DM) \cite{Bertone:2004pz,Cirelli:2024ssz}. The prevailing hypothesis considers that the DM particle is in thermal equilibrium with the Standard Model(SM) bath during the early universe, subsequently freezing out to yield the observed thermal relic density. This is known as the weakly interacting massive particle (WIMP) paradigm. Typically, the DM candidate is a singlet under the SM gauge symmetry, which then requires a portal to mediate the interactions between DM and SM \cite{DeSimone:2016fbz,Arcadi:2017kky}.  Based on the spin of the mediator, it can be categorized into the fermion portal \cite{Bai:2013iqa,Bai:2014osa,Escudero:2016ksa,Blennow:2019fhy,Coito:2022kif,Li:2022bpp}, the scalar portal \cite{Patt:2006fw,March-Russell:2008lng,Okada:2010wd,Djouadi:2011aa,Cline:2013gha,Arcadi:2019lka}, as well as the vector portal \cite{Alves:2013tqa,DEramo:2016gos,Okada:2018ktp,Blanco:2019hah,Fitzpatrick:2020vba}.

The gauged $U(1)_{B-L}$ symmetry is the simplest anomaly-free extension of the SM \cite{Mohapatra:1980qe}, in which three right-handed neutrinos $N$ are introduced for anomaly cancellation. These right-handed neutrinos could produce tiny neutrino mass via the seesaw mechanism \cite{Minkowski:1977sc,Mohapatra:1979ia,Schechter:1980gr,Schechter:1981cv}, and explain the baryon asymmetry via the leptogenesis mechanism \cite{Fukugita:1986hr, Davidson:2008bu,Iso:2010mv,Dev:2017xry,Das:2024gua}. Meanwhile, DM can be implemented in the $U(1)_{B-L}$ symmetry \cite{Okada:2012sg,Basak:2013cga,Escudero:2016tzx,Das:2019pua,Liu:2024esf}.  In this work, we focus on the phenomenon of DM. For simplicity, the influence of right-handed neutrino on DM is disregarded by assuming $m_N$ at the canonical seesaw scale $\mathcal{O}(10^{14})$ GeV.

In the traditional $Z^\prime$ portal Dirac DM scenario, the observed relic density of DM $\chi$ requires the gauge coupling $g'\sim\mathcal{O}(0.1)$  with new symmetry charge $Q_\chi=-1$ \cite{Abdallah:2015ter,Klasen:2016qux}.  This results in the spin-independent DM-nucleon scattering cross section mediated by $Z^\prime$ being strictly constrained by current DM direct detection experiments \cite{DarkSide-50:2023fcw,PandaX:2024qfu,LZ:2024zvo}. Furthermore, the direct searches of $Z^\prime$ on  various colliders favor exceedingly small $g^\prime$ for $m_{Z'}$ below the TeV-scale  \cite{BaBar:2014zli,BaBar:2017tiz,LHCb:2017trq,LHCb:2019vmc,ATLAS:2019erb}. To satisfy current constraints, one improvement approach is to treat $Q_\chi$ as a free parameter. In this way, the effective gauge coupling of DM  $g_\chi=Q_{\chi}g^\prime$ is expected to be much larger than $g^\prime$, which can alleviate the pressure from direct detection in both $Z^\prime$ resonance \cite{Nath:2021uqb} and secluded \cite{Mohapatra:2019ysk} schemes. An alternative pathway is the inelastic DM \cite{Tucker-Smith:2001myb,Filimonova:2022pkj,Foguel:2024lca}, in which the relic density is determined via coannihilation of the DM and the dark partner. With sufficiently large mass splitting, the inelastic DM-nucleon scattering is kinematically suppressed by the DM velocity \cite{Tucker-Smith:2001myb}.

Motivated by the above two methods, we investigate the $Z'$ portal inelastic Dirac DM with the $U(1)_{B-L}$ symmetry \cite{Zhang:2024sox}. This model  includes one $Z_2$ odd Dirac fermion $\tilde{\chi}_1$ and its partner $\tilde{\chi}_2$, which carry zero $Q_{\tilde{\chi}_1}=0$ and non-zero $Q_{\tilde{\chi}_2}\neq0$ charge under the $U(1)_{B-L}$ symmetry, respectively. In this way, the interaction between the DM candidate $\chi_1$ in the mass eigenstate and $Z^\prime$ occurs through the mixing $\theta$ between the two dark fermions, which arises from a mass term $\delta m \bar{\tilde{\chi}}_1\tilde{\chi}_2$. Considering that $g_\chi=Q_{\tilde{\chi}_2}g^\prime\sim1$,
the significant annihilation rate of $\chi_2$ pairs may lead to DM production governed by the traditional coannihilation mechanism \cite{Griest:1990kh}, or through the novel coscattering or conversion mechanism \cite{DAgnolo:2017dbv,Garny:2017rxs}. The prominent characteristic of the latter  is that the relic density of DM is determined by the inelastic conversions with its heavier dark partner \cite{Garny:2018icg,DAgnolo:2018wcn,Cheng:2018vaj,Junius:2019dci,DAgnolo:2019zkf,Maity:2019hre,Brummer:2019inq,Heeck:2022rep,Heisig:2024mwr,Heisig:2024xbh,DiazSaez:2024nrq,DiazSaez:2024dzx,Paul:2024prs,Chatterjee:2025vdz,Liu:2025swd,Paul:2025spm}. This transformation processes in our model primarily consists of coscattering $\chi_2f\to\chi_1f$ and conversion $\chi_2\chi_i\to\chi_1\chi_j$ processes with $f$ the SM fermion and $\{i,j\}=\{1,2\}$. This scenario imposes very lenient requirements on DM annihilation, i.e., allowing the relevant small values of mixing $\theta$ and gauge coupling $g^\prime$. Therefore, it is possible to obtain the viable parameter space allowed by  various experiments.

 Different from the minimal inelastic Dirac DM case with $Q_{\tilde{\chi}_2}=-1$ \cite{Zhang:2024sox}, we study the extended scenario with $Q_{\tilde{\chi}_2}$ as a free parameter in this paper. The different value of $Q_{\tilde{\chi}_2}$ leads to distinct mechanisms governing dark matter production. For instance, coannihilation mechanism is the only viable dominate channel in the minimal scenario \cite{Zhang:2024sox}. In contrast, coscattering and conversion mechanism could become the dominant channel in the extended case when $g_\chi=g'\times Q_{\tilde{\chi}_2}\sim \mathcal{O}(1)$. It is worth noting that previous study of inelastic Dirac DM mainly concentrates on the MeV mass scale \cite{Filimonova:2022pkj,Zhang:2024sox},  however our focus lies on the GeV to TeV mass scale.

The structure of this paper is organized as follows. In Section \ref{SEC:TM}, we provide a brief introduction to the theoretical  model employed in our study. The calculation of  relic density as well as the associated phenomenological constraints in the resonance scenario are discussed in Section \ref{SEC:RS}. Next, we investigate the secluded scenario in Section \ref{SEC:SS}. Finally, we summarize the results in Section \ref{SEC:CL}.

\section{The model}\label{SEC:TM}

We extend two vector-like Dirac fermions $\tilde{\chi}_1$ and $\tilde{\chi}_2$ beyond the SM.
Among them, only $\tilde{\chi}_2$ carries a non-zero charge $Q_{\tilde{\chi}_2}$ under the $U(1)_{B-L}$ symmetry, and $\tilde{\chi}_1$ is considered to be neutral.  Meanwhile,  these two dark fermions are stipulated to be $Z_2$ odd to  ensure the stability of DM. In this configuration, when a dark scalar $\phi$ with $U(1)_{B-L}$ charge $Q_{\phi}=-Q_{\tilde{\chi}_2}$ acquires a vacuum expectation value $\langle\phi\rangle=v_\phi$ \cite{Filimonova:2022pkj,Zhang:2024sox}, the Yukawa term $y\phi\bar{\tilde{\chi}}_1\tilde{\chi}_2$ will naturally induce mass mixing $\delta m \bar{\tilde{\chi}}_1\tilde{\chi}_2$ with $\delta m=y v_\phi$. 

The mass eigenstates $\chi_1$ and $\chi_2$ could be formed through a transformation:
\begin{equation}
	\begin{pmatrix}
		\chi_1 \\ \chi_2
	\end{pmatrix} =
	\begin{pmatrix}
		\cos\theta & -\sin\theta \\ \sin\theta & \cos\theta
	\end{pmatrix}
	\begin{pmatrix}
		\tilde{\chi}_1 \\ \tilde{\chi}_2
	\end{pmatrix}.
\end{equation}
Here, we consider that $m_{\chi_1}$ is slightly smaller than $m_{\chi_2}$, thus $\chi_1$ is regarded as the DM candidate. In principle, the scalar portal interactions also contribute to the DM relic abundance. However, the corresponding Yukawa coupling $y$ is subjected to dual suppression from small mixing $\theta$ and mass splitting $m_{\chi_{2}}-m_{\chi_{1}}$ \cite{Zhang:2024sox}, which is favored by the conversion mechanism in this study. For simplicity, we further assume that the mass of the new scalar $m_\phi$ is much larger than $m_{\chi_{1,2}}$, so the contribution resulting from $\phi$ portal interaction is significantly smaller than that of $Z^\prime$ portal \cite{Zhang:2024sox}. 

In the mass eigenstates, the $Z^\prime$ portal interactions could be expressed as
\begin{eqnarray}
	\mathcal{L}\supset +Z^\prime_{\mu}g_{\chi}\left(\cos^2\theta\bar{\chi_2}\gamma^\mu\chi_2-\frac{\sin2\theta}{2}\bar{\chi_2}\gamma^\mu\chi_1-\frac{\sin2\theta}{2}\bar{\chi_1}\gamma^\mu\chi_2+\sin^2\theta\bar{\chi_1}\gamma^\mu\chi_1\right)-Z^\prime_{\mu}g^{\prime}Q_f\bar{f}\gamma^\mu f
\end{eqnarray}
with $g^\prime$ the $U(1)_{B-L}$ gauge coupling and $Q_f$ the  $U(1)_{B-L}$ charge of SM fermion $f$.

In such an inelastic model, we investigate the DM phenomenology with mass above the GeV scale.  The specific execution based on the magnitude of  $m_{Z^\prime}$ and $m_{\chi_2}$ will be categorized into: (1) the $Z^\prime$ resonance \cite{Nath:2021uqb} and (2) secluded scenarios \cite{Mohapatra:2019ysk}.  The total free parameters involved in the subsequent calculations are as follows:
\begin{eqnarray}
	\{m_{\chi_1},\Delta_\chi,r_{Z^\prime},g^\prime,g_{\chi},\theta\}
\end{eqnarray}
where $\Delta_\chi=(m_{\chi_2}-m_{\chi_1})/m_{\chi_1}$ and $r_{Z^\prime}=m_{Z^\prime}/m_{\chi_2}$.

\section{Resonance Scenario}\label{SEC:RS}

\subsection{Relic Density}\label{RD-R}

 As will shown in Section \ref{PZ-R}, to evade the stringent collider constraints on $Z'$, the mass ratio $r_{Z'}$ has to be near the resonance region, i.e., $r_{Z'}\simeq2$. We set the mass ratio $r_{Z^\prime}=2$ in this resonance scenario, which maximizes the suppression of the gauge coupling $g^\prime$ under the requirement of reproducing the observed dark matter relic density.
 The reactions associated with the dark particles include two types: (co)annihilation and conversion.  The former primarily involves $s$-channel processes mediated by $Z^\prime$, i.e., $\chi_1\bar{\chi}_1\to f\bar{f}$,  $\chi_2\bar{\chi}_2\to f\bar{f}$ and $\chi_2\chi_1\to f\bar{f}$. The latter encompasses not only the inelastic scattering $\chi_2f\to\chi_1f$ and the three body decay $\chi_2\to\chi_1f\bar{f}$ which determine the coscattering mechanism, but also the self-interactions $\chi_2\chi_i\to\chi_1\chi_j$ within the dark sector, which corresponds to the conversion mechanism.

\allowdisplaybreaks

The abundances of $\chi_1$ and $\chi_2$ could be calculated by numerically solving the Boltzmann equations:
\begin{eqnarray}\label{Eqn:BE-R}
	\frac{dY_{\chi_1}}{dz} &= & -\frac{s}{\mathcal{H}z}\bigg[ \langle \sigma v\rangle_{\chi_1\bar{\chi}_1\to f\bar{f}}\Big(Y_{\chi_1}^2-(Y_{\chi_1}^{\eq})^2\Big)+\langle \sigma v\rangle_{\chi_2\chi_1\to f\bar{f} }\Big(Y_{\chi_2}Y_{\chi_1}-Y_{\chi_2}^{\eq}Y_{\chi_1}^{\eq}\Big)\nonumber \\
	&-& \langle \sigma v\rangle_{\chi_2 f\to\chi_1 f} \left(Y_{\chi_2}Y_f^{\eq}-\frac{Y_{\chi_2}^{\eq}}{Y_{\chi_1}^{\eq}}Y_{\chi_1}Y_f^{\eq}\right)-\langle \sigma v\rangle_{\chi_2\chi_2\to\chi_1\chi_1}\left(Y_{\chi_2}^2-\frac{(Y_{\chi_2}^{\eq})^2}{(Y_{\chi_1}^{\eq})^2}Y_{\chi_1}^2\right)
	\nonumber \\
	&-&\langle \sigma v\rangle_{\chi_1\chi_2\to\chi_1\chi_1}\left(Y_{\chi_1} Y_{\chi_2}-\frac{Y_{\chi_2}^{\eq}}{Y_{\chi_1}^{\eq}}Y_{\chi_1}^2\right)-\langle \sigma v\rangle_{\chi_2\chi_2\to\chi_1\chi_2}\left(Y_{\chi_2}^2-\frac{Y_{\chi_2}^{\eq}}{Y_{\chi_1}^{\eq}}Y_{\chi_1}Y_{\chi_2}\right)
	\nonumber \\
	&-&\frac{\Gamma_{\chi_2\to\chi_1 f\bar{f}}}{s}\left(Y_{\chi_2}-\frac{Y_{\chi_2}^{\eq}}{Y_{\chi_1}^{\eq}}Y_{\chi_1}\right)\bigg],\\	
	\frac{dY_{\chi_2}}{dz} &= & -\frac{s}{\mathcal{H}z}\bigg[ \langle \sigma v\rangle_{\chi_2 \bar{\chi}_2\to f\bar{f}}\Big(Y_{\chi_2}^2-(Y_{\chi_2}^{\eq})^2\Big)+\langle \sigma v\rangle_{\chi_2\chi_1\to f\bar{f}}\Big(Y_{\chi_2}Y_{\chi_1}-Y_{\chi_2}^{\eq}Y_{\chi_1}^{\eq}\Big)
	\nonumber \\
	&+& \langle \sigma v\rangle_{\chi_2 f\to\chi_1 f}\left(Y_{\chi_2}Y_{f}^{\eq}-\frac{Y_{\chi_2}^{\eq}}{Y_{\chi_1}^{\eq}}Y_{\chi_1}Y_{f}^{\eq}\right)+\langle \sigma v\rangle_{\chi_2\chi_2\to\chi_1\chi_1}\left(Y_{\chi_2}^2-\frac{(Y_{\chi_2}^{\eq})^2}{(Y_{\chi_1}^{\eq})^2}Y_{\chi_1}^2\right)
	\nonumber \\
	&+&\langle \sigma v\rangle_{\chi_1\chi_2\to\chi_1\chi_1}\left(Y_{\chi_1} Y_{\chi_2}-\frac{Y_{\chi_2}^{\eq}}{Y_{\chi_1}^{\eq}}Y_{\chi_1}^2\right)+\langle \sigma v\rangle_{\chi_2\chi_2\to\chi_1\chi_2}\left(Y_{\chi_2}^2-\frac{Y_{\chi_2}^{\eq}}{Y_{\chi_1}^{\eq}}Y_{\chi_1}Y_{\chi_2}\right)
	\nonumber \\
	&+&\frac{\Gamma_{\chi_2\to\chi_1 f\bar{f}}}{s}\left(Y_{\chi_2}-\frac{Y_{\chi_2}^{\eq}}{Y_{\chi_1}^{\eq}}Y_{\chi_1}\right)\bigg],
\end{eqnarray}
where $z=m_{\chi_1}/T$, the entropy density $s=2\pi^2 g_s T^3/45$, and the
Hubble expansion rate is defined as $\mathcal{H}=\sqrt{4\pi^3g_*/45} T^2/m_{pl}$ with the Planck mass $m_{pl}=1.22\times10^{19}~\GeV$.  $g_s$ and $g_\star$  are the number of relativistic degrees of freedom for the entropy density and energy density, respectively. The thermal average cross sections $\left<\sigma v\right>$ of various channels are calculated numerically by micrOMEGAs~\cite{Alguero:2022inz,Alguero:2023zol}.

Due to the kinematic prohibition of two-body decay $\chi_2\to \chi_1 Z^\prime$, the predominant decay process of $\chi_2$ is the three-body decay $\chi_2\to\chi_1{Z^\prime}^*\to\chi_1f \bar{f}$. The corresponding thermal decay width is written as
\begin{eqnarray}
	\Gamma_{\chi_2\to\chi_1 f\bar{f} }=\frac{\mathcal{K}_1\left(\frac{m_{\chi_2}}{m_{\chi_1}}x\right)}{\mathcal{K}_2\left(\frac{m_{\chi_2}}{m_{\chi_1}}x\right)}\tilde{\Gamma}_{\chi_2\to\chi_1 f\bar{f}},
\end{eqnarray}
where $\mathcal{K}_{1,2}$ are modified Bessel functions of the second kind. In the limit of small mass splitting, the three-body decay width can be estimated as \cite{Tsai:2019buq}
\begin{eqnarray}\label{Eq:Chi2Wd}
	\tilde{\Gamma}_{\chi_2\to\chi_1 f \bar{f}}\simeq\frac{N_c^f{g_\chi}^2{g^\prime}^2Q_f^2\Delta_{\chi}^5m_{\chi_1}^5\sin^2 2\theta}{120\pi^3 m_{Z^\prime}^4}\times\theta^\prime(\Delta_{\chi}-2m_f),
\end{eqnarray}
where $N_c^f$ is the color number of $f$, $Q_f=1/3$ and $-1$ for quarks and  leptons. $\theta^\prime$ is the Heaviside theta function. More precise results are obtained  numerically  through micrOMEGAs in this paper. The abundance of $\chi_1$ and $\chi_2$ at thermally equilibrium  can be expressed as \cite{Alguero:2022inz}
\begin{eqnarray}
	Y_{\chi_1}^{\eq}=\frac{45z^2}{2\pi^4g_s}\mathcal{K}_2(z),~Y_{\chi_2}^{\eq}=\frac{45z^2}{2\pi^4g_s}\left(\frac{m_{\chi_2}}{m_{\chi_1}}\right)^2\mathcal{K}_2\left(\frac{m_{\chi_2}}{m_{\chi_1}}z\right).
\end{eqnarray}
$Y_{f}^{\eq}$ takes the value of 0.238.

\begin{figure}
	\begin{center}
		\includegraphics[width=0.42\linewidth]{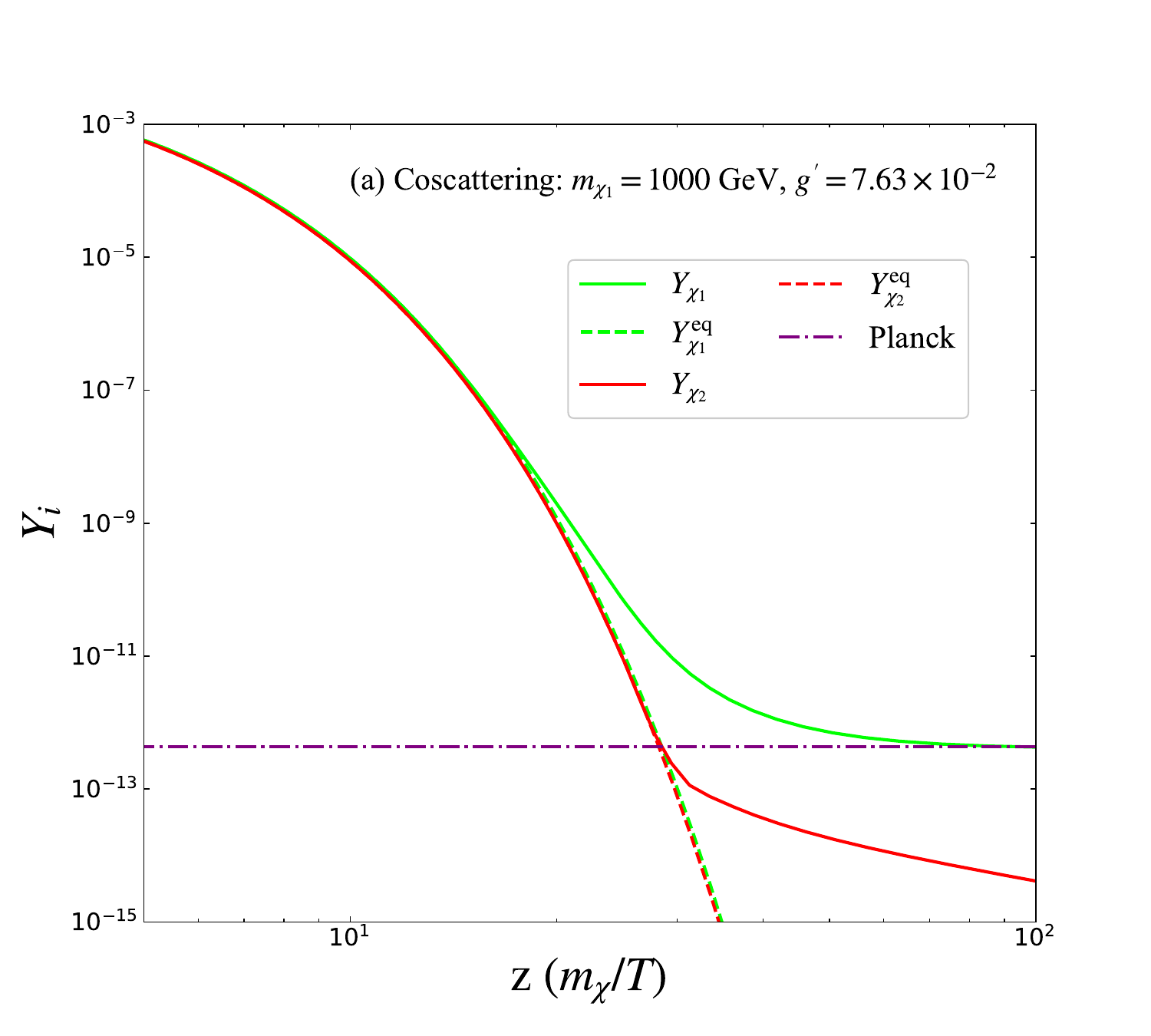}
		\includegraphics[width=0.42\linewidth]{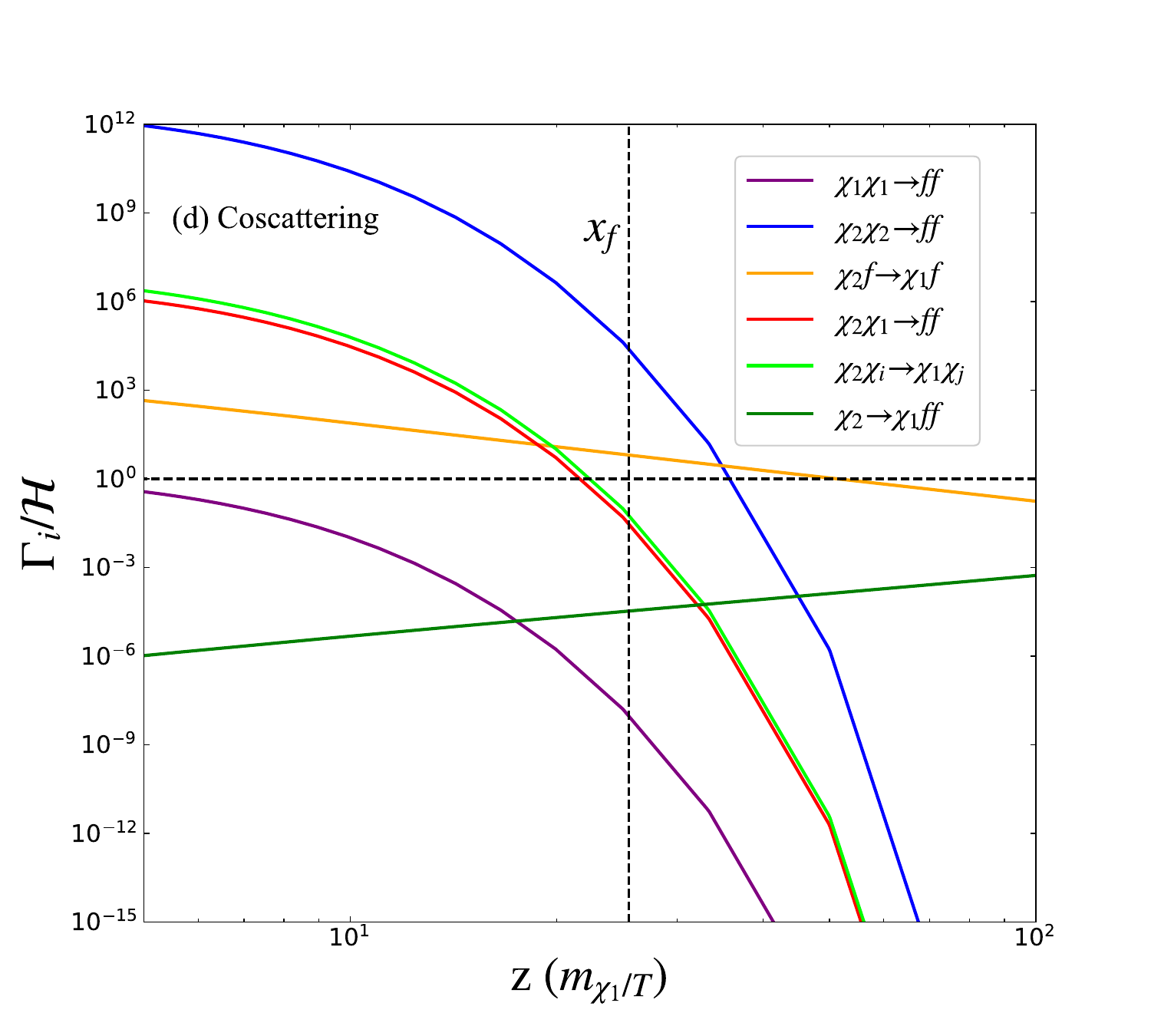}
		\includegraphics[width=0.42\linewidth]{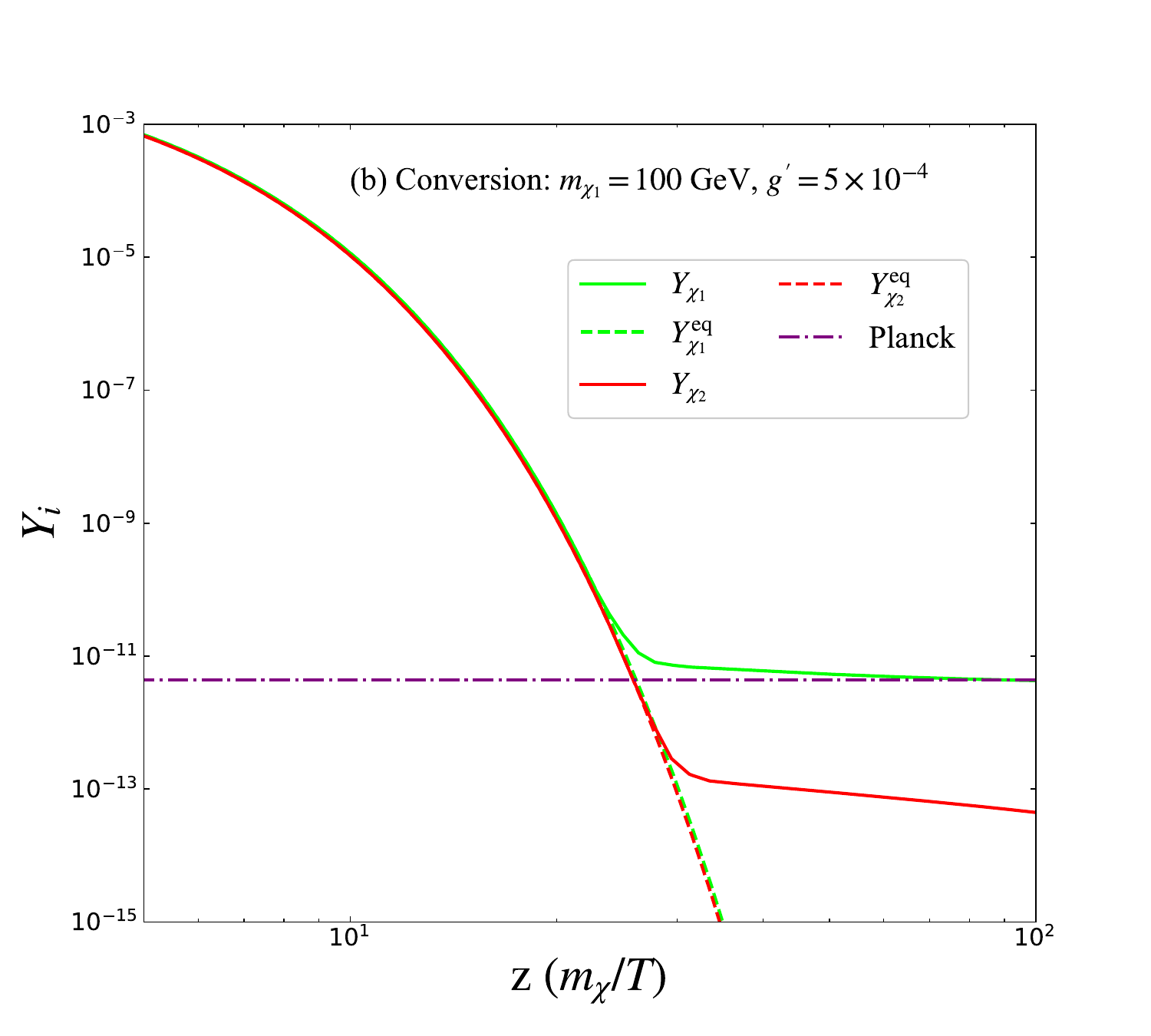}
		\includegraphics[width=0.42\linewidth]{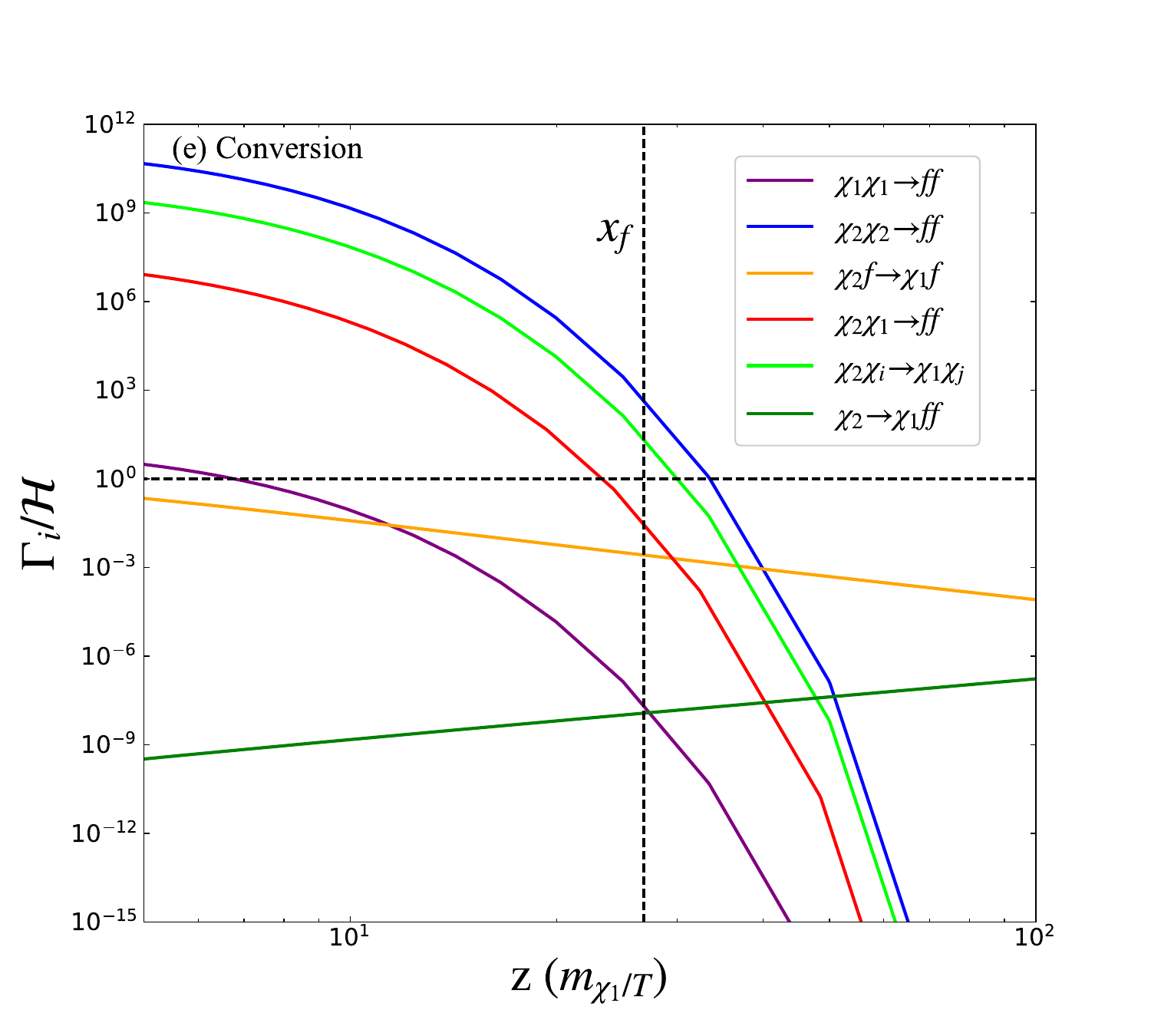}
		\includegraphics[width=0.42\linewidth]{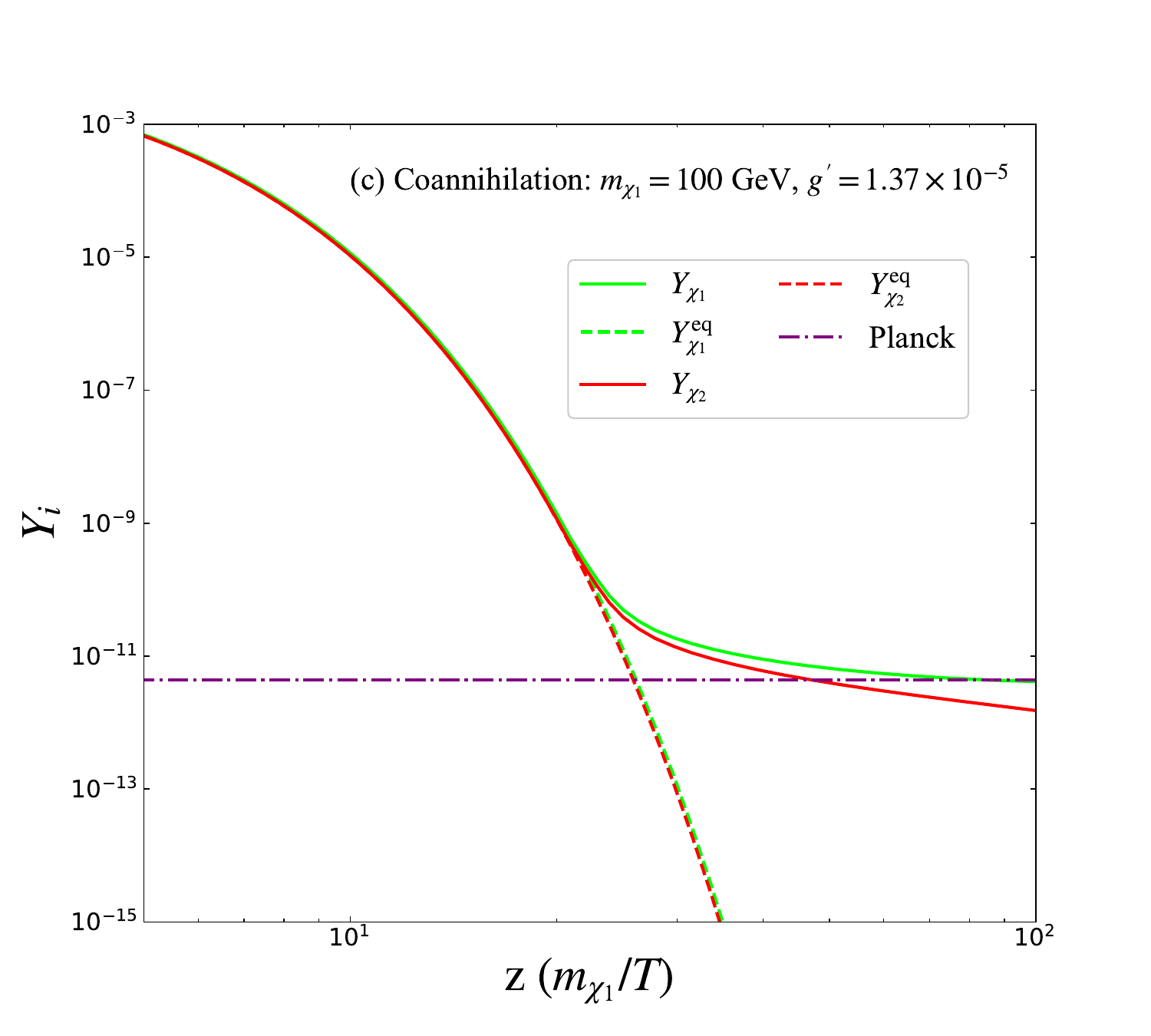}		
		\includegraphics[width=0.42\linewidth]{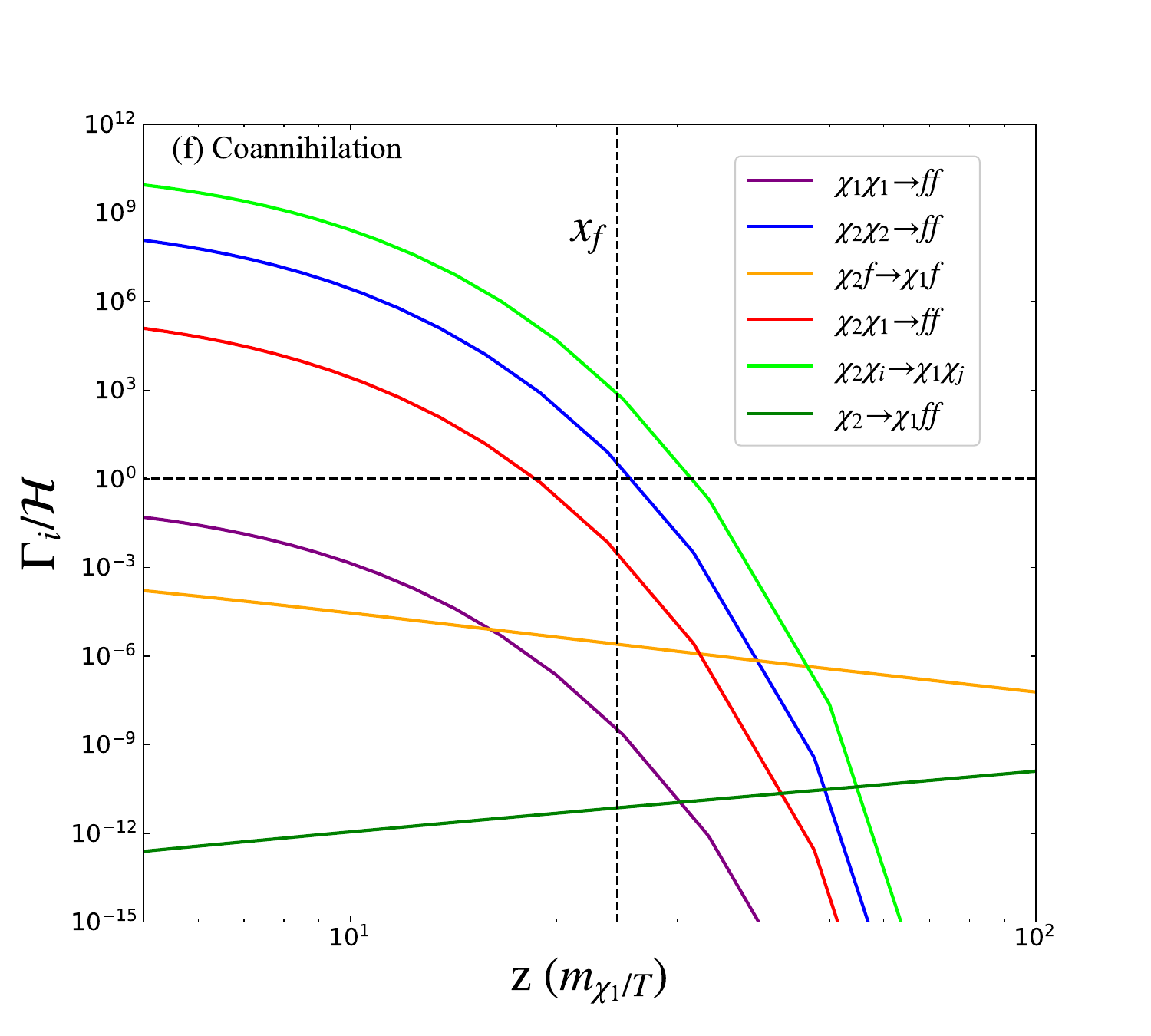}
	\end{center}
	\caption{The evolutions of various abundances $Y_i$ for (a) coscattering, (b) conversion , and (c) coannihilation benchmarks in the resonance scenario. The subfigures (d), (e), and (f) on the right correspond to the thermal rates of various reaction processes in three different phases. We fix $\Delta_\chi=10^{-2}$, $r_{Z^\prime}=2$ and $\theta=5\times10^{-4}$. The solid green and red lines in (a)-(c) represent the abundance  of $\chi_1$ and $\chi_2$, meanwhile the dashed lines are their thermal equilibrium. Purple dotdashed cruve is the observation of DM $\Omega_{\chi_1}h^2=0.12$~\cite{Planck:2018vyg}.   In subfigures (d)-(f), the black vertical dashed line corresponds to the thermal decoupling temperature when $Y_{\chi_1}/Y_{\chi_1}^\eq=2.5$, and the horizontal black line is $\Gamma_i=\mathcal{H}$. Moreover,  $\chi_i \chi_2\to \chi_j \chi_1$ is the sum of conversion channels $\chi_2\chi_2\to\chi_1\chi_1$, $\chi_1\chi_2\to\chi_1\chi_1$ and $\chi_2\chi_2\to\chi_1\chi_2$. 
	}
	\label{FIG:fig1}
\end{figure}

According to the classification in Ref.~\cite{DAgnolo:2019zkf}, a benchmark that falls within the coscattering regime  primarily satisfies three conditions at the freeze-out temperature: (1) $\chi_1$ is in kinetic equilibrium with the SM. (2) There is no chemical potential for $\chi_1$. (3) The last reaction to decouple, which changes the number density of $\chi_1$, is exchange reactions between $\chi_1$ and $\chi_2$.  Regarding the first condition,since the elastic reaction $\Gamma_{\chi_1f\to\chi_1f}$ in our scenario is suppressed by the small mixing $\theta$, we assume that $\chi_{1}$ stays in the kinetic equilibrium with SM via the intense inelastic scattering process $\chi_2f\to\chi_1f$, which can be parameterized by $\Gamma_{\chi_2f\to\chi_1f}\gg\mathcal{H}$~\cite{DAgnolo:2017dbv,DAgnolo:2019zkf}.  Obtaining precise results requires solving the full unintegrated Boltzmann equations, which may introduce an $\mathcal{O}(10\%)$ distinction\cite{Garny:2017rxs} compared to $\chi_1$ not being in kinetic equilibrium. The requirement (2) is satisfied through $\Gamma_{\chi_2\bar{\chi}_2\to f\bar{f}}>\Gamma_{\chi_2f\to\chi_1f}$.  The strong reaction of $\chi_2\bar{\chi}_2\to f\bar{f}$  rapidly consumes $\chi_2$, thereby disrupting the chemical equilibrium between $\chi_1$ and $\chi_2$ in the process $\chi_2f\leftrightarrow\chi_1f$.   As for condition (3),  it suffices that $\Gamma_{\chi_2f\to\chi_1f}$ is greater than $\Gamma_{\chi_1\bar{\chi}_1\to f\bar{f}}$. 

The process $\chi_2\to\chi_1f\bar{f}$ that is closely related to coscattering has an ignored contribution in this paper. The impact of $\chi_2 Z^\prime\to\chi_1 Z^\prime$ is also very small in the resonance scenario,  because $m_{Z^\prime}>m_{\chi_{1,2}}$ causes $Y_{Z^\prime}^{\eq}$ to undergo an exponential suppression earlier than $Y_{\chi_{1,2}}^{\eq}$. In this way,  $Y_{Z^\prime}^{\eq}$ is already diminished to a negligibly small value at the decoupling temperature of DM, naturally suppressing the contribution of process $\chi_2 Z^\prime\to\chi_1 Z^\prime$ in the Boltzmann equations. Therefore, we only need to focus on the inelastic scattering $\chi_2f\to\chi_1f$.  Moreover, in the coscattering regime, both the WIMP like pair annihilation $\chi_1\bar{\chi_1}\to f \bar{f}$ and coannihilation $\chi_{1}\chi_2\to f\bar{f}$  processes have the reaction rates smaller than $\mathcal{H}$ to ensure that $\chi_{1}$ has departed from thermal equilibrium at the freeze-out temperature. On the whole, the relevant processes approximately satisfy the relationship $\Gamma_{\chi_2\bar{\chi}_2\to f\bar{f}}>\Gamma_{\chi_2f\to\chi_1f}\gg\mathcal{H}\gtrsim\Gamma_{\chi_{1,2}\bar{\chi}_1\to f\bar{f}}$ at the freeze-out temperature. Regarding the conversion process $\chi_i \chi_2\to \chi_j \chi_1$, $\Gamma_{\chi_2f\to\chi_1f}>\Gamma_{\chi_i \chi_2\to \chi_j \chi_1}$ suffices to guarantee that the contribution of coscattering is dominant, meanwhile its relative magnitude compared to $\mathcal{H}$ becomes irrelevant.	 

By employing a similar judgment method, when the dominant $\chi_2f\to\chi_1f$  is exceeded by $\chi_i \chi_2\to \chi_j \chi_1$, it results in conversion regime.  At this point, the reaction rates of these processes satisfy the relationship $\Gamma_{\chi_2\bar{\chi}_2\to f\bar{f}}>\Gamma_{\chi_i \chi_2\to \chi_j \chi_1}\gg\mathcal{H}\gtrsim\Gamma_{\chi_{1,2}\bar{\chi}_1\to\bar{f}f}$  . In other cases where either $\chi_2\bar{\chi}_2\to f\bar{f}$ or $\chi_2\chi_1\to f\bar{f}$ makes the most significant contribution, this can be referred to as coannihilation,  which requires $\Gamma_{\chi_{1,2}\bar{\chi}_2\to f\bar{f}}\sim\mathcal{H}>\Gamma_{\chi_1\bar{\chi}_1\to\bar{f}f}$. As for the coannihilation phase, the respective reaction rates of  the coscattering process $\chi_2f\to\chi_1f$ and the conversion process $\chi_i \chi_2\to \chi_j \chi_1$ may be either greater than or less than $\Gamma_{\chi_{1,2}\bar{\chi}_2\to f\bar{f}}$ as long as the conditions for determining coscattering and conversion regimes are not satisfied.

We utilize three  benchmark points corresponding to the coscattering, the conversion and the coannihilation phases to present the evolutions of dark fermion abundances in Figure~\ref{FIG:fig1}, which also aims to elucidate how to distinguish them. The thermal rates of $2\to2$ processes in (d)-(f) are denoted as $\Gamma_i=n_{a}^{\eq}\left<\sigma v\right>_i$ with $n_a^{\eq}$ the number density of particle $a$ at the thermal equilibrium. In Figure~\ref{FIG:fig1}, panels (a) and (d), (b) and (e), (c) and (f) satisfy  the criteria for determining coscattering, conversion, and coannihilation, respectively. Among these phases, coscattering is the first to exhibit an obvious deviation from thermal equilibrium, occurring approximately  at $\Gamma_{\chi_2\chi_1\to ff}\sim \mathcal{H}$. Then $Y_{\chi_1}$ subsequently continues to show a rapid decline until the depletion of inelastic reaction $\chi_2f\to\chi_1f$~\cite{Alguero:2022inz}. In contrast, the decoupling of conversion and coannihilation  occurs at a later stage. And for the same $m_{\chi_1}$, significant differences are observed in $g^\prime$ due to variations in the defined conditions.

Furthermore, the inability to achieve freezing-out raises concerns with tiny $g'$ and $\theta$. Specifically, DM $\chi_1$ cannot reach  thermal equilibrium with the SM bath. This can be parameterized as $\Gamma_{\chi_2\chi_1\to ff}/\mathcal{H}<1$, as the contribution of pair annihilation $\chi_1\chi_1\to ff$ is deemed negligible due to its minimal impact. Here, we use micrOMEGAs to perform calculations over a broad range of $g^\prime\in[10^{-10},10^{-2}]$, $\Delta_\chi\in[10^{-3},10^{-1}]$ and $g_\chi\in[0.1,1]$. We report that the lower limit for freezing-out of $\chi_1$ is $\theta\gtrsim\mathcal{O}(10^{-6})$ when $m_{\chi_1}\sim$ TeV. For the GeV scale, this result can drop to $\mathcal{O}(10^{-8})$. Therefore, in the subsequent phenomenological study, the selection of parameters adheres to the freezing-out condition.

\subsection{Phenomenology of $Z^\prime$}\label{PZ-R}

\begin{figure}
	\begin{center}
		\includegraphics[width=0.45\linewidth]{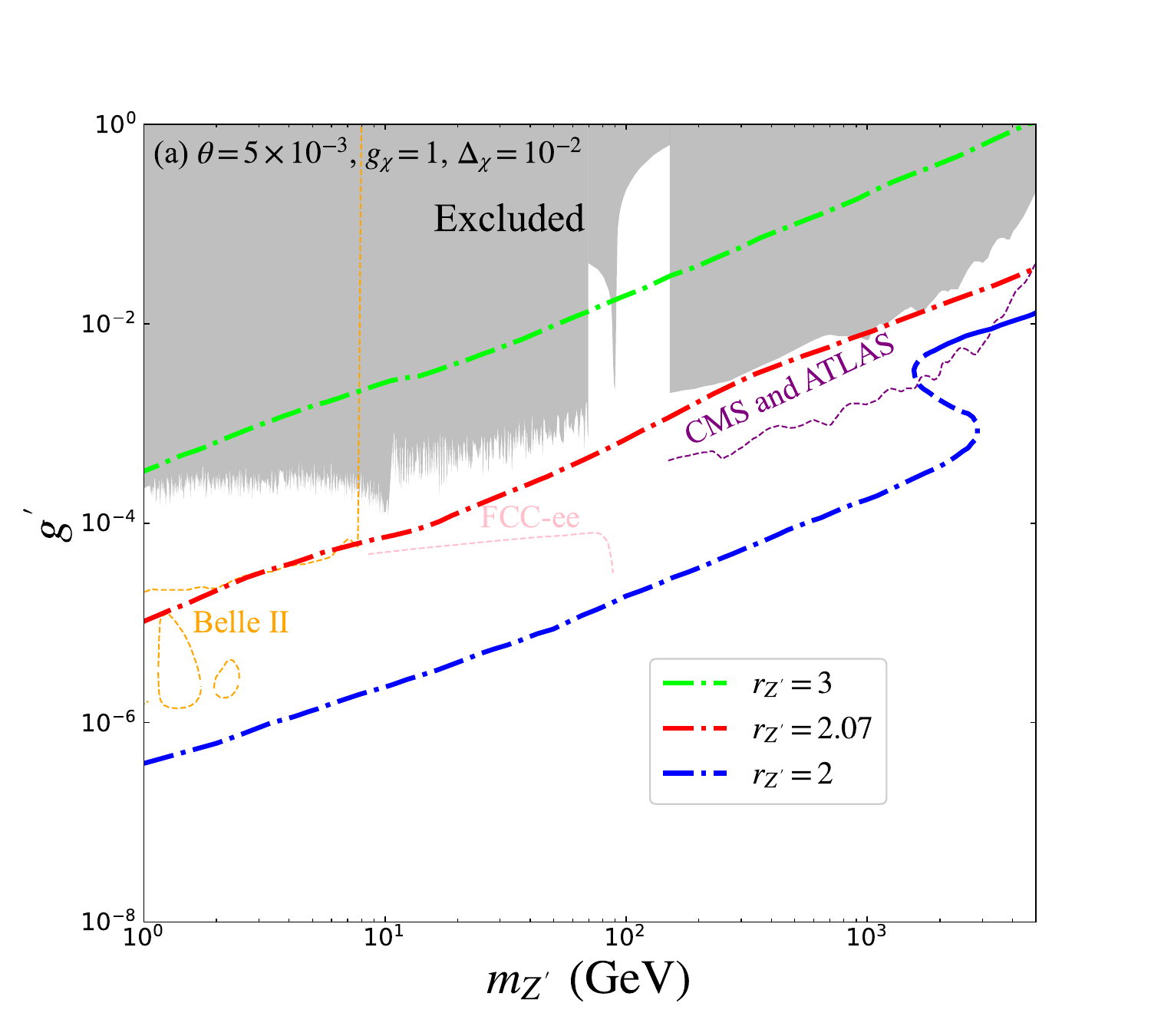}
		\includegraphics[width=0.45\linewidth]{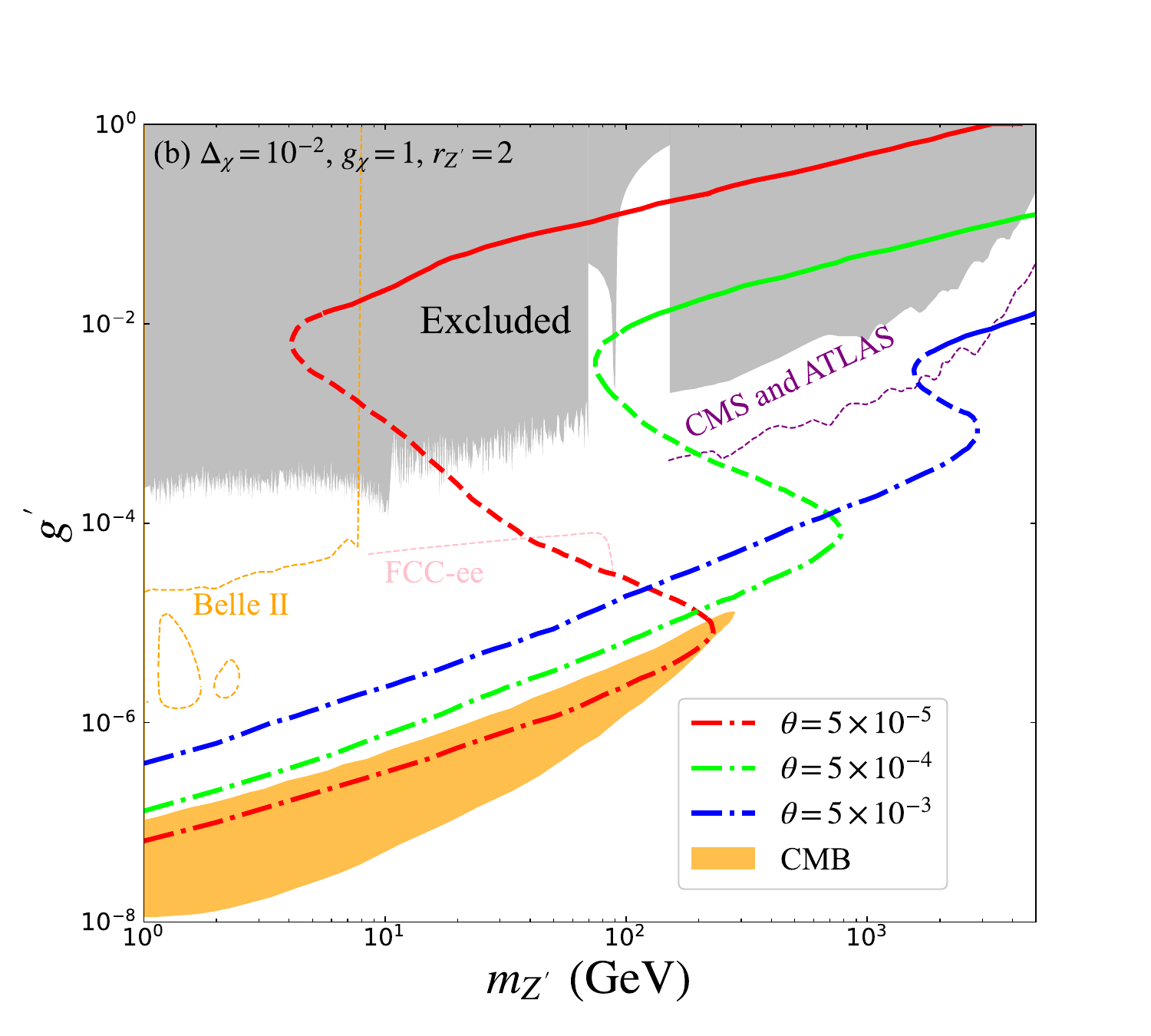}
		\includegraphics[width=0.45\linewidth]{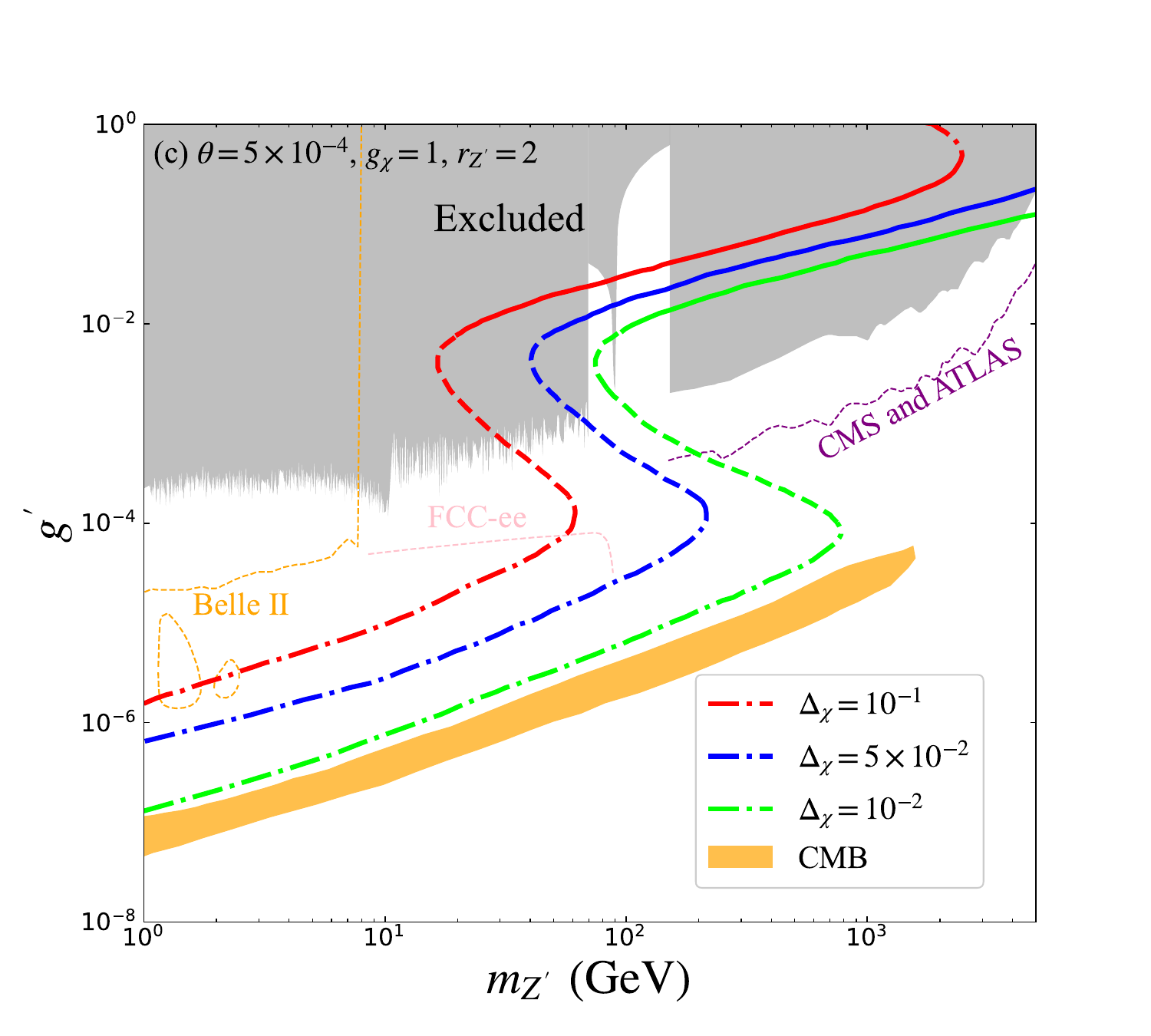}
		\includegraphics[width=0.45\linewidth]{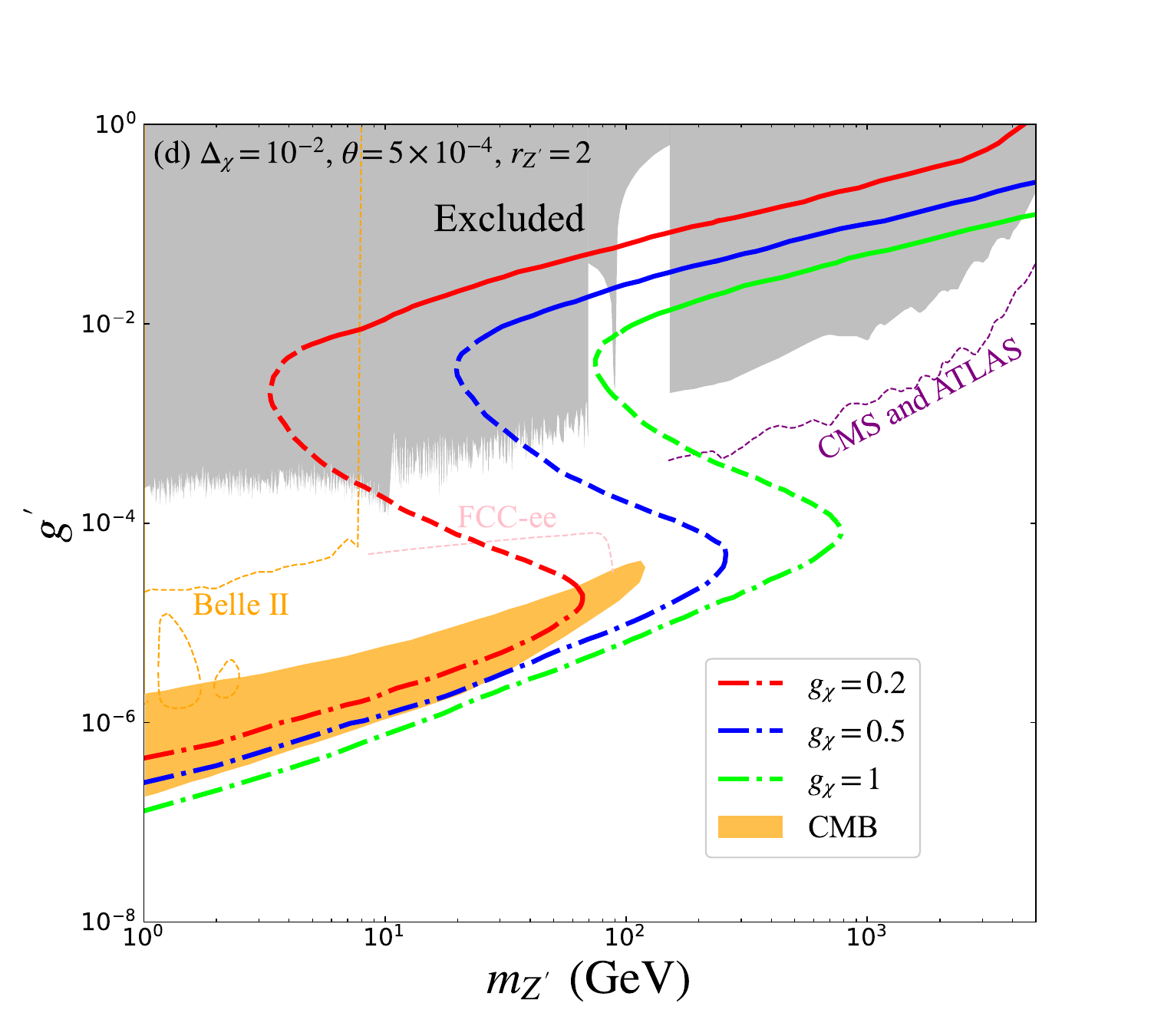}
	\end{center}
	\caption{Constraints on the $m_{Z^\prime}-g^\prime$ parameter space in the resonance scenario. Panels (a)-(d) correspond to different selections with fixed parameters. Each panel features red, green, and blue lines representing three distinct benchmarks, all of which are consistent with DM observation. The solid, dashed, and dot-dashed parts on each line correspond to coscattering, conversion, and coannihilation phases, respectively. The gray shaded area indicates the current exclusion on $Z^\prime$ by various colliders. The future sensitivities of Belle II, FCC-ee, CMS, and ATLAS are illustrated by the orange, pink, and purple dashed lines, respectively. The orange shaded areas appearing in (b), (c) and (d) indicate the promising coannihilation region that can be probed by future CMB on long lived $\chi_2$.
	}
	\label{FIG:fig2}
\end{figure}

The phenomenology in the aspect of $Z^\prime$ mainly arises from searches  at various colliders. At the scale above GeV that we are focusing on, constraints come from the probe of dark photon at current experiments BaBar \cite{BaBar:2014zli,BaBar:2017tiz} and LHCb \cite{LHCb:2017trq,LHCb:2019vmc}, dilepton searches of LEP at $Z$ peak \cite{KA:2023dyz,ALEPH:2013dgf}, as well as of CMS and ATLAS  \cite{CMS:2021ctt,ATLAS:2019erb}. These experimental limits collectively provide stringent constraints on coupling $g^\prime$ as illustrated in the shaded gray area of Figure~\ref{FIG:fig2}, in which the parameter space with $g^\prime\gtrsim3\times10^{-4}$ is excluded. The future Belle II sensitivity for long-lived \cite{Ferber:2022ewf} and invisible \cite{Dolan:2017osp} $Z^\prime$ are colored by an orange dashed line,  which could probe $m_{Z^\prime}\lesssim8$~GeV.  The projection sensitivity of searching for dark photos at FCC-ee  \cite{Karliner:2015tga} with $\sqrt{s}=90$ GeV is colored by a pink dashed line, which is likely to detect $m_{Z^\prime}\in[10,100]~\GeV$ with $g^\prime\sim6\times10^{-5}$.  For $m_{Z^\prime}\gtrsim150~\GeV$ and $g^\prime\gtrsim4.3\times10^{-4}$, the sensitivity of future CMS and ATLAS with a projected 3 $\rm{ab^{-1}}$ luminosity in the search for dileptons improves by approximately one order of magnitude compared to the current results \cite{KA:2023dyz}, which is marked as purple dashed curve.

Figure~\ref{FIG:fig2} illustrates the impact of constraints from $Z^\prime$ on various phases of benchmark cases. In panel (a), the parameters $\theta$, $g_\chi$ and $\Delta_\chi$ are fixed as $5\times10^{-3}$, 1 and $10^{-2}$ respectively, while the changing parameter is $r_{Z^\prime}$. The conventional  non-resonant case of $r_{Z^\prime}=3$ is dominated by coannihilation,  and almost all portions except for $m_{Z^\prime}\sim100$ GeV are excluded by the current constraints. When $r_{Z^\prime}$ decreases to the resonance region $2\lesssim r_{Z^\prime}\lesssim2.07$, as shown by the red and blue curves, it is indeed possible to overcome these limitations. Furthermore, as one approaches the  extreme resonance with $r_{Z'}=2$, the required $g^\prime$ decreases, which facilitates the emergence of conversion and coscattering phases. In the extreme resonance case $r_{Z^\prime}=2$, coannihilation  favors $g^\prime\lesssim10^{-3}$, while the corresponding $m_{Z^\prime}$ does not exceed 2850 GeV. A larger $g^\prime$ naturally leads to a more intense annihilation reaction of $\chi_2$ pairs. When $\chi_2\bar{\chi}_2\to f\bar{f}$ exceeding the conversion process $\chi_i \chi_2\to \chi_j \chi_1$, it belongs to the conversion phase based on the specified criteria, which distributes at $g^\prime\in[10^{-3},5.8\times10^{-3}]$ with $m_{Z^\prime}\in[1560,2850]$ GeV. For $g^\prime\gtrsim 5.8\times 10^{-3}$, the contribution brought by  inelastic process $\chi_2f\to\chi_1f$ is greater than that of $\chi_i \chi_2\to \chi_j \chi_1$, so it becomes the coscattering phase. In this phase, $m_{Z^\prime}$ continuously rises to 5000 GeV as $g^\prime$ increases to 0.013. Obviously, there is an approximate proportional relationship between $g^\prime$ and $m_{Z^\prime}$ in the coannihilation and coscattering phases, whereas the situation in conversion is quite the opposite. In the future, the projected sensitivity of CMS and ATLAS is expected to capture conversion and coscattering within  $1610~\GeV \lesssim m_{Z^\prime}\lesssim3300~\GeV$ when $g^\prime\sim\mathcal{O}(10^{-3})$.  Furthermore, by comparing the parameter $r_{Z^\prime}$, it can be observed that the current stringent constraints compel us to select a more optimistic $Z^\prime$ resonance scenario, which serves as the focal point for the subsequent three panels.

In panel (b) of Figure \ref{FIG:fig2}, we vary $\theta$ to obtain different benchmark cases, while keeping other parameters as $\Delta_\chi=10^{-2},g_\chi=1$ and $r_{Z'}=2$.  As $\theta$ increases from $5\times10^{-5}$ to $5\times10^{-3}$, coannihilation exhibits an upward trend on $g^\prime$, where its maximum achievable value at $(m_{Z^\prime},g^\prime)$ rises from $(225~\GeV,8.6\times10^{-6})$ to $(2850~\GeV,10^{-3})$. In contrast, both conversion and coscattering demonstrate a distinct rightward shift tendency, with the critical points of these two phases increasing from $(6.5~\GeV,1.42\times10^{-2})$ to $(1943~\GeV,5.8\times10^{-3})$.  For coannihilation, the required  $g^\prime$ is not subject to any exclusion at present, and the future Belle II is sensitive to it with $m_{Z^\prime}\sim\mathcal{O}(1)$ GeV when $\theta\sim\mathcal{O}(10^{-2})$.  The conversion of $\theta\lesssim5\times10^{-4}$ is excluded within the range of $g^\prime\gtrsim3\times10^{-4}$ and $m_{Z^\prime}\lesssim100$ GeV. However, under the exclusion limits, there is significant potential for  conversion to be captured by future Belle II and FCC-ee. Furthermore, when $m_{Z^\prime}\gtrsim150$ GeV with $g^\prime\sim\mathcal{O}(10^{-3})$, conversion is also promising with $\theta\in[5\times10^{-4},5\times10^{-3}]$ for CMS and ATLAS.  Due to the large magnitude of $g^\prime$, most coscattering region is not permitted. In the small unrestricted area, we anticipate that there is hope at $m_{Z^\prime}\gtrsim 1000$ GeV when $\theta\sim\mathcal{O}(10^{-3})$ and the narrow region slightly below 100 GeV.

Panel (c) of Figure \ref{FIG:fig2} researches the impact of $\Delta_\chi$, while $\theta=5\times10^{-4},g_\chi=1$ and $r_{Z'}=2$ are fixed.  The three phases shift towards larger $m_{Z^\prime}$ as $\Delta_\chi$ decreases, among them, coannihilation, conversion and coscattering are distributed at $g^\prime\lesssim10^{-4}$,  $10^{-4}\lesssim g^\prime\lesssim10^{-2}$ and $g^\prime\gtrsim 10^{-2}$, respectively.  However, when $\Delta_\chi=0.1$, the excessive mass splitting prevents coscattering from being valid within the range of $m_{Z^\prime}\gtrsim 2450$ GeV. Therefore, returning to small $m_{Z^\prime}$, $g^\prime\gtrsim 0.5$ can only satisfy dark matter observations through coannihilation.  Under current constraints,  it is regrettable that a small portion of conversion with $\Delta_\chi\gtrsim10^{-2}$ and the bulk of coscattering are  restricted at larger $g^\prime$. The remaining small $g^\prime$ that are likely to be detected by  future experiments include coannihilation when $\Delta_\chi\gtrsim10^{-1}$ and conversion with relatively free  $\Delta_\chi$.

In the final panel (d) of Figure \ref{FIG:fig2}, the increase of $g_\chi$ also leads to a  rightward shift of the curve. Similar to that in panel (c), a portion of the conversion will be excluded as long as $g_\chi\lesssim1$, while nearly all coscattering are destined to face exclusion.  When $g_\chi\ll 0.2$, coannihilation with $g^\prime\sim\mathcal{O}(10^{-6})$ has the potential to be identified by Belle II. Meanwhile, conversion is more encouraging, since future experiments hold greater promise for capturing it when $g_\chi\sim\mathcal{O}(0.1)$.   Too large $g_\chi$ may lead to non-perturbative issues,  therefore, such cases with $g\geq1$ are not within the scope of our consideration.

In general, within the resonance scenario, various promising phases exhibit distinct parameter selections. For instance, coannihilation favors $\theta\sim\mathcal{O}(10^{-2})$, $\Delta_\chi\gtrsim10^{-1}$ and $g_\chi\ll0.2$. Conversion with $\theta\lesssim5\times10^{-3}$, $g_\chi\sim\mathcal{O}(0.1)$ and relatively unlimited $\Delta_\chi$ demonstrates significant potential for future experiments. The coscattering situation  is not very optimistic owing to the limitations imposed by current constraints. In addition, when the gauge coupling $g'$ and the mixing angle $\theta$ are  small enough, the dark partner $\chi_2$ becomes long-lived, which can be probed by future CMB experiments. In Figure \ref{FIG:fig2}, we depict the CMB sensitive regions in orange, which will be discussed in detail in Subsection \ref{CD-R}.

\subsection{Phenomenology of $\chi_1$}\label{PC1-R}

\begin{figure}
	\begin{center}
		\includegraphics[width=0.45\linewidth]{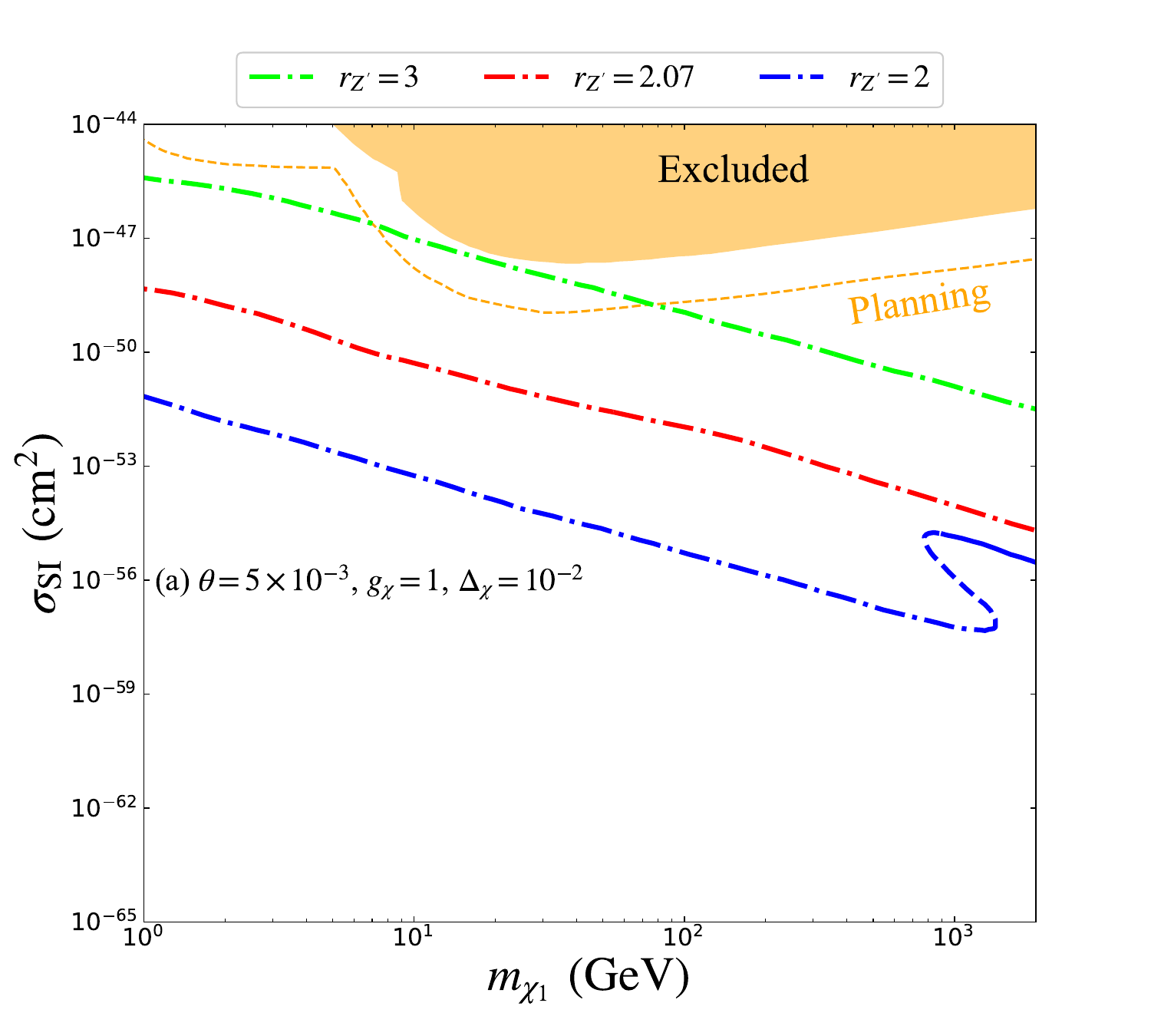}
		\includegraphics[width=0.45\linewidth]{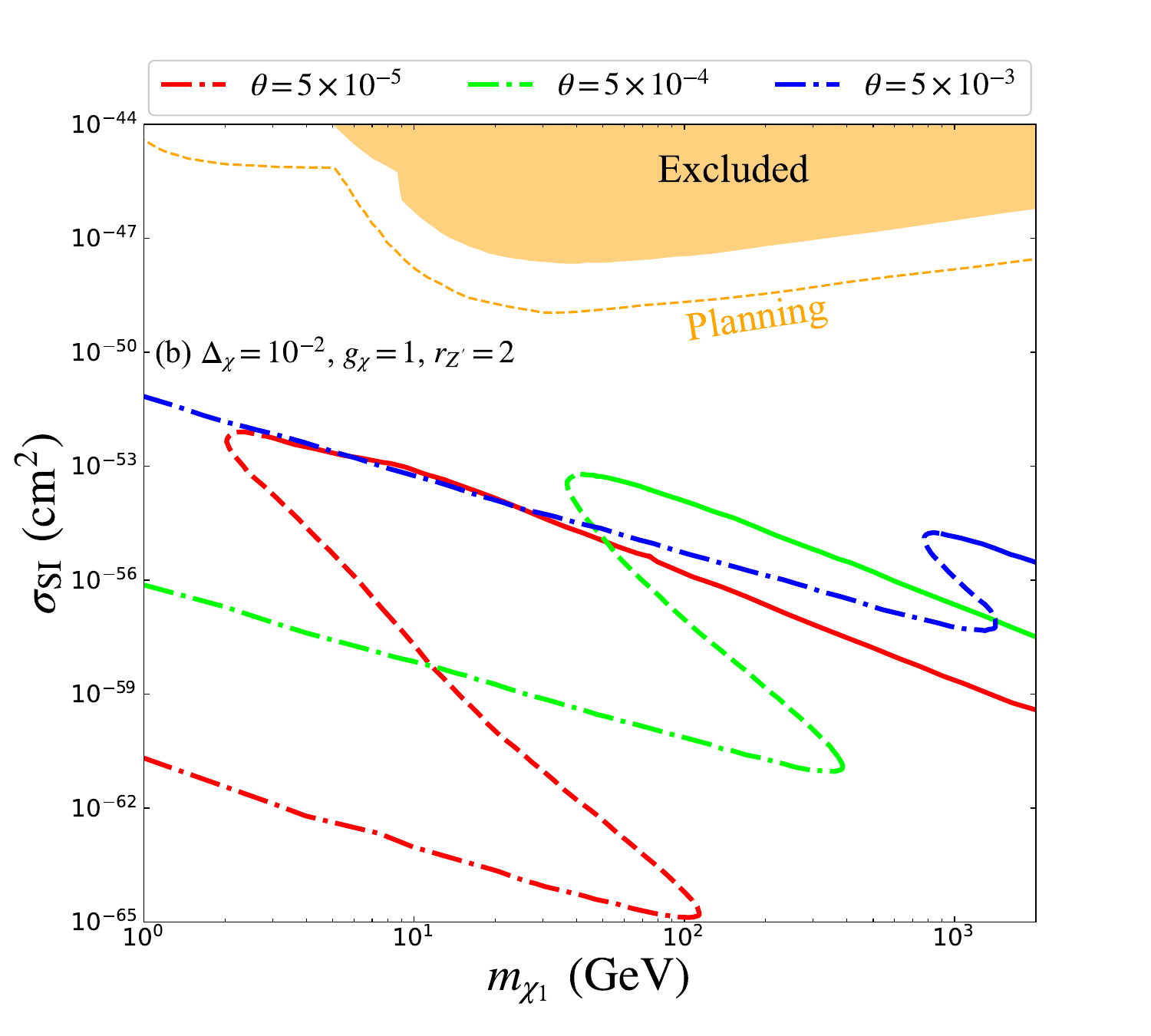}
		\includegraphics[width=0.45\linewidth]{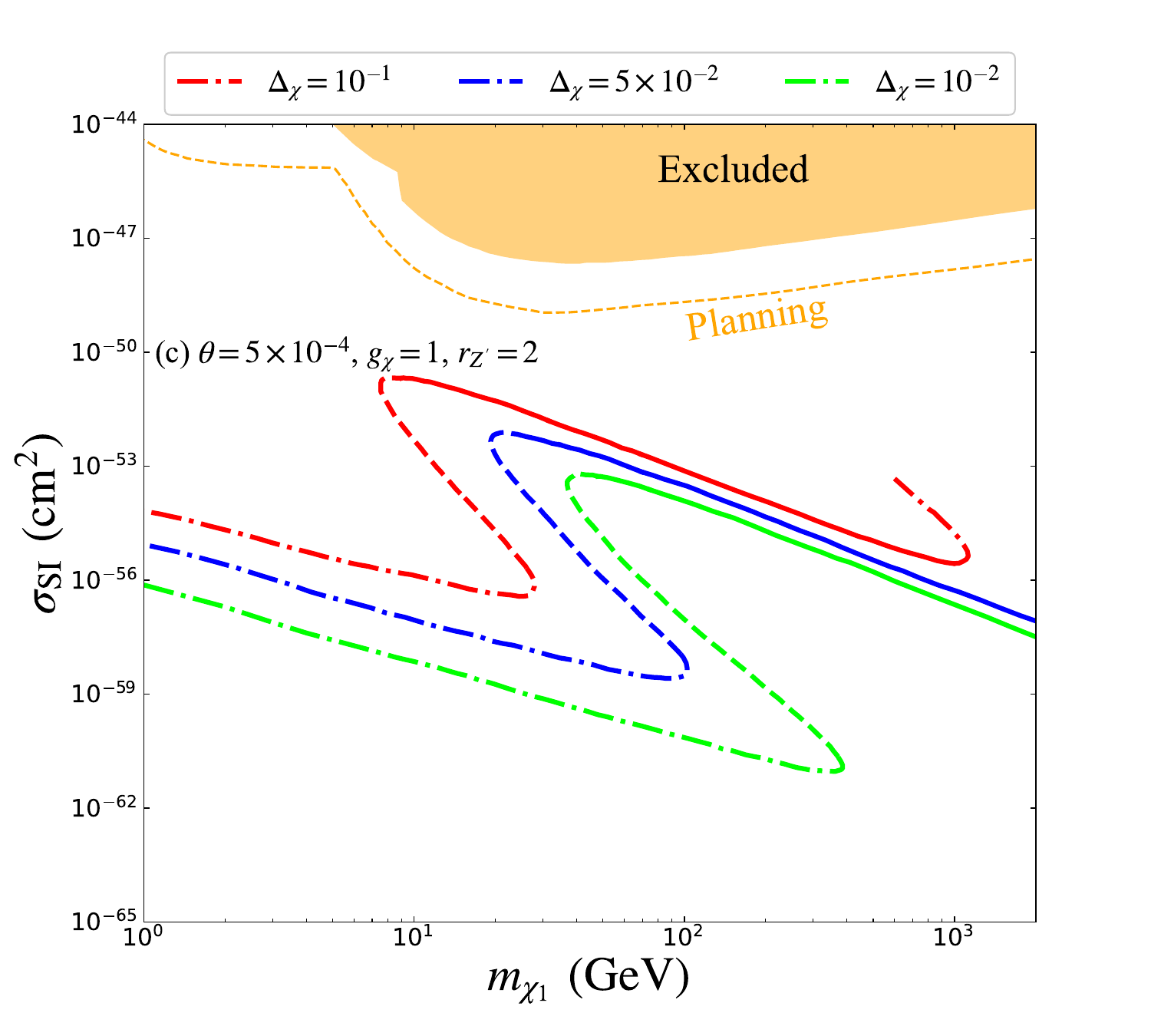}
		\includegraphics[width=0.45\linewidth]{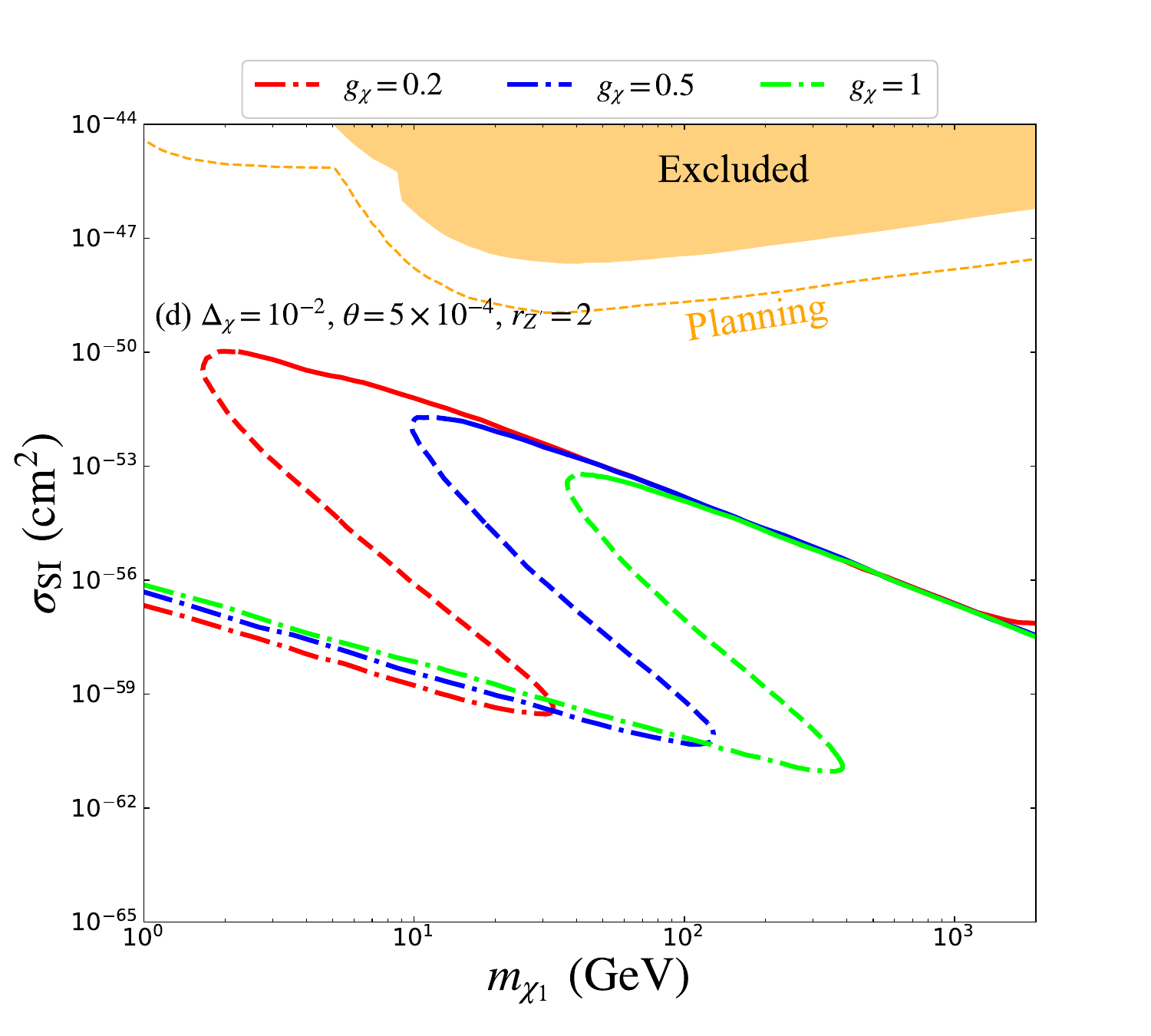}
	\end{center}
	\caption{The constraints of direct detection experiments in the resonance scenario. The fixed parameters in subfigures (a)-(d), as well as the selections of benchmarks in each panel, are consistent with those presented in Figure~\ref{FIG:fig2}. The solid, dashed, and dot-dashed components of each benchmark line still correspond to coscattering, conversion, and coannihilation phases. The orange region and  dashed line represent the results of current and future direct detection experiments.
	}
	\label{FIG:fig3}
\end{figure}

For DM above the GeV scale, the spin-independent scattering cross section with nucleons is strictly constrained by current direct detection experiments, such as DarkSide-50 \cite{DarkSide-50:2023fcw}, PandaX-4T \cite{PandaX:2024qfu}, and LZ \cite{LZ:2024zvo}. The combined exclusion region is represented by the orange shading region in Figure~\ref{FIG:fig3}, in which the strongest constraint is located at $m_{\chi_1}\sim30$ GeV with $\sigma_{\rm SI}\sim2\times10^{-48}\cm^2$. The future sensitivity represented by the orange dashed curve is provided by DarkSide-LowMass \cite{GlobalArgonDarkMatter:2022ppc}, SuperCDMS \cite{SuperCDMS:2016wui}, and LZ \cite{LZ:2015kxe},  which is nearly an order of magnitude lower than the current results.

In this model, the spin-independent cross section can be calculated by
\begin{eqnarray}\label{Eqn:dd}
	\sigma_{\rm SI}=\frac{m_{\chi_1}^2 m_n^2 \sin^4 \theta~g_\chi^2{g^\prime}^2}{\pi m_{Z^\prime}^4(m_{\chi_1}+m_n)^2},
\end{eqnarray}
where the mass of nucleons $m_n\simeq0.939$ GeV. Compared with the traditional $Z'$ portal DM, the cross section is further suppressed by the small mixing angle $\theta$. The predictions are shown in Figure~\ref{FIG:fig3} as red, blue, and green lines for the benchmark cases.

In panel (a) of Figure~\ref{FIG:fig3}, although the future outcomes are expected to capture $m_{\chi_1}\sim \mathcal{O}(10)~\GeV$ in the non-resonant case, the constraints on $Z'$ from the collider already disfavor such a region.  In the resonance cases, the $\sigma_{\rm SI}$ for three phases decreases as $m_{\chi_1}$ increases. Being suppressed by $\theta$, $\sigma_{\rm SI}$ reaches its maximum $6.8\times10^{-52}\cm^2$ for $r_{Z'}=2$ at 1 GeV, which is nearly six orders of magnitude lower than future sensitivity. Therefore, such a resonance case is challenging for the direct detection experiments.

In the subsequent three panels (b)-(d) of Figure~\ref{FIG:fig3}, all benchmarks predict $\sigma_{\rm SI}\lesssim10^{-50} \cm^2$. None of them is within the reach of future experiments. However, it is evident that adjusting certain parameters can alter this situation. For instance, when $\theta\gg5\times10^{-3}$ in panel (b), it can be inferred that coannihilation is completely dominant based on the trend in changing $\theta$. As $\theta$ increases,  $m_{\chi_1}$ may be detectable over a wide range. 

When the mass splitting $\delta=m_{\chi_2} - m_{\chi_1}$ is less than $\mathcal{O}(100)$ keV, the inelastic scattering $\chi_1 n\to \chi_2n$ in principle can yield observable signature \cite{Tucker-Smith:2001myb}. However, such an inelastic scattering cross section is also suppressed by the small mixing angle $\theta$, which makes it unpromising \cite{Filimonova:2022pkj}. In this paper, we consider $\delta= m_{\chi_1} \Delta_\chi\geq 10^{-3} $ GeV, so it is far above the experimental sensitive region \cite{CDEX:2025mgp}. 

\begin{figure}
	\begin{center}
		\includegraphics[width=0.45\linewidth]{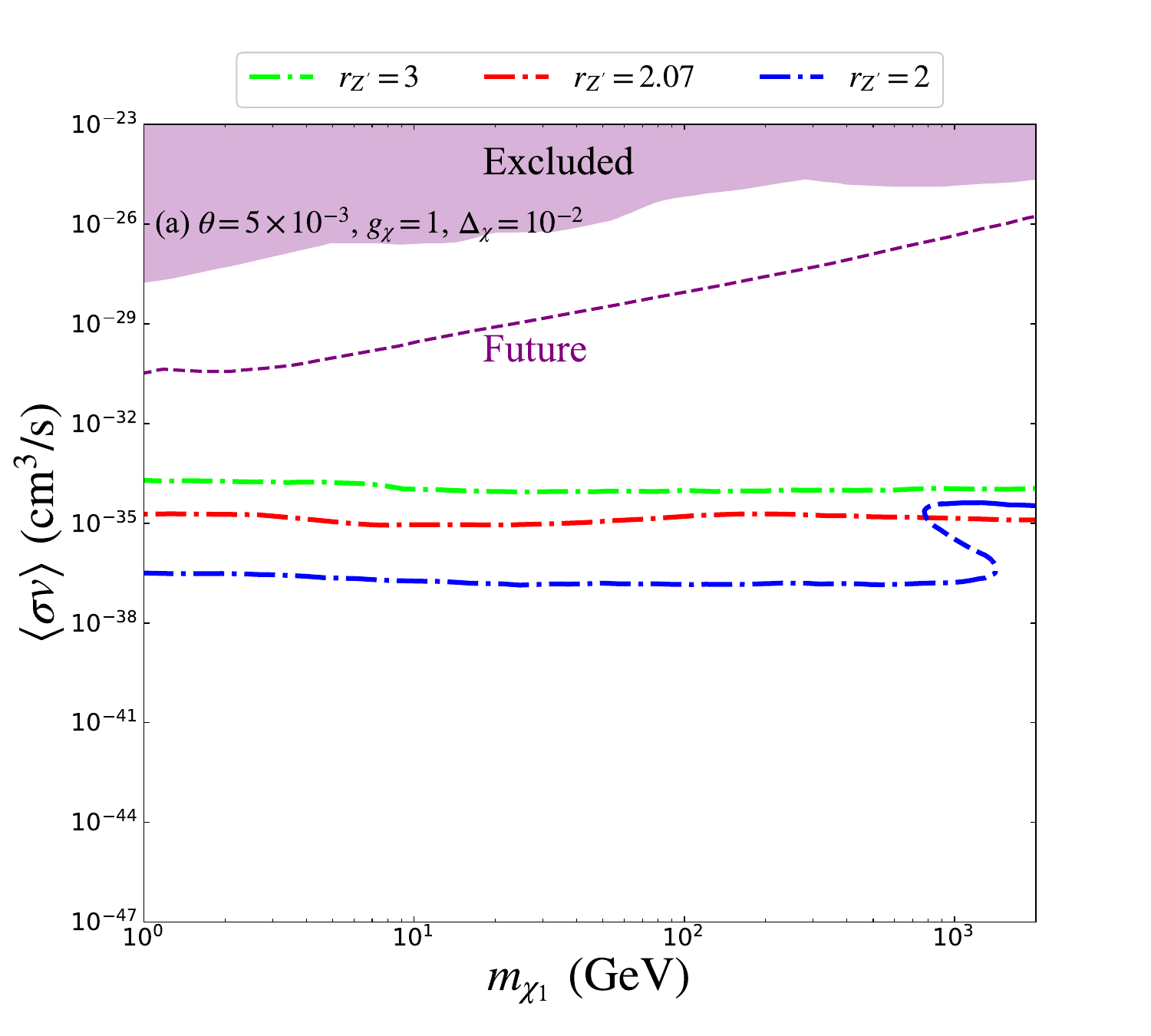}
		\includegraphics[width=0.45\linewidth]{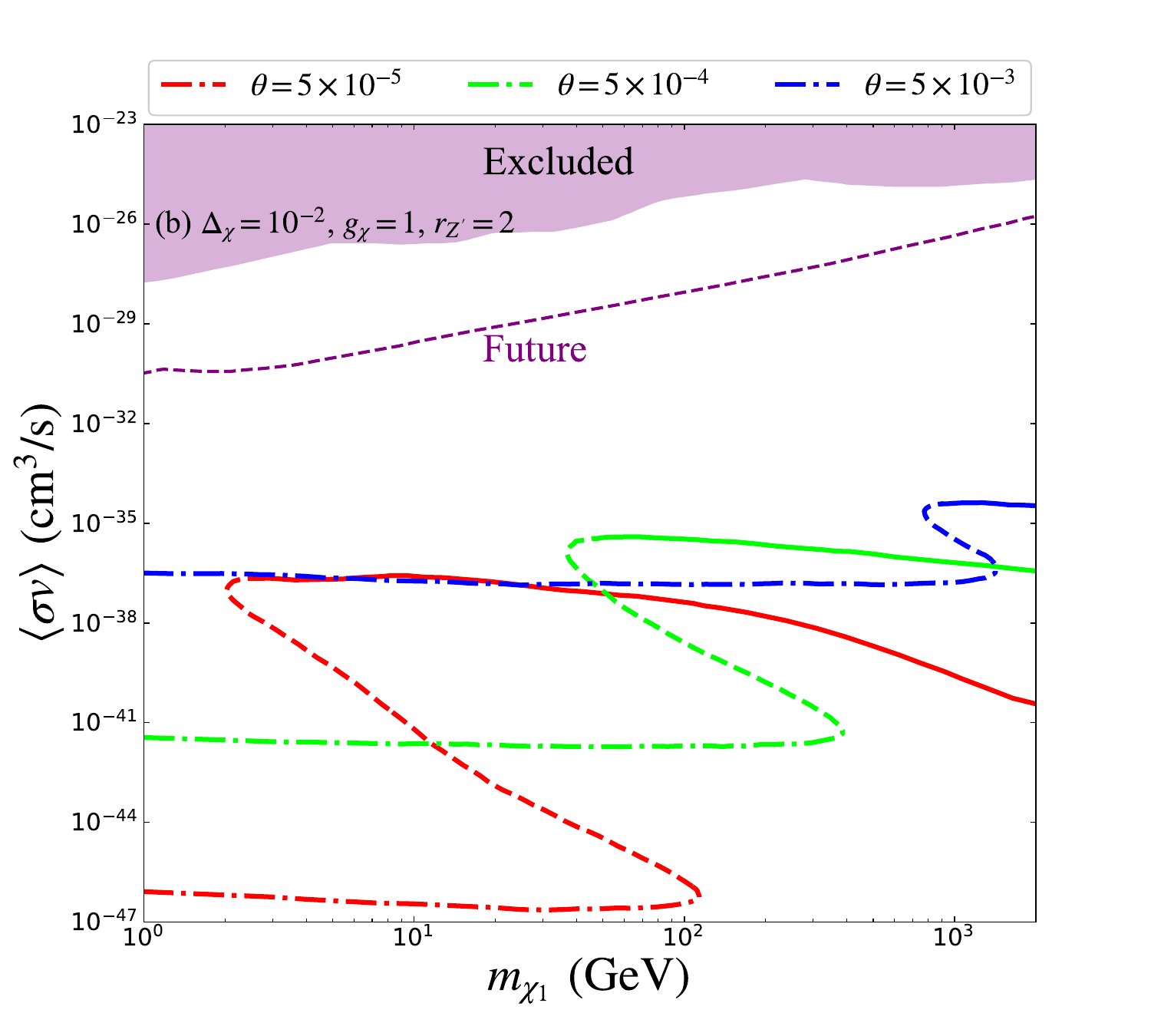}
		\includegraphics[width=0.45\linewidth]{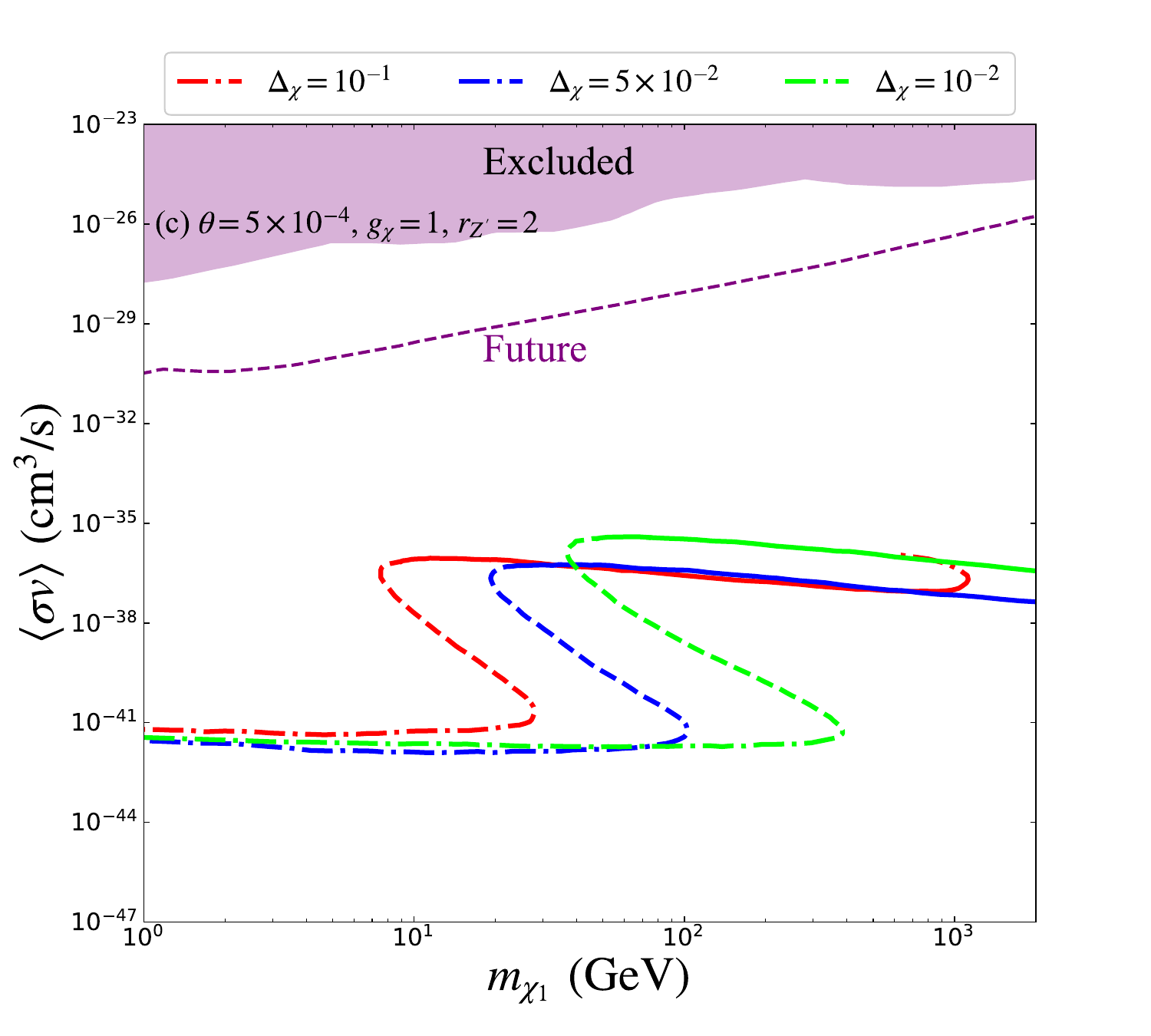}
		\includegraphics[width=0.45\linewidth]{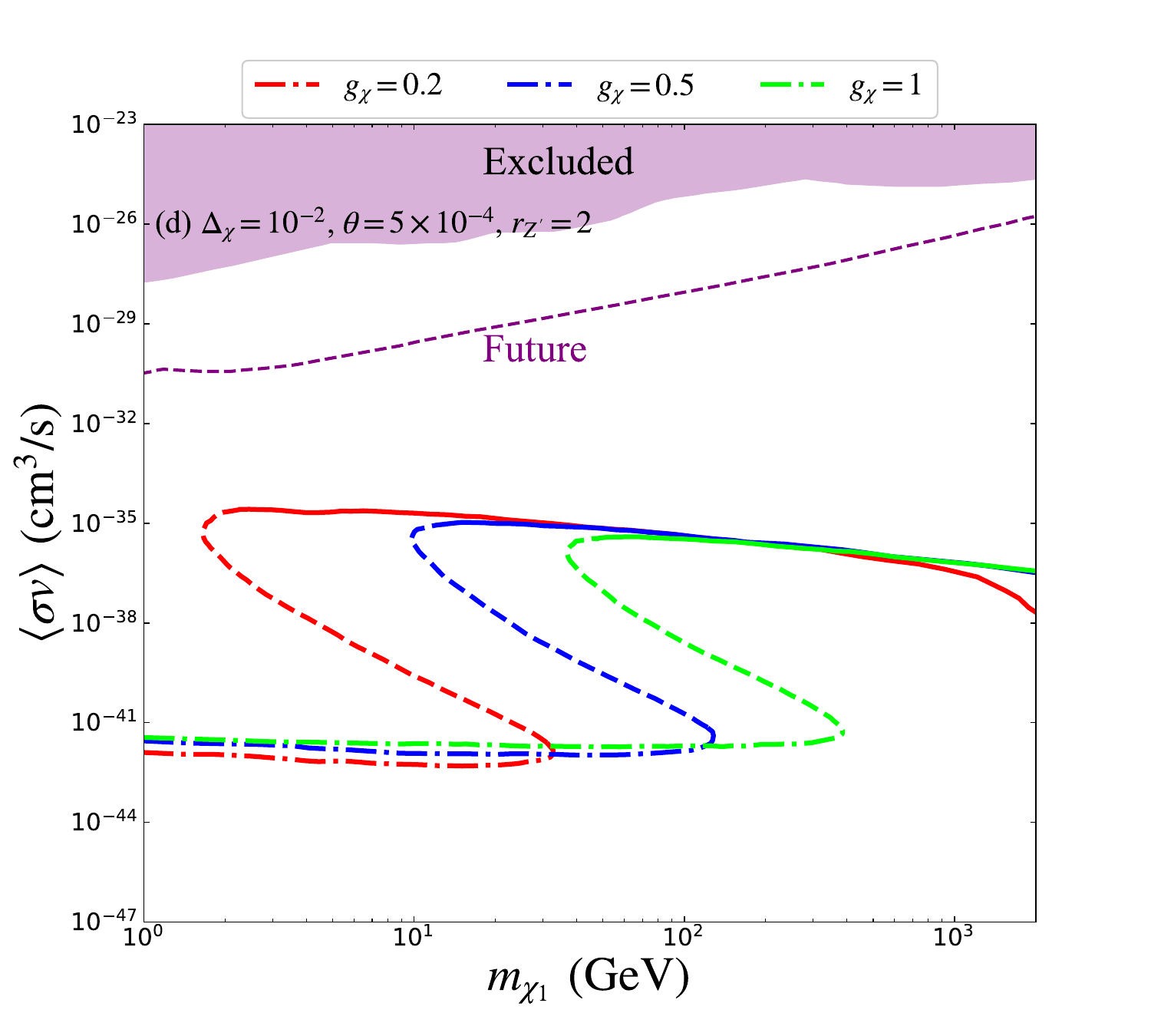}
	\end{center}
	\caption{The constraints of indirect detection experiments in the resonance scenario. The legends of panels (a)-(d) and markers of benchmark lines  are consistent with those in Figure~\ref{FIG:fig2}. The purple region is  not permitted by the existing constraints of indirect detection experiments, meanwhile the projected sensitivity of future experiments is represented as a purple dashed line.
	}
	\label{FIG:fig4}
\end{figure}

Within the resonance scenario, DM pairs could annihilate into SM fermions $f\bar{f}$ via $Z^\prime$ mediator. The present annihilation cross section $\langle \sigma v\rangle$ is constrained by outcomes from indirect detection experiments. As the leptonic final sate  $\ell^+\ell^-$ is the dominant annihilation channel, we take the limits of the $e^+e^-$ final state as an example, which is illustrated in Figure~\ref{FIG:fig4}. The existing constraints with $m_{\chi_1}\lesssim5$ GeV come from experiments involving XMM-NEWTON $X$-rays \cite{Cirelli:2023tnx} and CMB (s-wave) \cite{Lopez-Honorez:2013cua,Slatyer:2015jla} observations.  While the results for larger $m_{\chi_1}$ are taken from literatures \cite{Leane:2018kjk,Dutta:2022wdi}, which are the convolutions of the bounds from AMS positron \cite{AMS:2014xys,AMS:2019rhg}, Fermi-LAT dwarfs \cite{Fermi-LAT:2016uux} and
H.E.S.S. GC observations \cite{HESS:2016mib,HESS:2022ygk}. These constraints collectively exclude the purple shaded area with $\langle \sigma v\rangle\gtrsim10^{-28}~\cm^3/s$. The purple dashed line represents the sensitivities of the future MeV telescopes AMEGO \cite{AMEGO:2019gny,Kierans:2020otl,Caputo:2022xpx},
E-ASTROGAM \cite{e-ASTROGAM:2016bph,e-ASTROGAM:2017pxr} and MAST \cite{Dzhatdoev:2019kay} in probing weak-scale DM, which is derived from \cite{Cirelli:2025qxx}. The future limit is roughly two orders of magnitude lower than the current one. The maximum detection capability is observed at the GeV scale with $\langle \sigma v\rangle\sim 10^{-31}~\cm^3/\text{s}$.

In this model, $\langle \sigma v\rangle$ could be numerically calculated through \cite{Mohapatra:2019ysk}:
\begin{eqnarray}\label{Eqn:id-R}
	\langle \sigma v\rangle_{\chi_1\bar{\chi}_1\to f\bar{f}}\simeq\frac{\sin^4\theta~g_\chi^2{g^\prime}^2}{2 \pi} \sum_f N_c^f Q_f^2 \frac{2m_{\chi_1}^2+m_f^2}{(4m_{\chi_1}^2-m_{Z^\prime}^2)^2+m_{Z^\prime}^2 \tilde{\Gamma}_{Z^\prime}^2} \sqrt{1-\frac{m_f^2}{m_{\chi_1}^2}},
\end{eqnarray}
where  $\tilde{\Gamma}_{Z^\prime}$ represents the total decay width of $Z^\prime$, which  can be decomposed into
\begin{eqnarray}
\tilde{\Gamma}_{Z^\prime\to f\bar{f}}& = &\sum_f \frac{N_c^f {g^\prime}^2 Q_f^2 m_{Z^\prime}}{12\pi}\left(1+\frac{2 m_f^2}{m_{Z^\prime}^2}\right)\sqrt{1-\frac{4 m_f^2}{m_{Z^\prime}^2}}, 
\end{eqnarray}
\begin{eqnarray}
\tilde{\Gamma}_{Z^\prime\to \chi_1\bar{\chi}_1}&=&\frac{{g_\chi}^2 \sin^4\theta~m_{Z^\prime}}{12\pi}\left(1+\frac{2 m_{\chi_1}^2}{m_{Z^\prime}^2}\right)\sqrt{1-\frac{4 m_{\chi_1}^2}{m_{Z^\prime}^2}},
\end{eqnarray}
\begin{eqnarray}
\tilde{\Gamma}_{Z^\prime\to \chi_2\bar{\chi}_2}&=&\frac{{g_\chi}^2 \cos^4\theta~m_{Z^\prime}}{12\pi}\left(1+\frac{2 m_{\chi_2}^2}{m_{Z^\prime}^2}\right)\sqrt{1-\frac{4 m_{\chi_2}^2}{m_{Z^\prime}^2}},\\
\tilde{\Gamma}_{Z^\prime\to \chi_1\chi_2}&=&\frac{g_\chi^2\sin^2 2\theta}{48\pi m_{Z^\prime}^5}\left((m_{\chi_2}^2-m_{\chi_1}^2)^2+m_{Z^\prime}^2(m_{\chi_1}^2+m_{\chi_2}^2-6m_{\chi_1}m_{\chi_2}-2m_{Z^\prime}^2)\right)\\ \nonumber
&\times&\sqrt{m_{\chi_1}^2(m_{\chi_1}^2-2m_{\chi_2}^2-2m_{Z^\prime}^2)+(m_{Z^\prime}^2-m_{\chi_2}^2)^2}.
\end{eqnarray}

The present $\langle \sigma v\rangle$ of DM annihilation  is represented by the red, blue, and green curves in Figure~\ref{FIG:fig4}.  Similar to the DM-nucleon scattering cross-section in direct detection, the current $\langle \sigma v\rangle$ is also suppressed by the mixing angle $\theta$. Consequently, all benchmarks exhibit $\langle \sigma v\rangle\lesssim10^{-34}~\rm cm^3/s$, which represents a detection range far beyond what future experiments can achieve. But then again, this relationship with $\theta$ implies that as long as $\theta\gg5\times10^{-3}$, coannihilation at the GeV scale firstly remains promising.

\subsection{Phenomenology of $\chi_2$}\label{PC2-R}

\begin{figure}
	\begin{center}
		\includegraphics[width=0.45\linewidth]{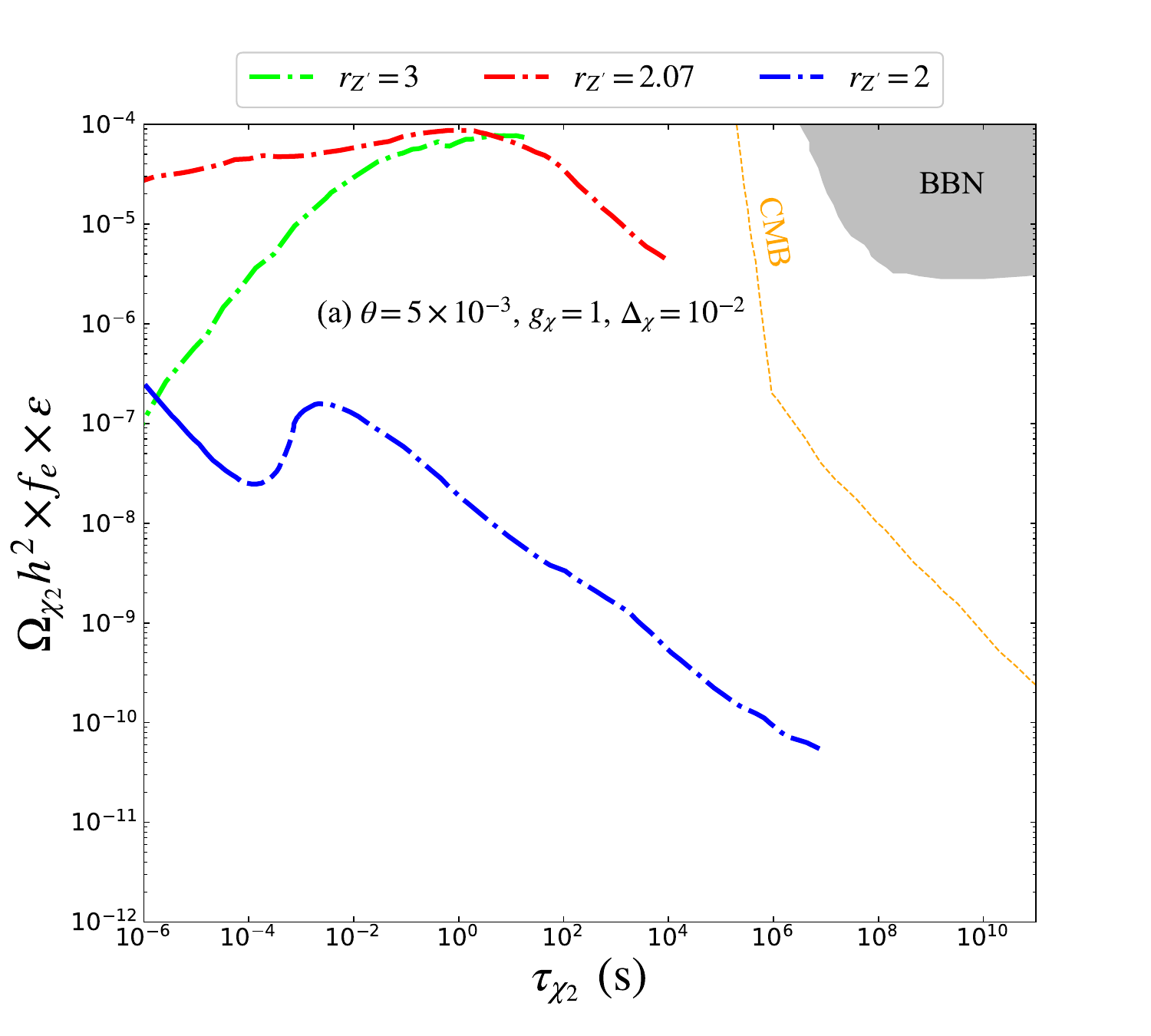}
		\includegraphics[width=0.45\linewidth]{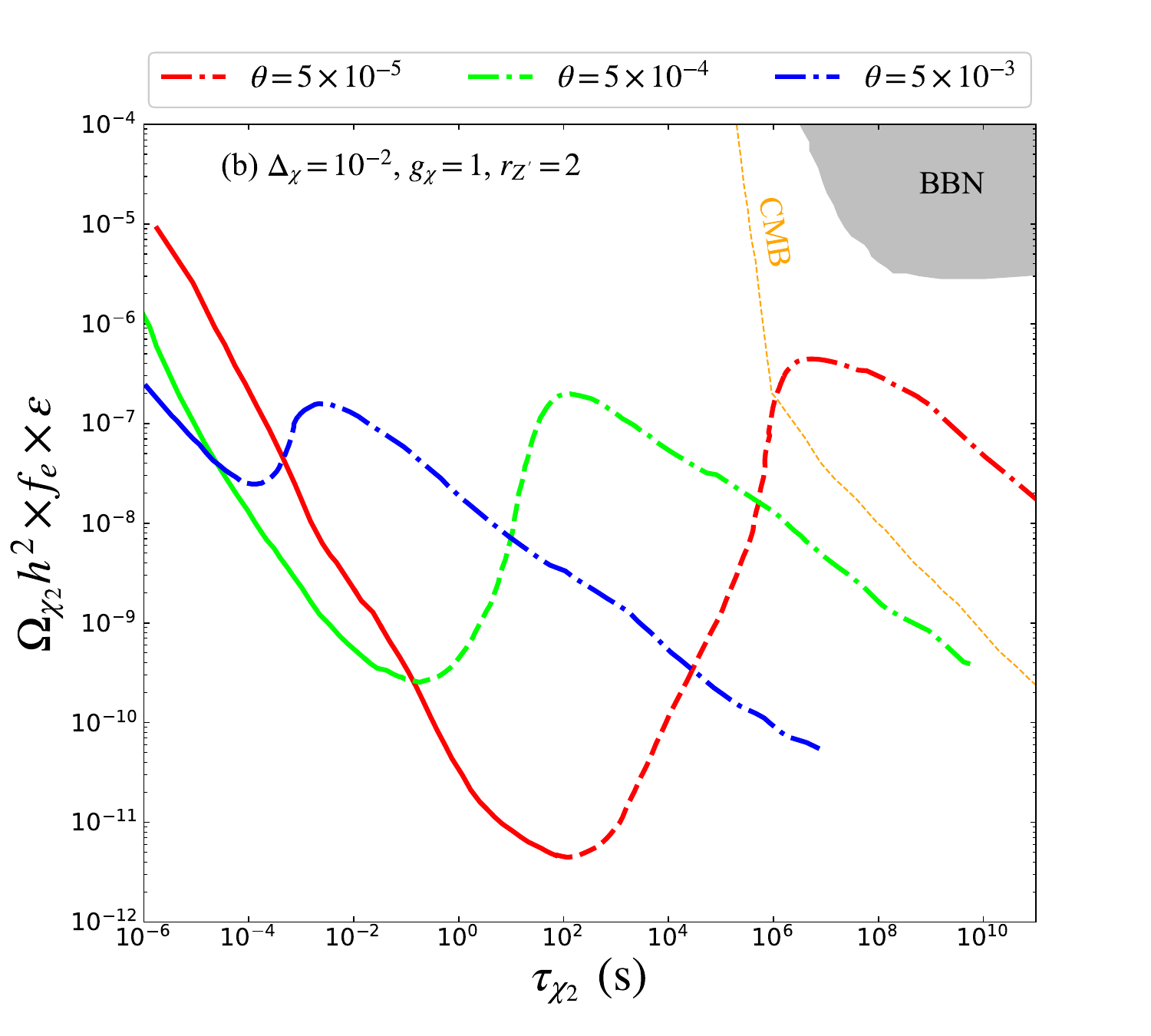}
		\includegraphics[width=0.45\linewidth]{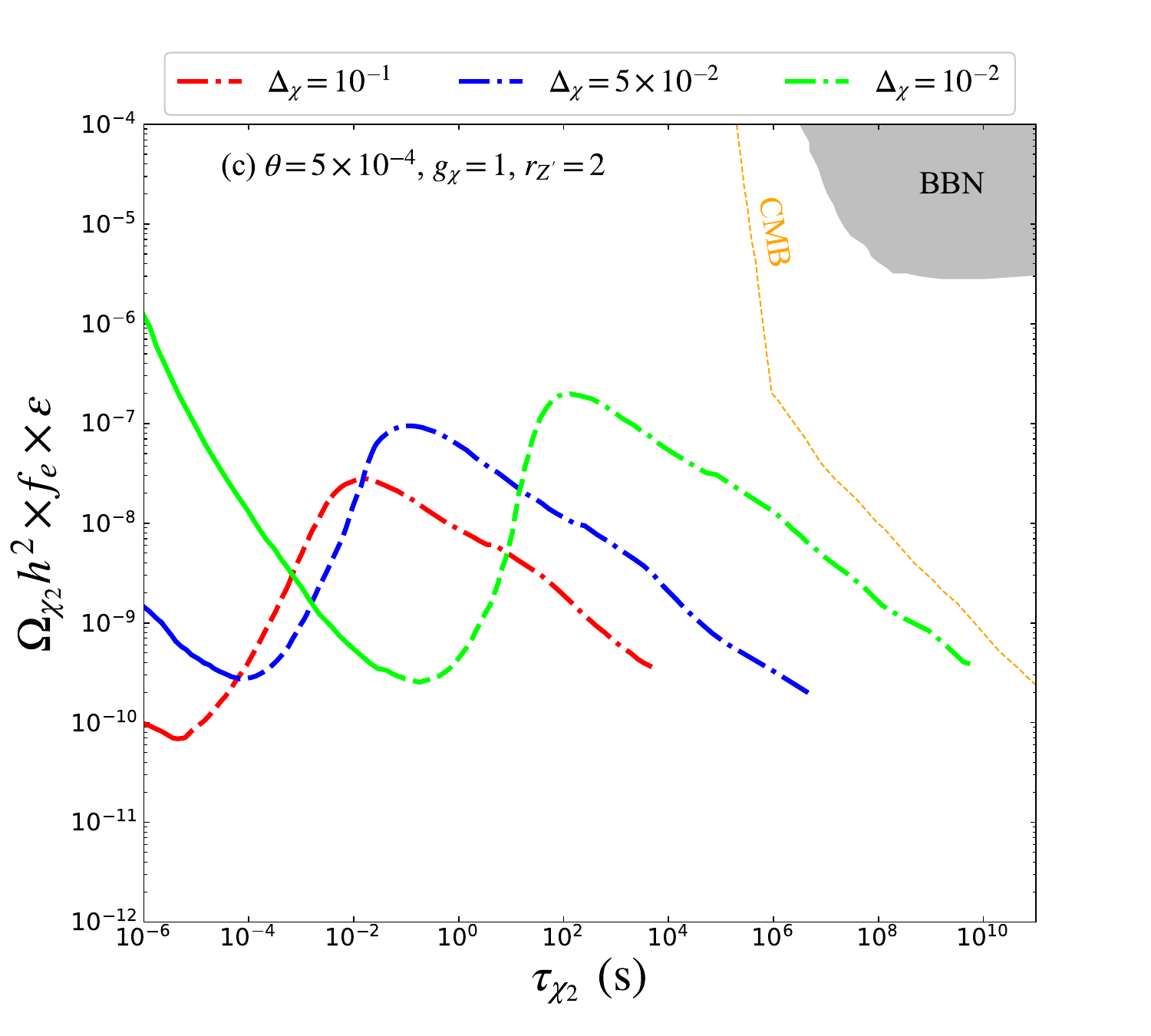}
		\includegraphics[width=0.45\linewidth]{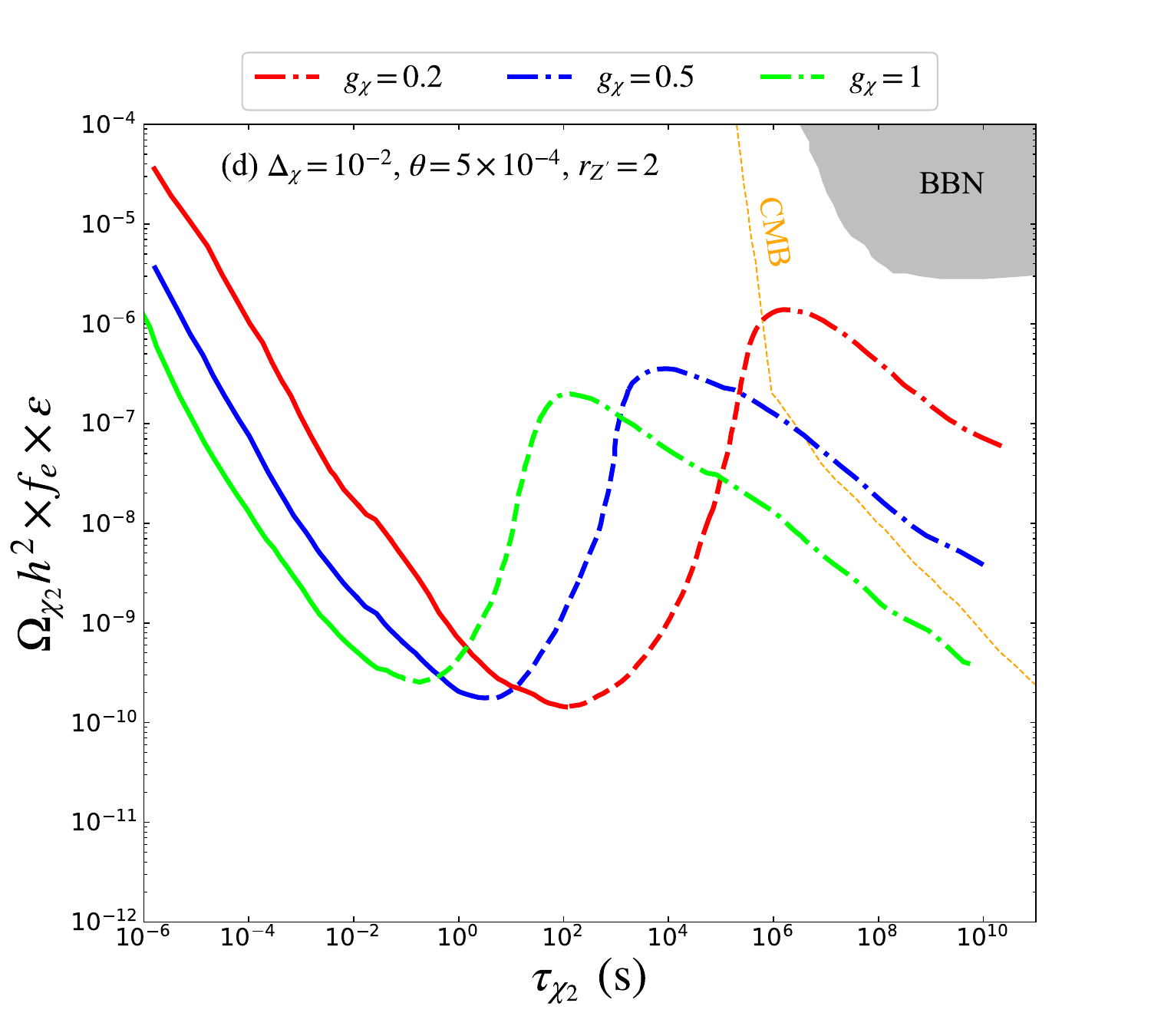}
	\end{center}
	\caption{Cosmological constraints on long-lived $\chi_2$ in the resonance scenario. The horizontal axis represents the lifespan of $\chi_2$, while the vertical axis indicates the relative relic density. Here, $f_e$  is the branching ratio of $\chi_2$ decay to $e^+e^-$ final state,  and $\epsilon=(m_{\chi_2}^2-m_{\chi_1}^2)/2m_{\chi_2}^2$ is the fraction of the energy of $\chi_2$ that has been transferred to electron. The legends of panels (a)-(d) and markers of benchmark lines are consistent with those in Figure~\ref{FIG:fig2}. The gray area is excluded by BBN, and the orange dashed line represents the future CMB results.
	}
	\label{FIG:fig5}
\end{figure}

In this scenario, the phenomena induced by the three-body decay $\chi_2\to\chi_1f\bar{f}$ primarily arises from two aspects. Firstly, it is the collider  signature. As long as the decay length of $\chi_2$ does not exceed the detection range of  colliders \cite{Berlin:2018jbm}, the signals  of prompt or displaced  have the potential to be observed.  For instance,  $pp\to Z^\prime\to \chi_1+\chi_2 \to \chi_1 \chi_1 \ell^+\ell^-$  at LHC \cite{CMS:2023bay}, and $e^+e^-\to Z^\prime+Z/\gamma$, $Z^\prime\to\chi_1+\chi_2\to\chi_1 \chi_1 \ell^+\ell^-/\chi_1 \chi_1 jj$ at lepton collider \cite{Liu:2025abt}. Certainly, since vertex $Z^\prime\chi_1\chi_2$ is suppressed in this model, multi-lepton or two  displaced vertices induced by  $Z^\prime\to\chi_2\bar{\chi}_2$ are more appropriate once kinematically allowed.  Furthermore,  when $\chi_2$ decays outside the detector or the decay final states are too soft to be detected, $\chi_2$ becomes invisible. Then, the expectation shifts to monophoton signal $e^+e^-\to Z'+\gamma \to \chi_2\bar{\chi}_2  \gamma$ \cite{BaBar:2017tiz}.

According to the cosmological constraints on $\chi_2$ in Figure~\ref{FIG:fig5}  , the decay length $d_{\chi_2}=c\tau_{\chi_2}$ can be roughly estimated via  the horizontal axis $\tau_{\chi_2}$. It is evident that searching for displaced vertices signals is more suitable within the coscattering regime, the other two phases can  study the monophoton signals because of too long-lived $\chi_2$. In the extreme resonance scenario with $r_{Z'}=2$, the decay $Z^\prime\to\chi_2\bar{\chi}_2$ is forbidden, thus the missing energy comes from the decay $Z'\to \chi_1\chi_2$. But the decay width of $Z'\to \chi_1\chi_2$ is also suppressed by the small mixing $\theta$ and phase space, so the collider signature for the case with $r_{Z'}=2$ is not promising.

For the long-lived dark partner, the additional energetic injection of  $\chi_2\to\chi_1f\bar{f}$ will affect the big bang nucleosynthesis (BBN) predictions and the cosmic microwave background (CMB) anisotropy power spectra. As illustrated in Figure~\ref{FIG:fig5}, since the area with  larger $\tau_{\chi_2}$  where BBN and CMB take effect is primarily distributed by coannihilation, the minor mass splitting results in $\chi_2\to \chi_1 e^+e^-$ being the dominant process.  Thus we consider the limits of the electron final state as an example. The gray area represents the current constraints imposed by BBN on the $e^+e^-$ final state \cite{Kawasaki:2017bqm}, which excludes regions for $\tau_{\chi_2}\gtrsim3\times10^6$ s with  relative relic density $\Omega_{\chi_2}h^2\times f_{e}\times\varepsilon\gtrsim3\times10^{-6}$. The orange dashed line stands for the upcoming CMB results of purely electromagnetic decay \cite{Lucca:2019rxf},  which is capable of detecting regions for $\tau_{\chi_2}\gtrsim2\times10^5$ s with  $\Omega_{\chi_2}h^2\times f_{e}\times\varepsilon\gtrsim2\times10^{-10}$. 

In Figure~\ref{FIG:fig5}, the maximum $\tau_{\chi_2}$ of all benchmarks is cut off at $m_{\chi_1}=1$ GeV. The detectability for  $m_{\chi_1}\leq1$ GeV can be  seen in previous study \cite{Zhang:2024sox}.  It is evident that the current BBN constraints \cite{Kawasaki:2017bqm} do not pose any threat to benchmarks due to the small $\tau_{\chi_2}$ and $\Omega_{\chi_2}h^2\times f_{e}\times\varepsilon$.  Coannihilation  with $\theta\lesssim5\times10^{-4}$, $\Delta_\chi\lesssim10^{-2}$ and $g_\chi\lesssim1$ is likely to be detected by future CMB \cite{Lucca:2019rxf}, which corresponds to $m_{\chi_1}$ typically below TeV. Furthermore, it is noteworthy that the evolutionary trends of coannihilation in the resonance and non-resonance conditions depicted in panel (a) of Figure~\ref{FIG:fig5}  are completely opposite.  This phenomenon arises because the conversion reaction of $\chi_i \chi_2\to \chi_j \chi_1$ at the resonance is significantly more intense, resulting in a continued decline in the abundance of $\chi_2$ after freezing-out, until this reaction is depleted. In more extreme cases of $\theta\lesssim5\times10^{-5}$ and $g_\chi\lesssim0.2$, conversion with $m_{\chi_1}\lesssim100$ GeV also become increasingly optimistic for future CMB.

Additionally, $\chi_2\to\chi_1\nu\bar{\nu}$ also exhibits a considerable branching ratio. For $\tau_{\chi_2}\gtrsim\mathcal{O}(10^{5})$ s with $m_{\chi_1}$ below 100 GeV,  the CMB observation set no limit due to too small fraction of energy injection into SM plasma from neutrinos ~\cite{Hambye:2021moy}. The neutrinos from delayed decay also contribute to the effective number of relativistic neutrino species $N_{\rm eff}$ ~\cite{Liu:2022cct}. The benchmarks  predict  $(f_\nu\epsilon~\Omega_{\chi_2}/\Omega_{\chi_1})^2\tau_{\chi_2}\lesssim\mathcal{O}(0.1)$ s, which is far below the current Planck limit $(f_\nu\epsilon~\Omega_{\chi_2}/\Omega_{\chi_1})^2\tau_{\chi_2}\lesssim5\times10^{9}$ s ~\cite{Hambye:2021moy}.

\subsection{Combination and Discussion}\label{CD-R}

In the resonance scenario,  the direct and indirect detection experiments can hardly detect the benchmarks. A relatively large gauge coupling $g'$ could induce a detectable collider signature of $Z'$. On the other hand, a sufficiently small $g'$ leads to long-lived $\chi_2$, which could affect CMB observables due to delayed decay $\chi_2\to\chi_1 f\bar{f}$. As shown in Figure~\ref{FIG:fig2} and Figure~\ref{FIG:fig5}, the future CMB experiment is primarily sensitive to coannihilation with $g'\lesssim10^{-4}$.  Therefore, the collider search for $Z^\prime$ and the CMB  observables provide complementary pathways to probe such a scenario. 

In Figure~\ref{FIG:fig2}, we depict the combined results. For case (a) with $\theta=5\times10^{-3}, g_\chi=1$, and $\Delta_{\chi}=10^{-2}$, the resonance benchmark line with $r_{Z'}=2$ is already the lowest. We report that the insufficient $\tau_{\chi_2}$ at  coannihilation fails to meet the sensitivity of CMB as shown in panel (a) of Figure~\ref{FIG:fig5}.  In contrast, coannihilation located at $m_{Z^\prime}\lesssim\mathcal{O}(100)$ GeV and $g^\prime\lesssim\mathcal{O}(10^{-6})$ often corresponds to smaller $\theta$ in panel~(b) of Figure~\ref{FIG:fig2}. Consequently, very large $\tau_{\chi_2}$ can easily fall within the detection range of CMB. However, excessively long $\tau_{\chi_2}$ may result in $\chi_2$ behaving as a decaying DM. To avoid this issue, we set a lower limit of $\theta\gtrsim\mathcal{O}(10^{-6})$, which also aligns with the effective range of freeze out.  Similar in panel (c) of Figure~\ref{FIG:fig2}, reducing $\Delta_\chi$ leads to an increase of $\tau_{\chi_2}$. An overly small $\Delta_\chi$ not only results in  $\chi_2$  having a lifetime longer than that of the current universe but also restricts electromagnetic decay processes for $\chi_2$. Considering these factors comprehensively, we adopt a lower limit of $\Delta_\chi\gtrsim\mathcal{O}(10^{-3})$.  The CMB detectable range can reach TeV scale with corresponding $g^\prime$ between $\mathcal{O}(10^{-7})$ and $\mathcal{O}(10^{-5})$.   Finally, examining changes of $g_\chi$ illustrated in case (d) of Figure~\ref{FIG:fig2} reveals that  coannihilation is distributed over larger $g^\prime$ as $g_\chi$ decreases. Although the detectable range of $m_{Z^\prime}$ remains similar to that in case (b), both being below $\mathcal{O}(100)$ GeV, the overall distribution of $g^\prime$ is elevated by an order of magnitude compared to that observed in case (b).

In summary, the current constraints from colliders have excluded most of the coscattering. Future colliders with sensitivities at $\mathcal{O}(10^{-5})\lesssim g^\prime\lesssim\mathcal{O}(10^{-3})$ hold promise for conversion. However, for a lower coannihilation dominated region $\mathcal{O}(10^{-8})\lesssim g^\prime\lesssim\mathcal{O}(10^{-5})$ beyond future colliders' reach, the CMB observation can play a significant role.

\section{Secluded Scenario}\label{SEC:SS}
\subsection{Relic Density}\label{RD-S}

\allowdisplaybreaks

We now consider the secluded scenario with $m_{Z'}<m_{\chi_{1,2}}$ \cite{Pospelov:2007mp}.
In comparison with the resonance scenario, the most notable difference of the secluded scenario is the emergence of $\chi_{1,2}~\chi_{1,2}\to Z^\prime Z^\prime$, which is the dominant process with $g_\chi\gg g^\prime$. For simplicity, we  denote the thermal bath particles $f$ and $Z^\prime$ as $\zeta$ in the following discussion. The Boltzmann equations related to the dark fermions are as follows:
\begin{eqnarray}\label{Eqn:BE-S}
	\frac{dY_{\chi_1}}{dz} &= & -\frac{s}{\mathcal{H}z}\bigg[ \langle \sigma v\rangle_{\chi_1\bar{\chi}_1\to \zeta\bar{\zeta}}\Big(Y_{\chi_1}^2-(Y_{\chi_1}^{\eq})^2\Big)+\langle \sigma v\rangle_{\chi_2\chi_1\to \zeta\zeta }\Big(Y_{\chi_2}Y_{\chi_1}-Y_{\chi_2}^{\eq}Y_{\chi_1}^{\eq}\Big)\nonumber \\
	&-& \langle \sigma v\rangle_{\chi_2 \zeta\to\chi_1 \zeta} \left(Y_{\chi_2}Y_\zeta^{\eq}-\frac{Y_{\chi_2}^{\eq}}{Y_{\chi_1}^{\eq}}Y_{\chi_1}Y_\zeta^{\eq}\right)-\langle \sigma v\rangle_{\chi_2\chi_2\to\chi_1\chi_1}\left(Y_{\chi_2}^2-\frac{(Y_{\chi_2}^{\eq})^2}{(Y_{\chi_1}^{\eq})^2}Y_{\chi_1}^2\right)
	\nonumber \\
	&-&\langle \sigma v\rangle_{\chi_1\chi_2\to\chi_1\chi_1}\left(Y_{\chi_1} Y_{\chi_2}-\frac{Y_{\chi_2}^{\eq}}{Y_{\chi_1}^{\eq}}Y_{\chi_1}^2\right)-\langle \sigma v\rangle_{\chi_2\chi_2\to\chi_1\chi_2}\left(Y_{\chi_2}^2-\frac{Y_{\chi_2}^{\eq}}{Y_{\chi_1}^{\eq}}Y_{\chi_1}Y_{\chi_2}\right)
	\nonumber \\
	&-&\frac{\Gamma_{\chi_2\to\chi_1 f\bar{f}}}{s}\left(Y_{\chi_2}-\frac{Y_{\chi_2}^{\eq}}{Y_{\chi_1}^{\eq}}Y_{\chi_1}\right)\bigg],\\	
	\frac{dY_{\chi_2}}{dz} &= & -\frac{s}{\mathcal{H}z}\bigg[ \langle \sigma v\rangle_{\chi_2\bar{\chi}_2\to \zeta\bar{\zeta}}\Big(Y_{\chi_2}^2-(Y_{\chi_2}^{\eq})^2\Big)+\langle \sigma v\rangle_{\chi_2\chi_1\to \zeta\zeta}\Big(Y_{\chi_2}Y_{\chi_1}-Y_{\chi_2}^{\eq}Y_{\chi_1}^{\eq}\Big)
	\nonumber \\
	&+& \langle \sigma v\rangle_{\chi_2 \zeta\to\chi_1\zeta }\left(Y_{\chi_2}Y_{\zeta}^{\eq}-\frac{Y_{\chi_2}^{\eq}}{Y_{\chi_1}^{\eq}}Y_{\chi_1}Y_{\zeta}^{\eq}\right)+\langle \sigma v\rangle_{\chi_2\chi_2\to\chi_1\chi_1}\left(Y_{\chi_2}^2-\frac{(Y_{\chi_2}^{\eq})^2}{(Y_{\chi_1}^{\eq})^2}Y_{\chi_1}^2\right)
	\nonumber \\
	&+&\langle \sigma v\rangle_{\chi_1\chi_2\to\chi_1\chi_1}\left(Y_{\chi_1} Y_{\chi_2}-\frac{Y_{\chi_2}^{\eq}}{Y_{\chi_1}^{\eq}}Y_{\chi_1}^2\right)+\langle \sigma v\rangle_{\chi_2\chi_2\to\chi_1\chi_2}\left(Y_{\chi_2}^2-\frac{Y_{\chi_2}^{\eq}}{Y_{\chi_1}^{\eq}}Y_{\chi_1}Y_{\chi_2}\right)
	\nonumber \\
	&+&\frac{\Gamma_{\chi_2\to\chi_1 f\bar{f}}}{s}\left(Y_{\chi_2}-\frac{Y_{\chi_2}^{\eq}}{Y_{\chi_1}^{\eq}}Y_{\chi_1}\right)\bigg],
\end{eqnarray}
where the various parameters  are consistent with those in Equation~\eqref{Eqn:BE-R}. The thermal average cross-section $\left<\sigma v\right>$  is also calculated using the micrOMEGAs~\cite{Alguero:2022inz,Alguero:2023zol}. Furthermore, the newly identified $Y_\zeta^{\eq}$ can be categorized into $Y_\zeta^{\eq}=0.238$ for $\zeta=f$  and
\begin{eqnarray}
	Y_{\zeta}^{\eq}=\frac{45z^2}{4\pi^4g_s}\left(\frac{m_{Z^\prime}}{m_{\chi_1}}\right)^2\mathcal{K}_2\left(\frac{m_{Z^\prime}}{m_{\chi_1}}z\right),
\end{eqnarray}
when $\zeta$ represents the massive $Z^\prime$.

\begin{figure}
	\begin{center}
		\includegraphics[width=0.42\linewidth]{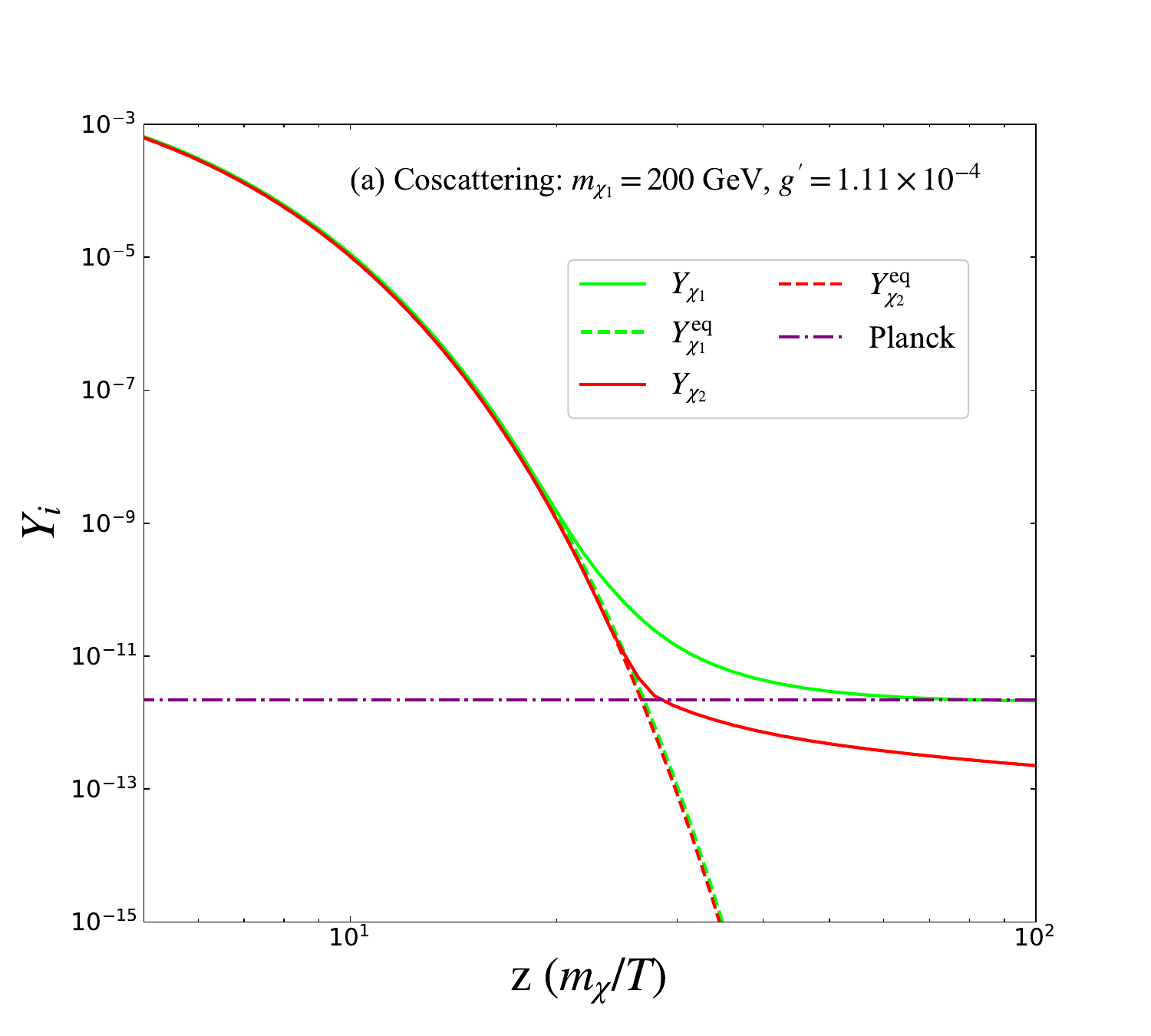}
		\includegraphics[width=0.42\linewidth]{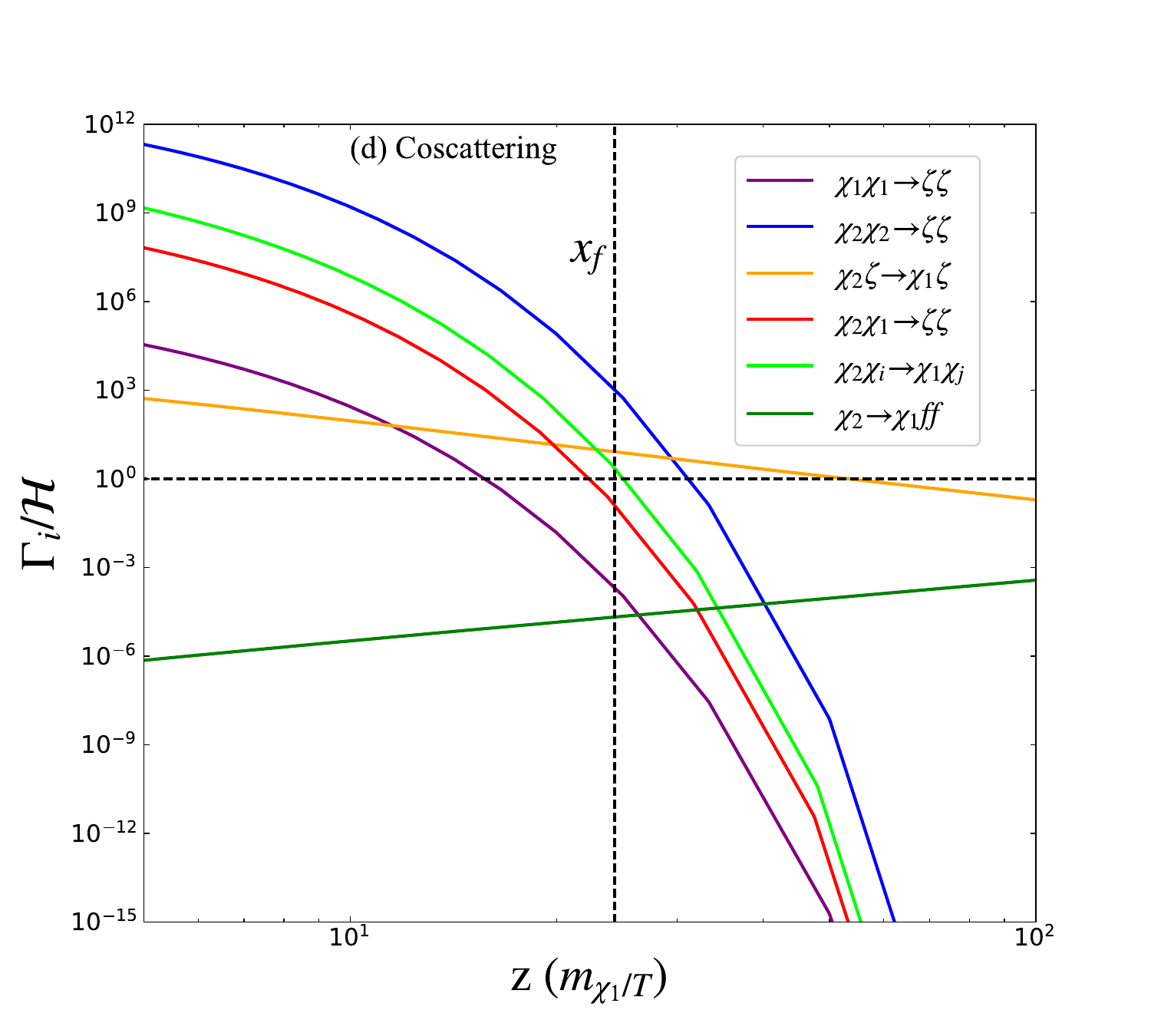}
		\includegraphics[width=0.42\linewidth]{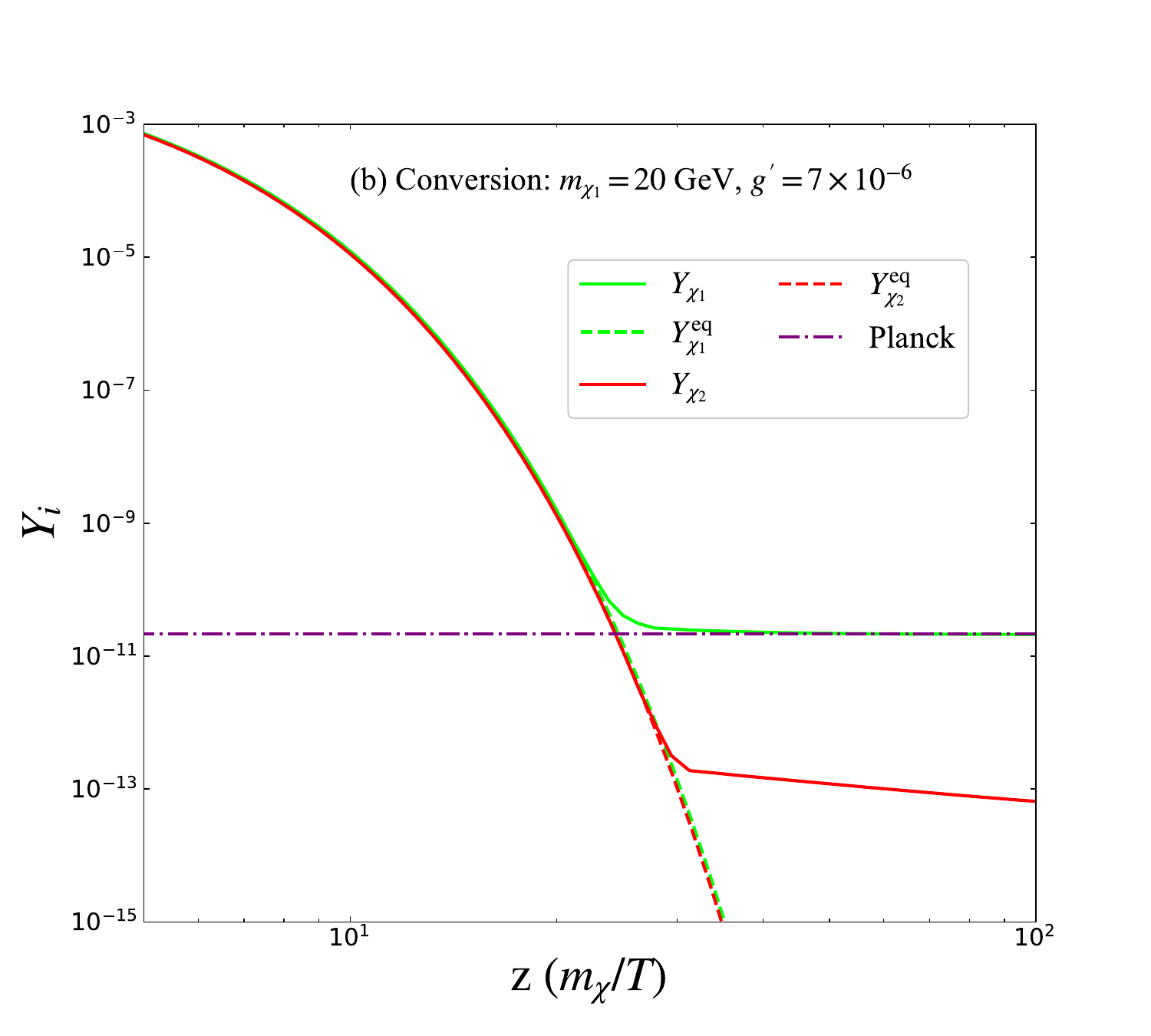}
		\includegraphics[width=0.42\linewidth]{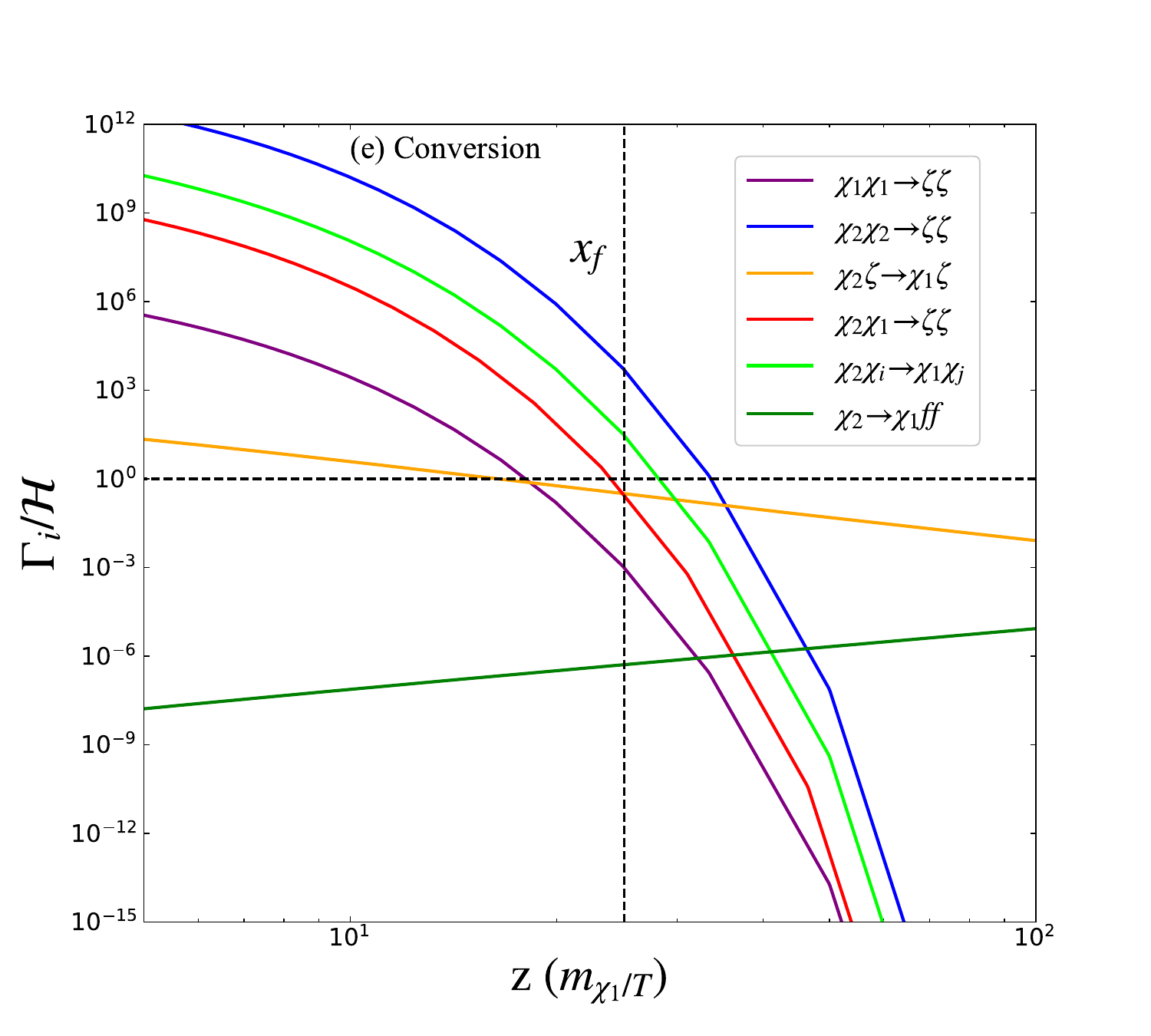}
		\includegraphics[width=0.42\linewidth]{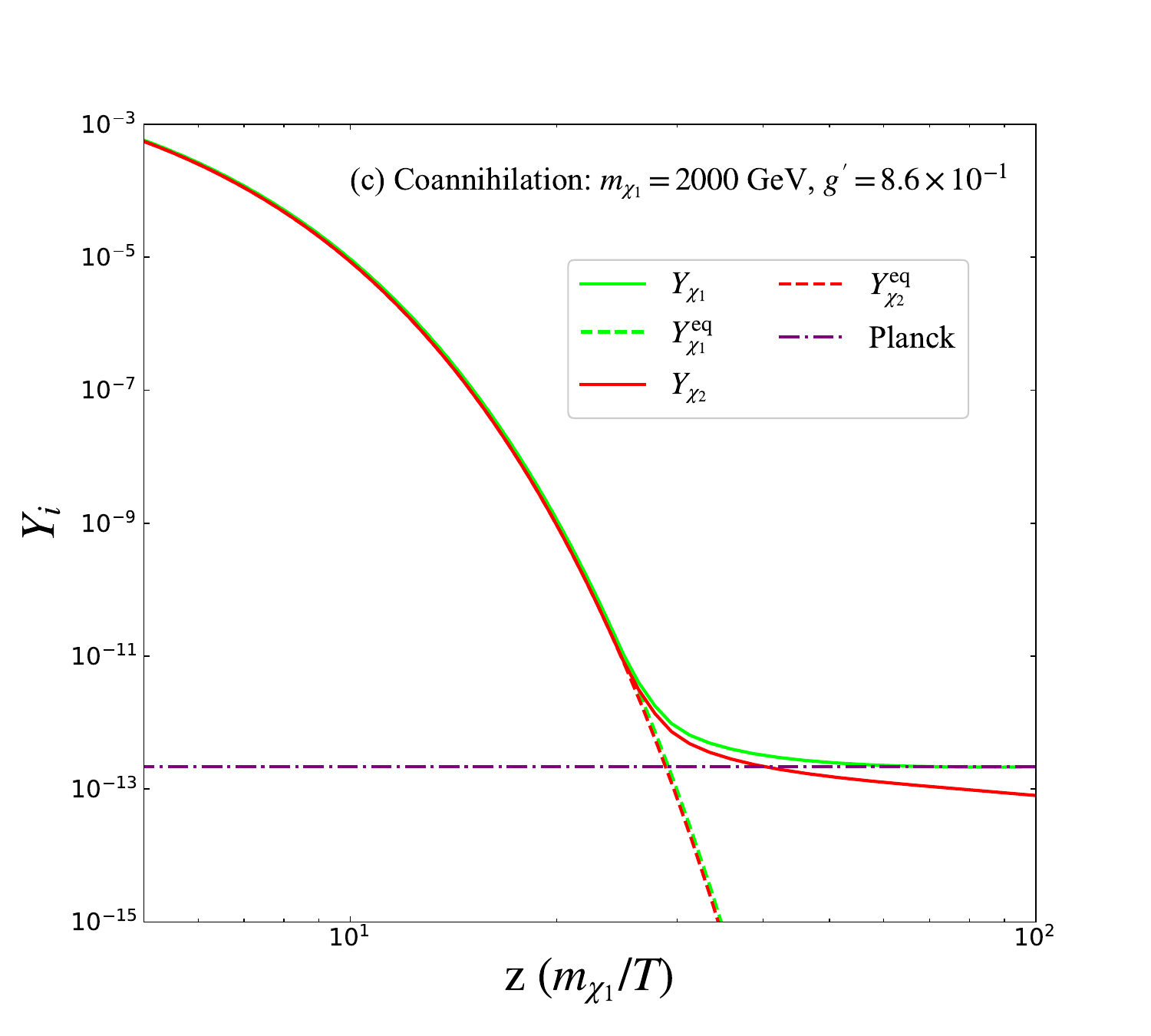}		
		\includegraphics[width=0.42\linewidth]{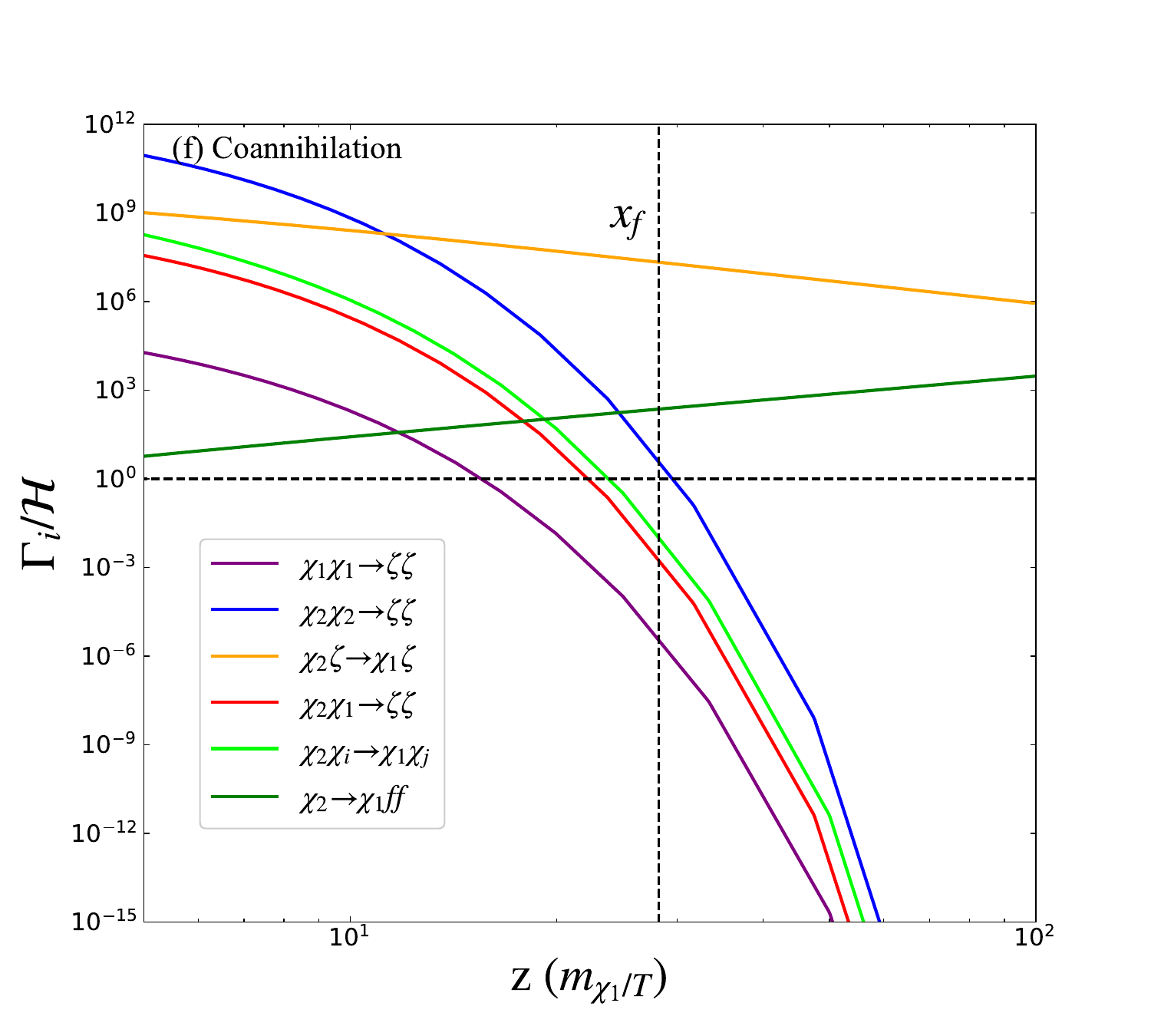}
	\end{center}
	\caption{The evolutions of various abundances $Y_i$ for (a) coscattering, (b) conversion, and (c) coannihilation benchmarks in the secluded scenario. Panels (d), (e), and (f) correspond to the thermal rates of various processes in  mechanisms described by (a), (b), and (c), respectively.  Here we fix $\Delta_\chi=10^{-2}$, $r_{Z^\prime}=0.75$ and $\theta=2\times10^{-2}$. The meanings of  different colored curves  can be referenced in Figure~\ref{FIG:fig1}.
	}
	\label{FIG:fig6}
\end{figure}

In Figure~\ref{FIG:fig6}, we introduce all relevant dark matter production mechanisms through the benchmarks, specifically coscattering, conversion, and coannihilation. The criteria for evaluating each mechanism are consistent with those outlined in the resonance scenario. It can be observed  from Figure~\ref{FIG:fig6} that conversion is located at light $m_{\chi_1}$ and small $g^\prime$.  As these two parameters gradually increase, the dominant phase transits towards coscattering, ultimately becomes entirely governed by coannihilation at $m_{\chi_1}\sim$ TeV and $g^\prime\sim\mathcal{O}(1)$. The overall distribution of these three phases on $g^\prime$ exhibits a monotonic relationship with $m_{\chi_1}$. This is significantly different from the resonance scenario,  where  the overall monotonic relationship is no longer exists. Such correlation is mainly due to the processes that dominate the generation of dark fermions as $\chi_{1,2}\chi_{1,2}\to Z'Z'$, which are almost unaffected by $g^\prime$ in the secluded scheme. However, the situation is quite the opposite in the resonance scenario.

Additionally, it is noteworthy that we incorporate the coscattering process $\chi_2 Z^\prime\to\chi_1 Z^\prime$ into Equation~\eqref{Eqn:BE-S}, whose contribution is comparable to that of $\chi_2 f\to\chi_1 f$ when $m_{Z^\prime}\ll m_{\chi_{1,2}}$. We explore the demand for freezing-out within a broad range of $r_{Z^\prime}\in[0.1,1]$, $g^\prime\in[10^{-10},10^{-2}]$, $\Delta_\chi\in[10^{-3},10^{-1}]$ and $g_\chi\in[0.1,1]$. We report that it is sufficient for $\theta$ to be greater than $\mathcal{O}(10^{-8})$ with $m_{\chi_1}$ from GeV to TeV scale.

 \subsection{Phenomenology of $Z^\prime$}\label{PZ-S}

\begin{figure}
	\begin{center}
		\includegraphics[width=0.45\linewidth]{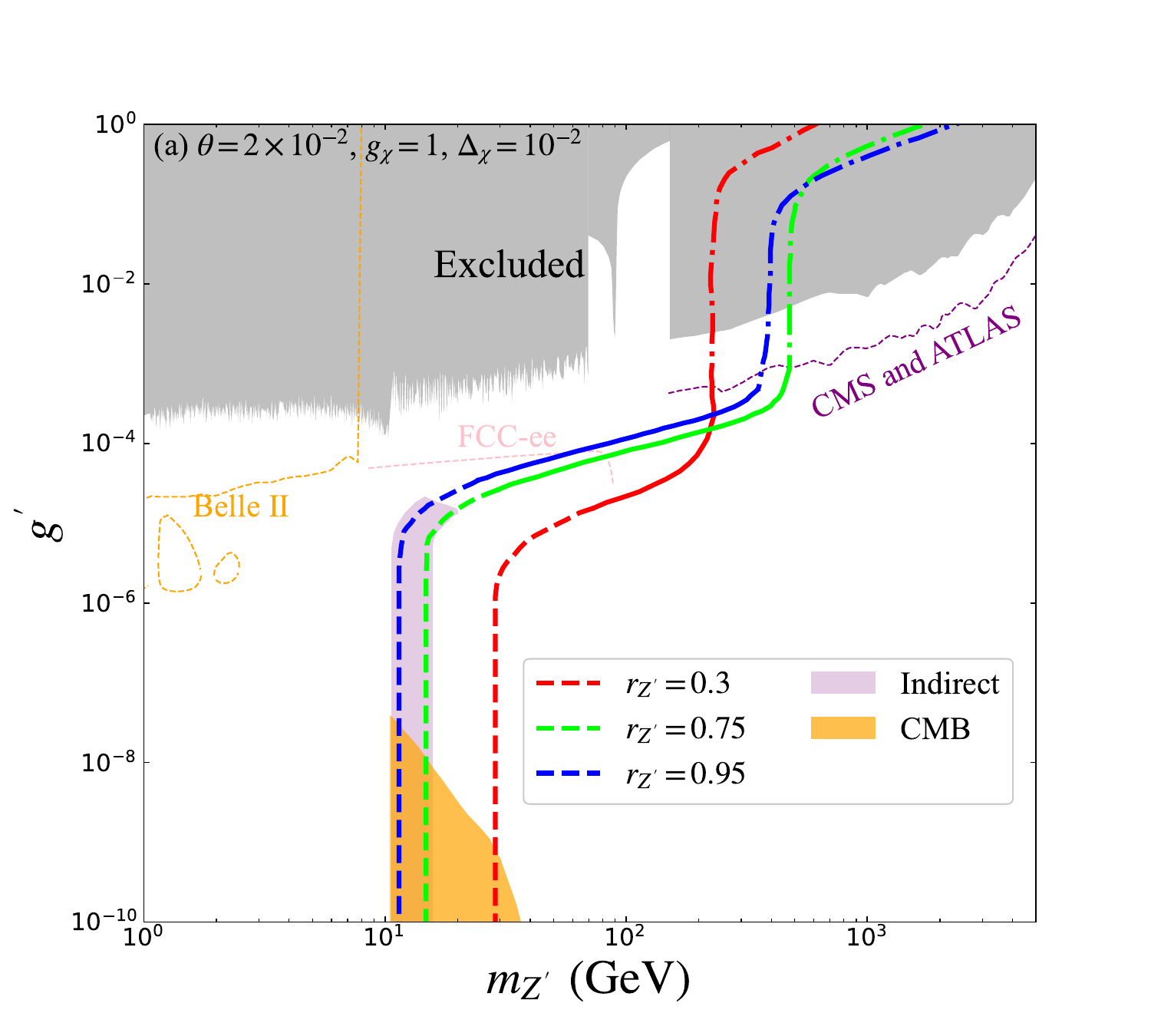}
		\includegraphics[width=0.45\linewidth]{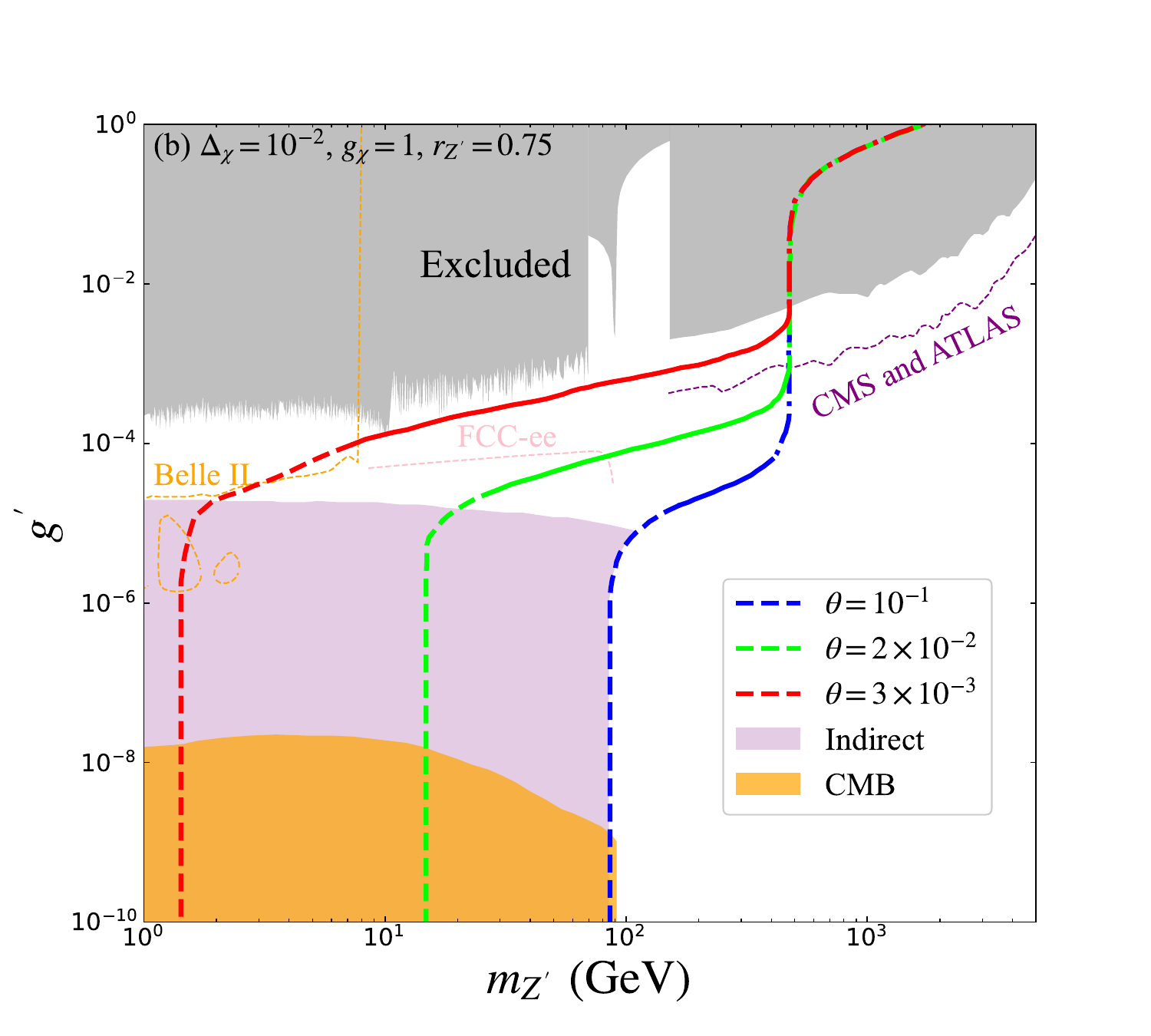}
		\includegraphics[width=0.45\linewidth]{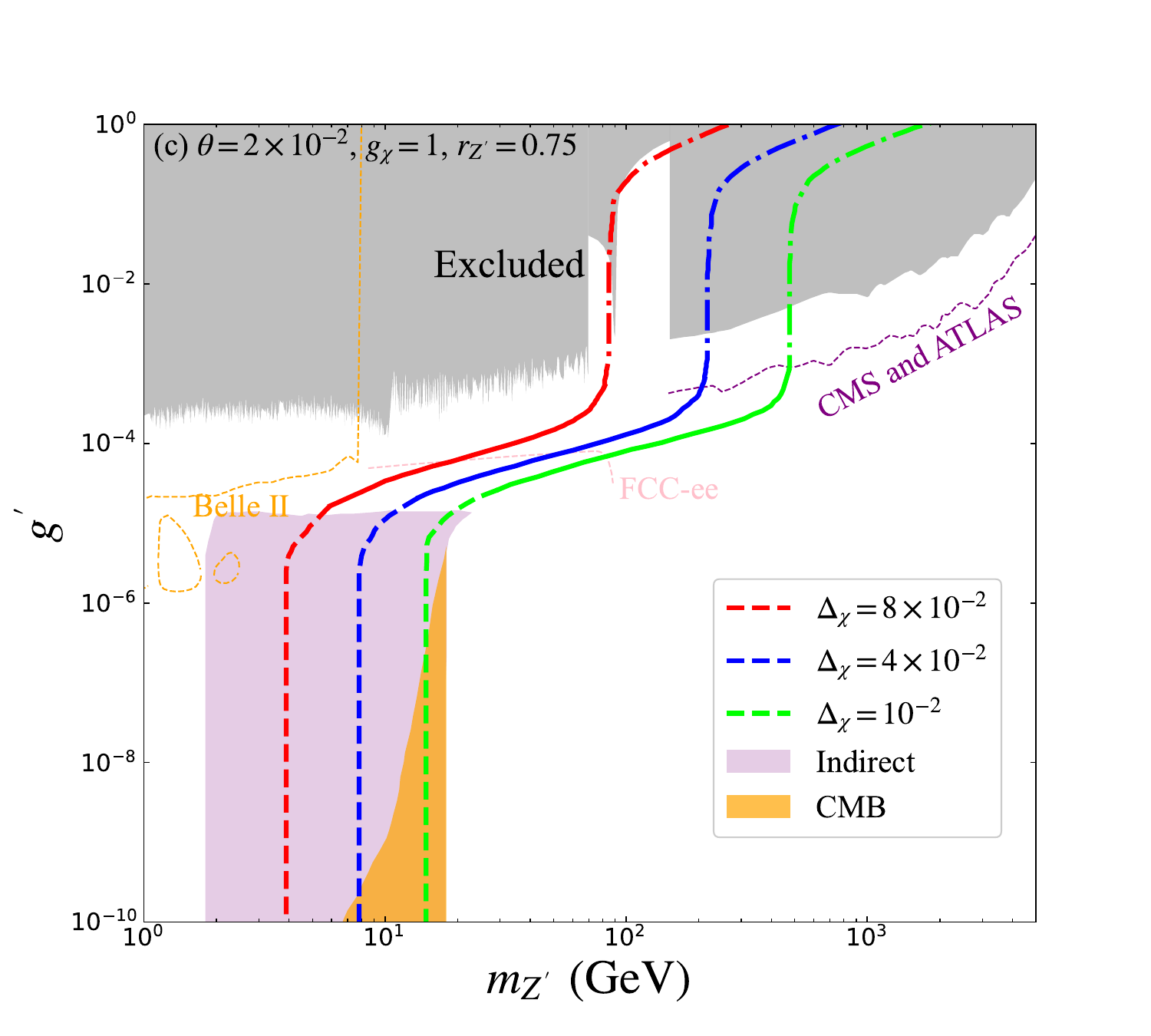}
		\includegraphics[width=0.45\linewidth]{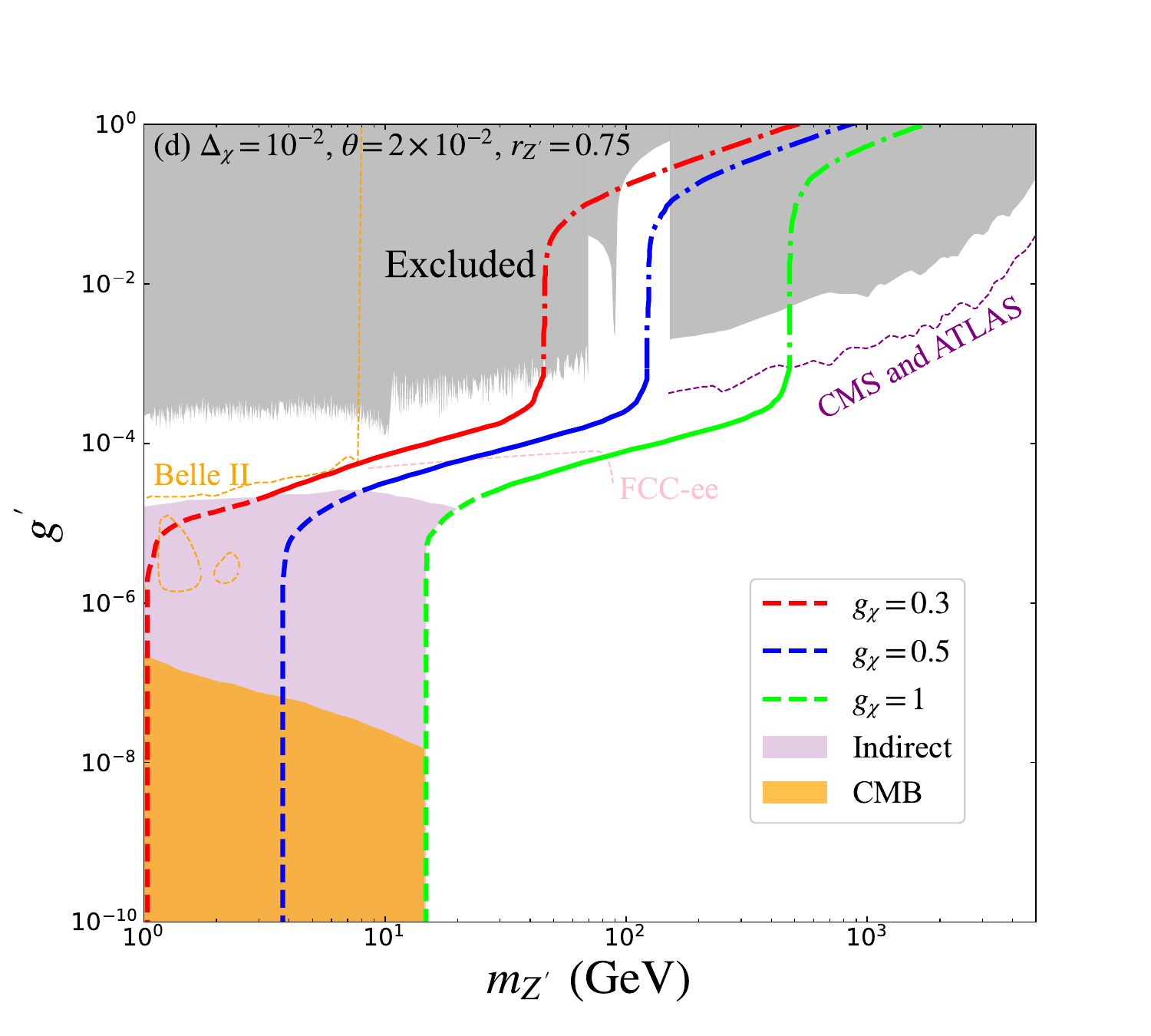}
	\end{center}
	\caption{Constraints on the $m_{Z^\prime}-g^\prime$ parameter space in the secluded scenario.  The three fixed parameters and one varying benchmarks marked by red, green, and blue lines  in each panel are consistent with those in the resonance scenario of Figure \ref{FIG:fig2}, although some parameters have different numerical values. Similarly,  coscattering, conversion, and coannihilation phases are  designated as solid, dashed and dotdashed lines, respectively. The gray shaded region are excluded by the current limits on $Z^\prime$ at colliders, while the future sensitivities are indicated by orange, pink, and purple dashed lines.
	}
	\label{FIG:fig7}
\end{figure}

In the secluded scenario, the constraints applied on $Z^\prime$ are consistent with those in the resonance scenario, encompassing the excluded region by the current experiments  BaBar \cite{BaBar:2014zli,BaBar:2017tiz}, LHCb \cite{LHCb:2017trq,LHCb:2019vmc},  LEP \cite{ALEPH:2013dgf}, CMS and ATLAS \cite{CMS:2021ctt,ATLAS:2019erb}, and the sensitivities of future Belle II~\cite{Ferber:2022ewf,Dolan:2017osp}, FCC-ee \cite{Karliner:2015tga}, high luminosity CMS and ATLAS \cite{KA:2023dyz}. They are depicted in Figure~\ref{FIG:fig7} as shaded gray areas for the current excluded limits as well as orange, pink, and  purple dashed lines for the future reaches.

In Figure~\ref{FIG:fig7}, we illustrate the impact of $Z^\prime$ induced constraints  on various parameters. The mass ratio  $r_{Z^\prime}$ varies from 0.3 to 0.95 in  panel (a) with fixed $\theta=2\times10^{-2}$, $g_\chi=1$ and $\Delta_\chi=10^{-2}$. From the perspective of $g^\prime$, all conversion regimes are distributed at $g^\prime\lesssim3\times10^{-5}$. This is understandable, since among all relevant reactions, the conversion process $\chi_i \chi_2\to \chi_j \chi_1$ is not sensitive to the gauge coupling $g'$. For such small $g'$,  the thermal rate of the coscattering process $\chi_2f\to\chi_1f$ is suppressed, resulting in a contribution that is less than that of the conversion process $\chi_i \chi_2\to \chi_j \chi_1$, which is illustrated in panels (b) and (e) in Figure~\ref{FIG:fig6} for specific details. At larger $g^\prime\in[3\times10^{-5},1.8\times10^{-3}]$, the contribution from $\chi_2f\to\chi_1f$ surpasses that of $\chi_i \chi_2\to \chi_j \chi_1$ and becomes dominant, corresponding to the coscattering phase.   As $g^\prime$ continues to increase, the thermal rate of  $\chi_2f\to\chi_1f$ exceeds those of the other production processes, thus violating the criteria for coscattering.  Concurrently,  contributions from $\chi_2f\to\chi_1f$ gradually become insufficient as $m_{Z^\prime}$ increases, hence entering into the domain of coannihilation. Moreover, it is evident that when $g^\prime\gtrsim0.1$, the production process of $f\bar{f}$ final state mediated by $Z^\prime$  begins to exert significant influence, causing $m_{Z^\prime}$ to once again push towards larger values.

From the viewpoint of $m_{Z^\prime}$, conversion, coscattering, and coannihilation correspond respectively at the orders of magnitude around $\mathcal{O}(10)$ GeV, $\mathcal{O}(100)$ GeV, and $\mathcal{O}(1000)$ GeV. Additionally, for smaller $r_{Z^\prime}$, the dominance within the coscattering channel shifts towards $\chi_2Z^\prime\to\chi_1Z^\prime$. This particular pathway is solely influenced by $\theta$, and only becomes effective when $\theta$ approaches  $\mathcal{O}(10^{-6})$. However, such a low magnitude of $\theta$ does not attract the interest of DM-related phenomenology, so this aspect is not considered in the following discussion.  Overall, any benchmark starts from the conversion with  minimum $g^\prime$ and $m_{Z^\prime}$. The distribution of conversion is minimally affected by $g^\prime$, because the key factors $\Gamma_{\chi_2\bar{\chi}_2\to Z^\prime Z^\prime}$, $\Gamma_{\chi_i \chi_2\to \chi_j \chi_1}$, and $\Gamma_{\chi_1\bar{\chi}_1\to Z^\prime Z^\prime}$ in the conversion determination criteria are all independent of $g^\prime$. When $g^\prime$ increases to $\mathcal{O}(10^{-5})$, it triggers a transition from the conversion phase to the coscattering phase because the reaction rate of the coscattering process is proportional to $g^\prime$. Subsequently, $m_{Z^\prime}$ must increase in synchrony with $g^\prime$ to ensure the judgment condition of coscattering $\Gamma_{\chi_2\bar{\chi}_2\to Z^\prime Z^\prime}>\Gamma_{\chi_2 f\to \chi_1 f}$. 
However, since $\Gamma_{\chi_2 f\to \chi_1 f}$ increases at an exceptionally rapid rate with increasing $g^\prime$, the criterion governing coscattering breaks down when $g^\prime\sim\mathcal{O}(10^{-3})$. Then it enters the coannihilation phase dominated by $\chi_2\bar{\chi}_2\to Z^\prime Z^\prime$ that is insensitive to $g^\prime$,  so a vertical line like the conversion phase exhibits until the emergence of coannihilation phase dominated by $\chi_2\bar{\chi}_2\to  f\bar{f}$. Subsequently,  similar to the resonance scenario, $g^\prime$ is proportional to $m_{Z^\prime}$. The benchmarks corresponding to different parameter choices discussed below exhibit nearly identical evolutionary trends, differing only in their initial and final $m_{Z^\prime}$. From a phenomenological perspective,  under the constraints from $Z'$, the results in the secluded scenario are markedly different from those obtained in the resonance scenario. The coannihilation distributed in large $g^\prime$ is almost entirely excluded by the current  $Z^\prime$ constraints. In the future, the collider experiments might probe the secluded coscattering phase with $Z'$ around the electroweak scale and $r_{Z^\prime}\sim\mathcal{O}(0.1)$.

In panel (b) of Figure \ref{FIG:fig7}, we choose to change $\theta$. For the conversion phase, the corresponding $m_{Z^\prime}$ rises from GeV to 100 GeV as $\theta$ increases, with upper limit on $g'$ decreasing from $10^{-4}$ to $10^{-5}$. Meanwhile, the distribution of coscattering continuously shrinks from an initial $m_{Z^\prime}\in[6.9,480]$ GeV with  $g^{\prime}\in[8.4\times10^{-5},10^{-2}]$ for $\theta=3\times10^{-3}$ to the final $m_{Z^\prime}\in[172.1,438.2]$ GeV with  $g^{\prime}\in[1.8\times10^{-5},9.2\times10^{-5}]$ for $\theta=10^{-1}$.  Ultimately, all benchmark lines converge at $m_{Z^\prime}\sim480$ GeV with the coannihilation phase, primarily due to the fact that the dominant process occurring at $\chi_2\chi_2\to\zeta\zeta$ within coannihilation shows little dependence on the mixing $\theta$.  The Belle II, which is sensitive to light $m_{Z^\prime}$, shows a preference for conversion and coscattering when $\theta\lesssim3\times10^{-3}$. The FCC-ee, CMS, and ATLAS further have the potential to capture coscattering as well as  coannihilation that has not yet been excluded when $\theta \gtrsim 3\times10^{-3}$.

In the subsequent panels (c) and (d) of Figure \ref{FIG:fig7}, it is obvious that as either $\Delta_\chi$ decreases or $g_\chi$ increases, the benchmark curve shifts towards higher $m_{Z^\prime}$.  For the three phases of conversion, coscattering and coannihilation, the corresponding $m_{Z^\prime}$  increase from GeV to $\mathcal{O}(10)$ GeV, from $\mathcal{O}(10)$ GeV to $\mathcal{O}(100)$ GeV and from $\mathcal{O}(100)$ GeV to TeV, respectively. In contrast, the distribution of certain phase on $g^{\prime}$ is not significantly influenced by parameters $\Delta_\chi$ or $g_\chi$, such as $g^\prime\lesssim3\times10^{-5}$ for conversion, $3\times10^{-5}\lesssim g^\prime\lesssim 1.8\times10^{-3}$ for coscattering and $g^\prime\gtrsim 1.8\times10^{-3}$ for coannihilation. In the future, Belle II is sensitive to conversion and coscattering with $\Delta_\chi\gtrsim8\times10^{-2}$ and $g_\chi\lesssim0.5$, while FCC-ee, CMS and ATLAS favor coscattering with $g_\chi\sim\mathcal{O}(0.1)$ and relatively free $\Delta_\chi$, and coannihilation with $\Delta_\chi\lesssim4\times10^{-2}$ and $g_\chi\gtrsim0.5$. It should be mentioned that the current constraints on $Z^\prime$ are lacking when $m_{Z^\prime}\sim100$ GeV. This allows a portion of coannihilation to persist at $\Delta_\chi\sim8\times10^{-2}$ and $g_\chi\sim0.5$, which is highly appealing for the subsequent discussion of DM phenomenology.

In short, the secluded and resonance schemes yield different results for the correct relic density under constraints of $Z'$.  For example, promising conversions appear at the parameter space with $\theta \lesssim 3\times10^{-3}$, $\Delta_\chi\gtrsim8\times10^{-2}$ and $g_\chi\lesssim0.5$ in the secluded scenario. While hopeful coscattering requires $\theta \gtrsim 3\times10^{-3}$ and $g_\chi\sim\mathcal{O}(0.1)$, as long as the value of $\Delta_{\chi}$ is not excessively large. The vast majority of coannihilation is excluded, but a small portion with  $\theta \gtrsim 3\times10^{-3}$, $\Delta_\chi\lesssim4\times10^{-2}$ and  $g_\chi\gtrsim0.5$ may potentially be detected by future CMS and ATLAS. In the conversion region with $g'\lesssim10^{-5}$, it is promising at future indirect and CMB experiments, which will be considered in the following discussion.

\subsection{Phenomenology of $\chi_1$}\label{PC1-S}

\begin{figure}
	\begin{center}
		\includegraphics[width=0.45\linewidth]{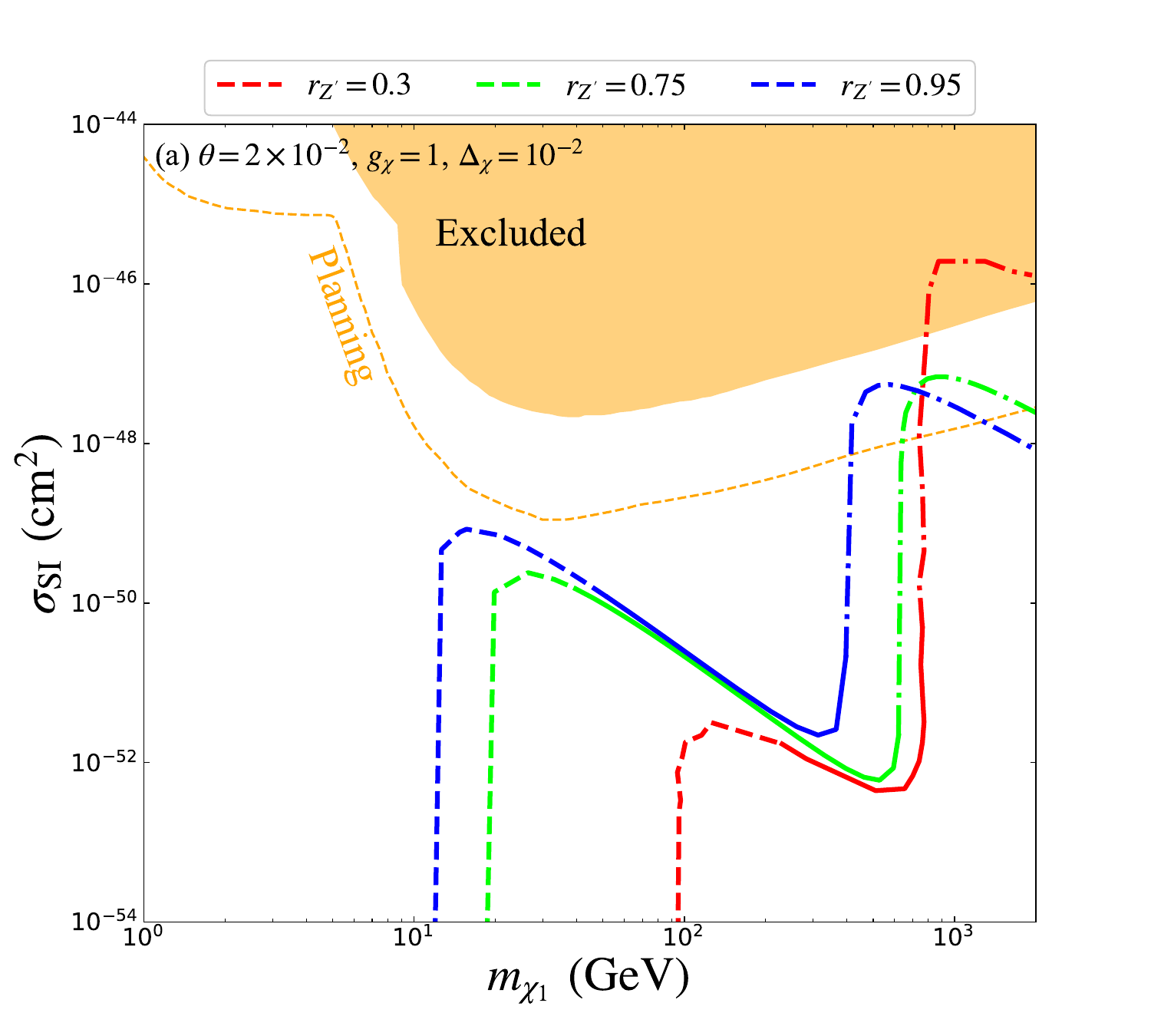}
		\includegraphics[width=0.45\linewidth]{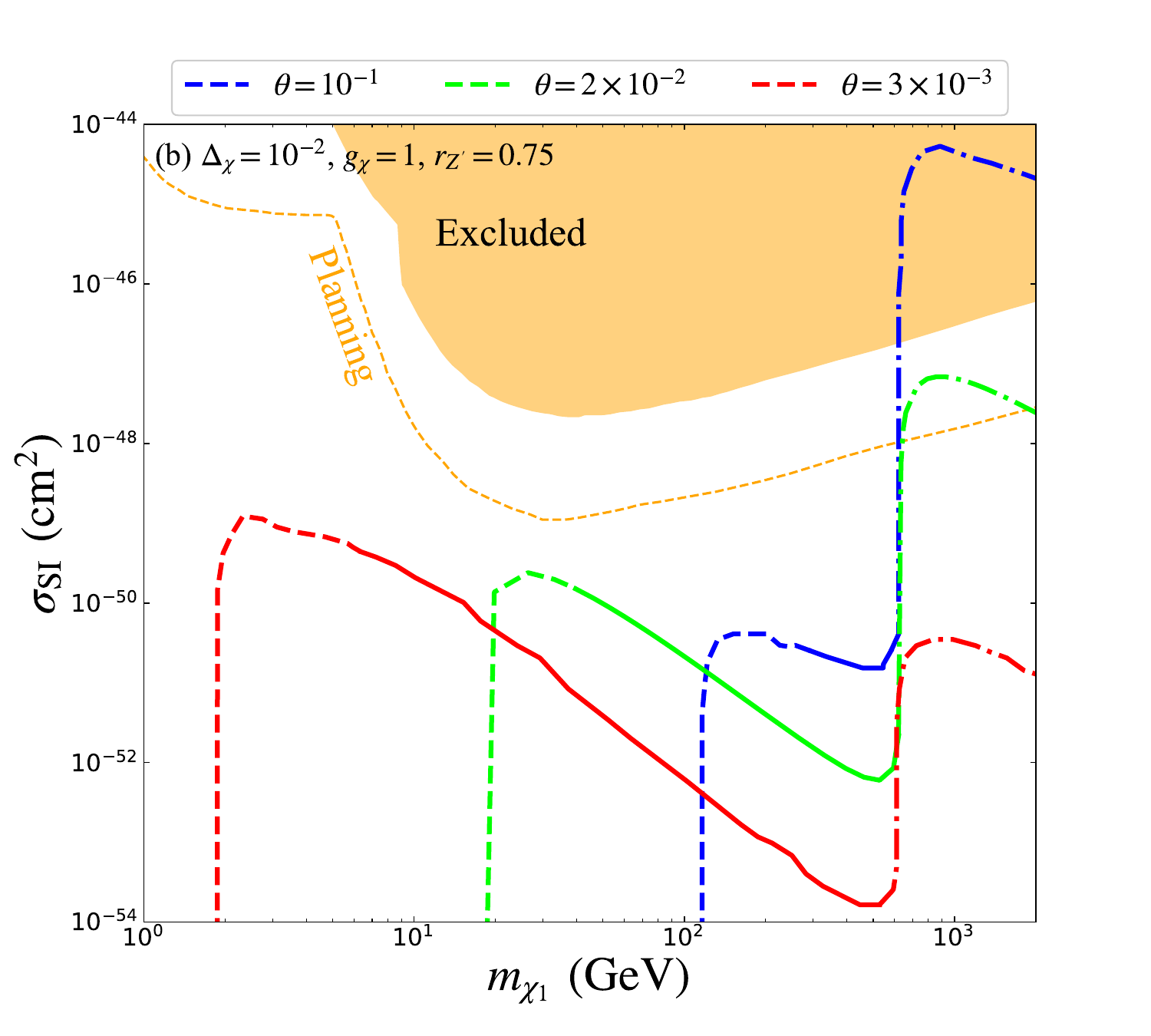}
		\includegraphics[width=0.45\linewidth]{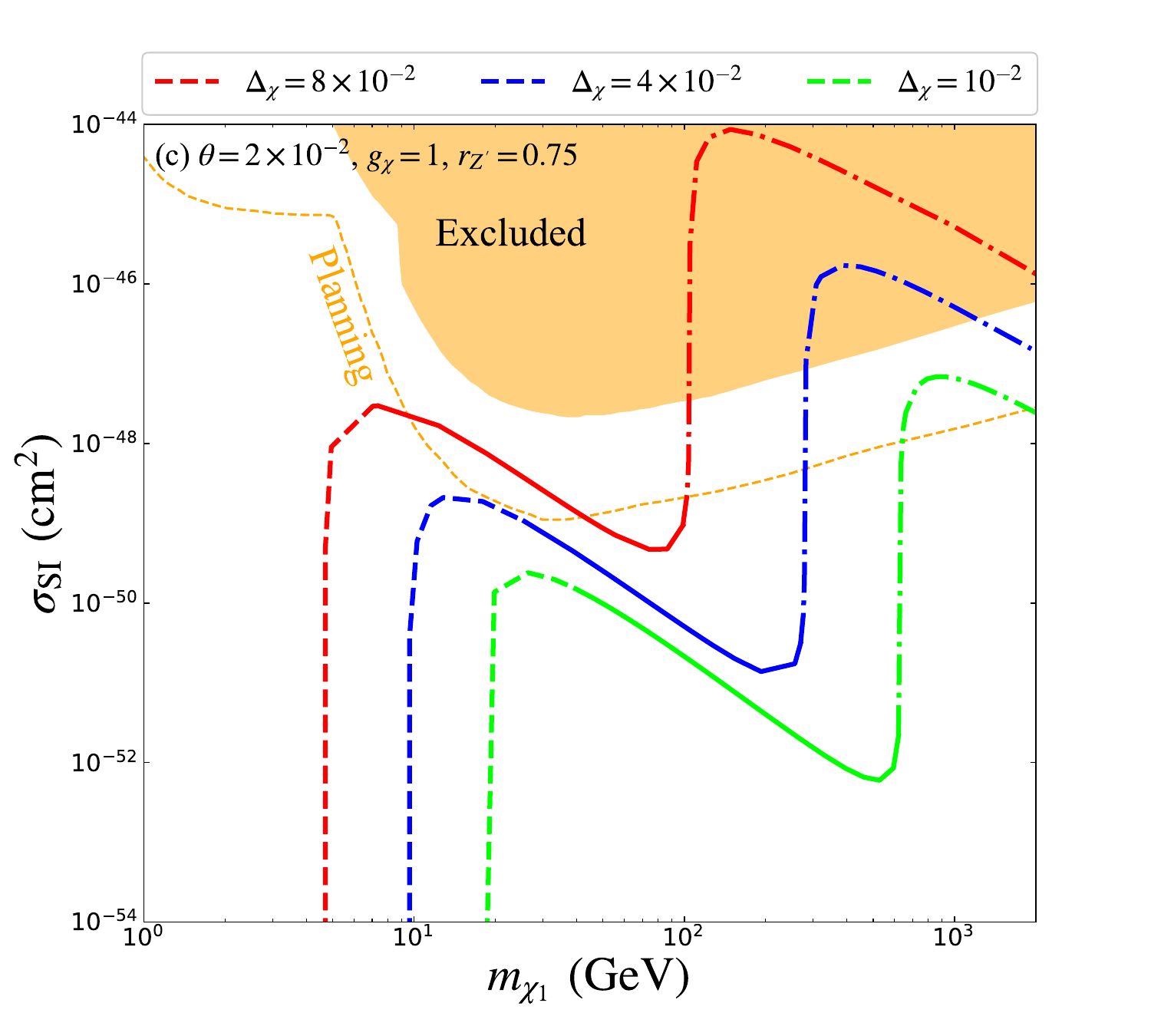}
		\includegraphics[width=0.45\linewidth]{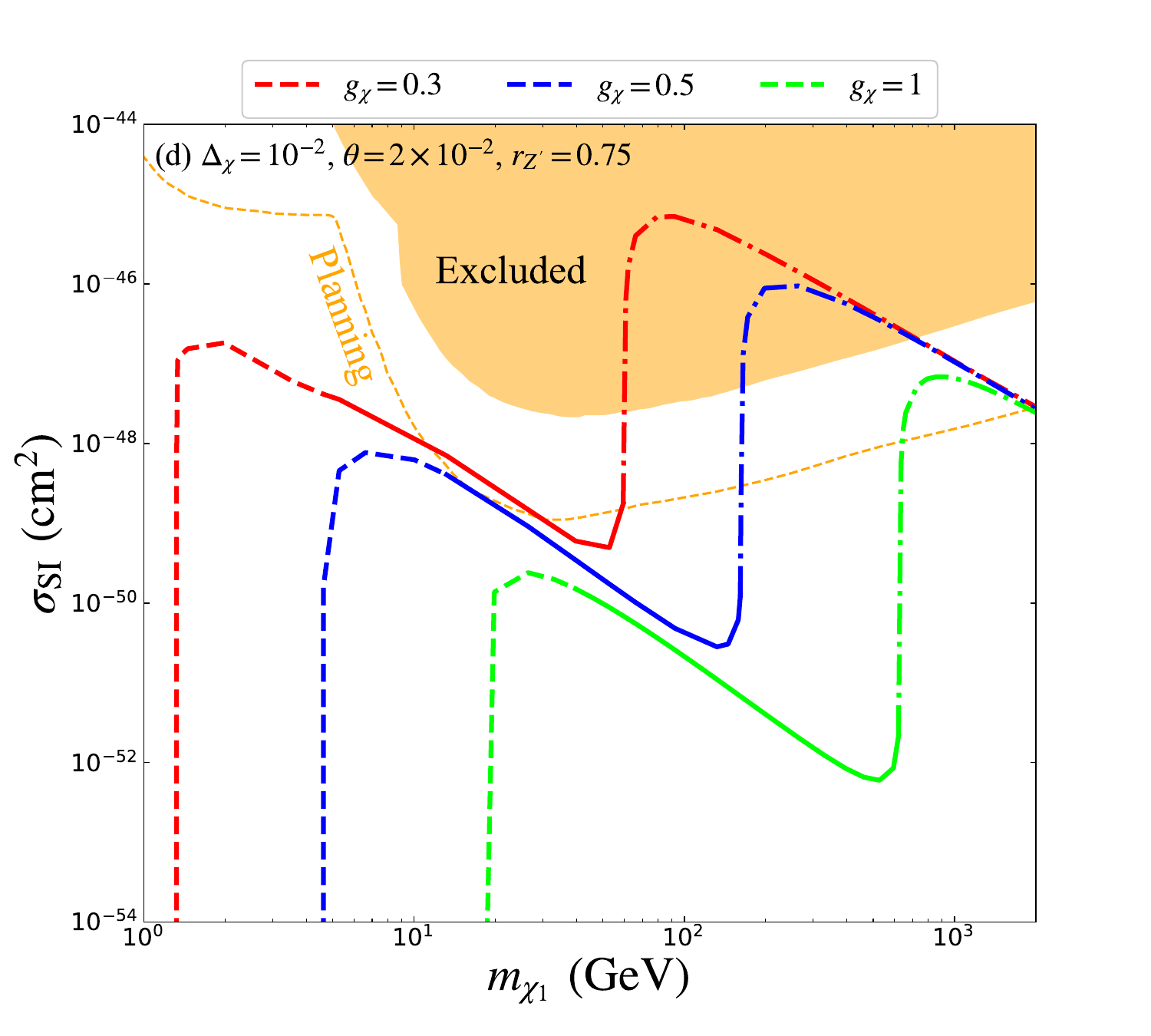}
	\end{center}
	\caption{The constraints of direct detection experiments in the secluded scenario. Legends and markers occurring in each panel correspond directly to those presented in Figure~\ref{FIG:fig7}. The current and future constraints are represented by the orange shaded region and dashed line, respectively.
	}
	\label{FIG:fig8}
\end{figure}

We first consider the direct detection  of dark matter in the secluded scenario, where the constraints on the spin-independent scattering cross section of DM-nucleons are identical to those presented in the resonance scenario depicted in Figure~\ref{FIG:fig3}. Meanwhile, the numerical result of the scattering cross section can also be calculated through Equation~\eqref{Eqn:dd}.

The constraints from direct detection experiments on the secluded scenario  are presented in Figure~\ref{FIG:fig8}.  Here, the results of the benchmarks are obtained by substituting the corresponding ones from Figure~\ref{FIG:fig7} into Equation~\eqref{Eqn:dd}. Consequently, all curves exhibit nearly identical shapes. However, unlike the shape shown in Figure~\ref{FIG:fig7}, the $\sigma_{\rm SI}$ of the coscattering and final coannihilation components are inversely proportional to $m_{\chi_{1}}$ due to the suppression  of $m_{Z^\prime}$.  In panel (a) with varying $r_{Z^\prime}$, conversion exhibits the highest $\sigma_{\rm SI}\sim2.2\times10^{-49}~\cm^2$ at $m_{\chi_1}\sim17$ GeV when $r_{Z^\prime}=0.95$, however this value remains slightly lower than the result  $\sigma_{\rm SI}\sim8.3\times10^{-50}~\cm^2$ from future direct detection experiments. Coscattering usually predicts $\sigma_{\rm SI}\lesssim10^{-50}~\cm^2$, thus it is also beyond the reach of direct detection.  Therefore, only coannihilation holds promise for validation by future direct detection. However, such a coannihilation phase is not permitted by the constraints of $Z^\prime$. In the subsequent panel (b) of Figure~\ref{FIG:fig8}, the variations in $\theta$ give rise to conversion and coscattering having $\sigma_{\rm SI}\lesssim1.2\times10^{-49}~\cm^2$. Meanwhile, the coannihilation that fall within the detection range are  excluded by $Z^\prime$ constraints. Therefore, similar conclusions can be drawn as those presented in panel (a). The situations have been improved in panel (c) and (d), where coscattering below 60 GeV  can be easily detected by the projected experiments when $\Delta_\chi\gtrsim 4\times10^{-2}$ and $g_\chi\lesssim 0.5$. Furthermore, due to the less stringent constraints imposed by $Z^\prime$ on $\Delta_\chi\sim8\times10^{-2}$ and $g_\chi\sim0.5$, this small portion of coannihilation near 100 GeV  also shows considerable promise.

\begin{figure}
	\begin{center}
		\includegraphics[width=0.45\linewidth]{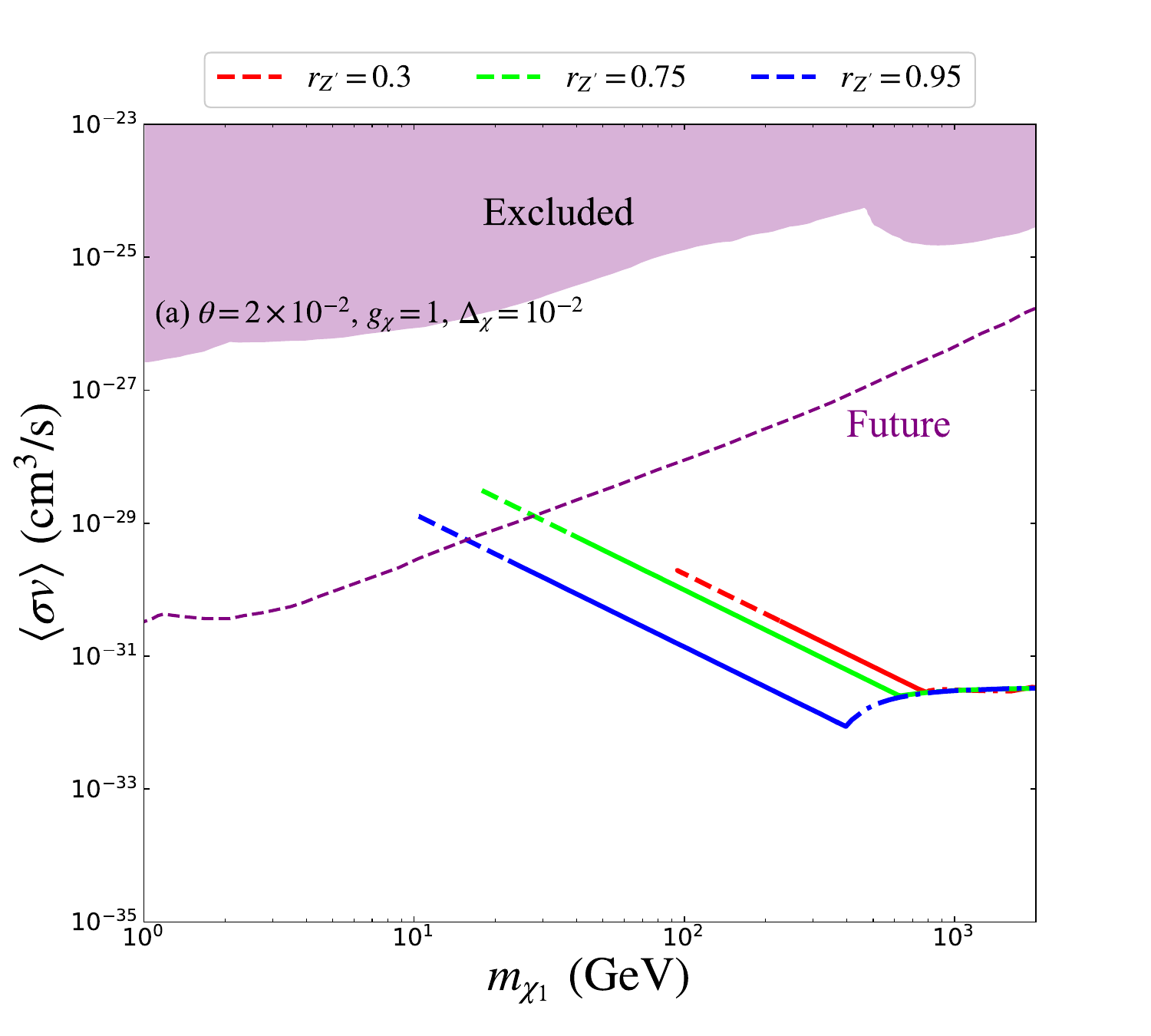}
		\includegraphics[width=0.45\linewidth]{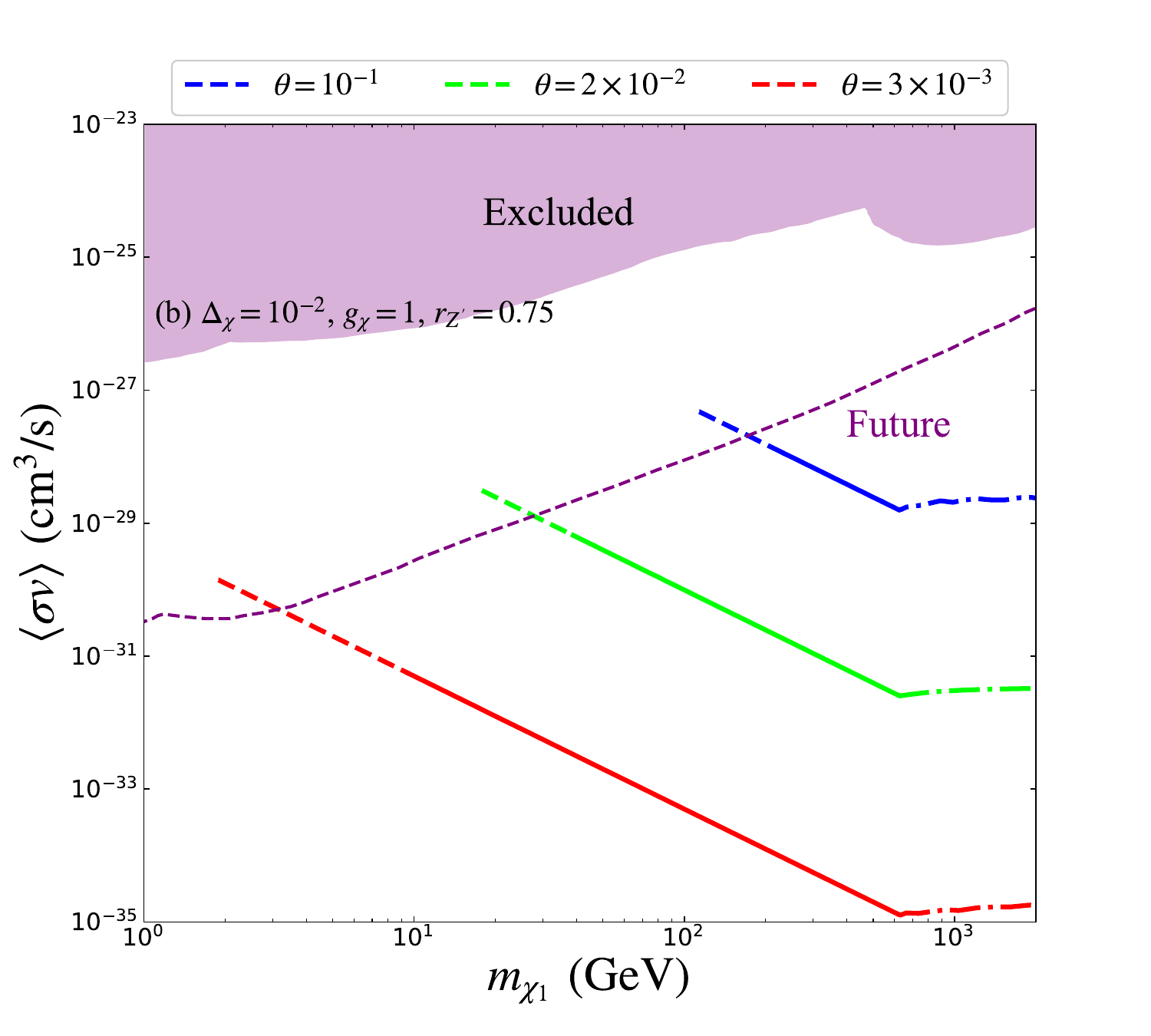}
		\includegraphics[width=0.45\linewidth]{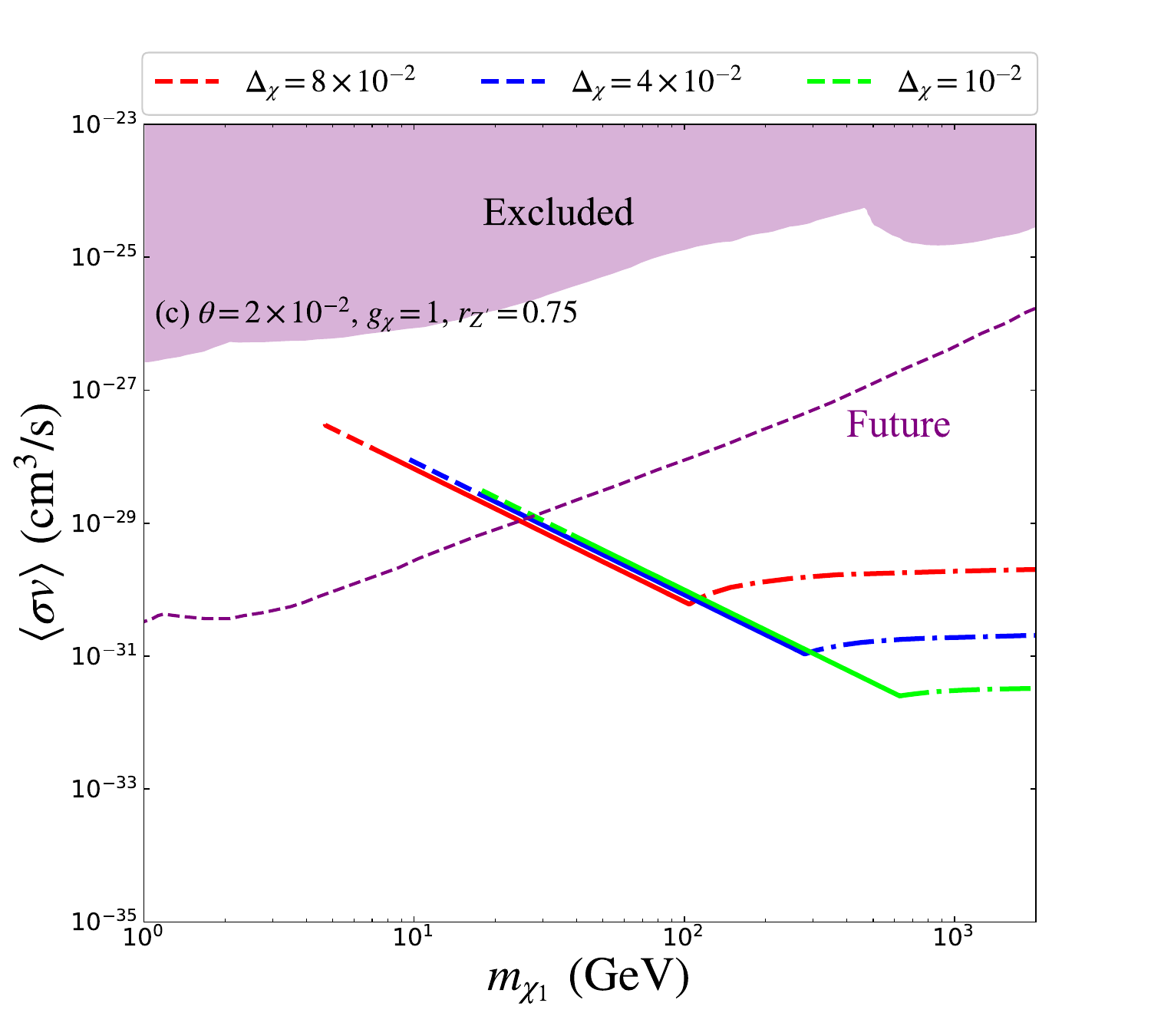}
		\includegraphics[width=0.45\linewidth]{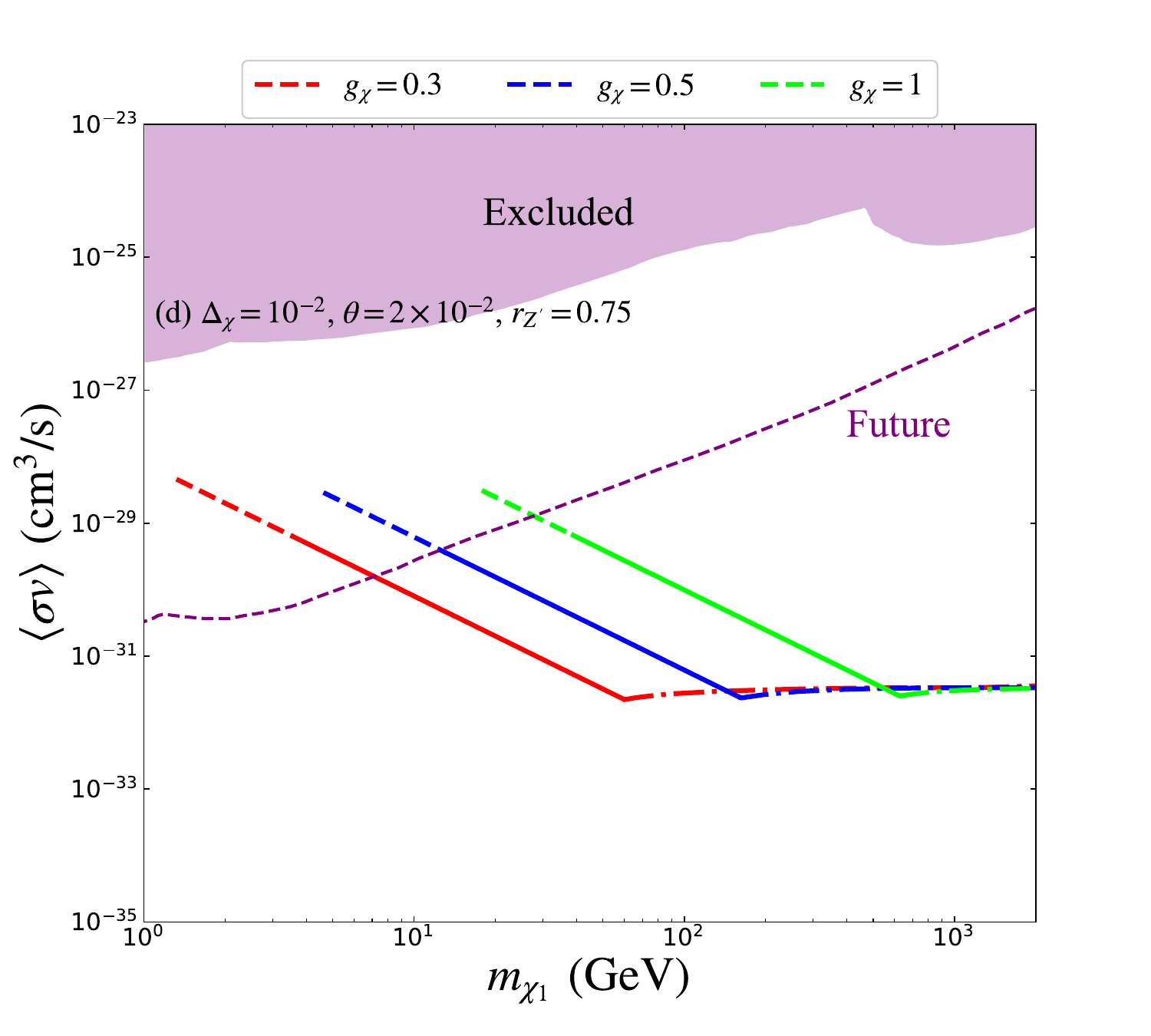}
	\end{center}
	\caption{The constraints of indirect detection experiments in the secluded scenario.  The benchmarks selected in panels (a)-(d) are consistent with those presented in Figure~\ref{FIG:fig7}. The purple shadow region and dashed line represent the existing constraints \cite{Profumo:2017obk} as well as the sensitivities of future experiments \cite{Cirelli:2025qxx}, respectively.
	}
	\label{FIG:fig9}
\end{figure}

One particularly bright aspect of traditional secluded DM arises from the indirect detection experiments, as the typical secluded annihilation cross section $\langle \sigma v \rangle$ is still around the benchmark WIMP value $2\times10^{-26}~\text{cm}^3/\text{s}$ \cite{Pospelov:2007mp}. In the secluded scenario of this model, the primary annihilation process of $\chi_1$ is denoted as $\chi_1\bar{\chi}_1\to Z^\prime Z^\prime$ followed by the decay $Z'\to f\bar{f}$. Because the direct $\chi_1\chi_1Z'$ coupling is heavily suppressed by $\sin^2\theta$, while the $\chi_1\chi_2Z'$ coupling is less suppressed by $\sin2\theta$, the dominant contribution of secluded annihilation  $\chi_1\bar{\chi}_1\to Z^\prime Z^\prime$ is mediated by the dark partner $\chi_2$ through the $t/u$-channel.  The current thermal average cross section with nearly degenerate masses of dark fermions for a small mixing angle $\theta$ can be approximated as  \cite{Mohapatra:2019ysk}:
\begin{eqnarray}\label{Eqn:id-S}
	\langle \sigma v\rangle_{\chi_1\bar{\chi}_1\to Z'Z'}\simeq\frac{\sin^4 2\theta~g_\chi^4}{256 \pi m_{\chi_1}^2}\left(1-\frac{m_{Z^\prime}^2}{m_{\chi_1}^2}\right)^{3/2}\left(1-\frac{m_{Z^\prime}^2}{2m_{\chi_1}^2}\right)^{-2}.
\end{eqnarray}

Obviously, compared to the traditional secluded DM, the secluded annihilation cross section of this model is suppressed by the mixing angle $\theta$. Therefore, the predicted value is expected to be much smaller than the benchmark WIMP value $2\times10^{-26}~\text{cm}^3/\text{s}$. The experimental limits from indirect detection on secluded DM also depend on the final states \cite{Profumo:2017obk}. In this paper, we consider the electron final state $\chi_{1}\bar{\chi}_1\to Z'Z'\to 4e$ for illustration, which is presented in Figure~\ref{FIG:fig9} accompanied by the predictions of $\langle \sigma v\rangle$ for various benchmarks. Here, the current limits are the joined results from Fermi-LAT dwarfs, CMB, and H.E.S.S. GC observations \cite{Profumo:2017obk}.

Limited by relic density, we report that lighter DM can be obtained with larger mass ratio $r_{Z'}$, smaller mixing angle $\theta$, smaller mass splitting ratio $\Delta_{\chi}$, and smaller dark coupling $g_\chi$.  In  Figure~\ref{FIG:fig9},  all benchmark lines exhibit nearly identical shapes. In the conversion and coscattering regions, the coupling $g^\prime$ is relatively small, the total $\langle \sigma v\rangle$ is well approximated by Equation~\eqref{Eqn:id-S} and is inversely proportional to $m_{\chi_1}$. However, in the final coannihilation segment, the influence of $\chi_1\bar{\chi_1}\to f\bar{f}$ must be taken into account, as $\langle \sigma v\rangle_{\chi_1\bar{\chi_1}\to f\bar{f}}$ could be larger than $\langle \sigma v\rangle_{\chi_1 \bar{\chi}_1\to Z^\prime Z^\prime}$. This part asymptotically approaches a constant value due to the constraint imposed by observed relic density.  In panel (a), the benchmark lines are cut off at the minimum $m_{\chi_1}$, i.e., conversion regime, since the secluded process $\chi_1\bar{\chi}_1\to Z^\prime Z^\prime$ remains unaffected by $g^\prime$.  When $r_{Z^\prime}\gtrsim0.75$, conversion exhibits a relatively large annihilation cross section $\langle \sigma v\rangle\sim\mathcal{O}(10^{-29})~\rm cm^3/s$ below 30 GeV, naturally falling within the detection range of future experiments. However, the $\langle \sigma v\rangle$ for coscattering and coannihilation is suppressed by heavier $m_{\chi_1}$, with most values dropping to a pessimistic level below  $\mathcal{O}(10^{-31})~\rm cm^3/s$.  In the subsequent panels (b)-(d) of Figure~\ref{FIG:fig9}, all benchmark lines possess  small enough $\langle \sigma v\rangle$ that are not excluded by the current constraints, and the most promising one is still conversion. Depending on  different parameters, the corresponding $m_{\chi_1}$ ranges from $\mathcal{O}(1)$ GeV to $\mathcal{O}(100)$  GeV.  Certainly, as observed in panels (c) and (d), when the $\Delta_\chi$ is relatively large or $g_\chi$ is relatively smaller, e.g., $\Delta_\chi\gtrsim0.04$ or $g_\chi\lesssim0.5$, the coscattering below about 20 GeV will also be subjected to examination.

On the whole, for $m_{\chi_1}\lesssim\mathcal{O}(100)$ GeV, future direct detection experiments are more inclined towards coscattering and heavily rely on $\Delta_\chi$  and $g_\chi$, requiring them to be greater than $4\times10^{-2}$ and less than $0.5$, respectively. In terms of indirect detection, it is clear that conversion demonstrates superior performance, with the necessary condition $m_{\chi_{1,2}}\sim m_{Z'}$.

\subsection{Phenomenology of $\chi_2$}\label{PC2-S}

\begin{figure}
	\begin{center}
		\includegraphics[width=0.45\linewidth]{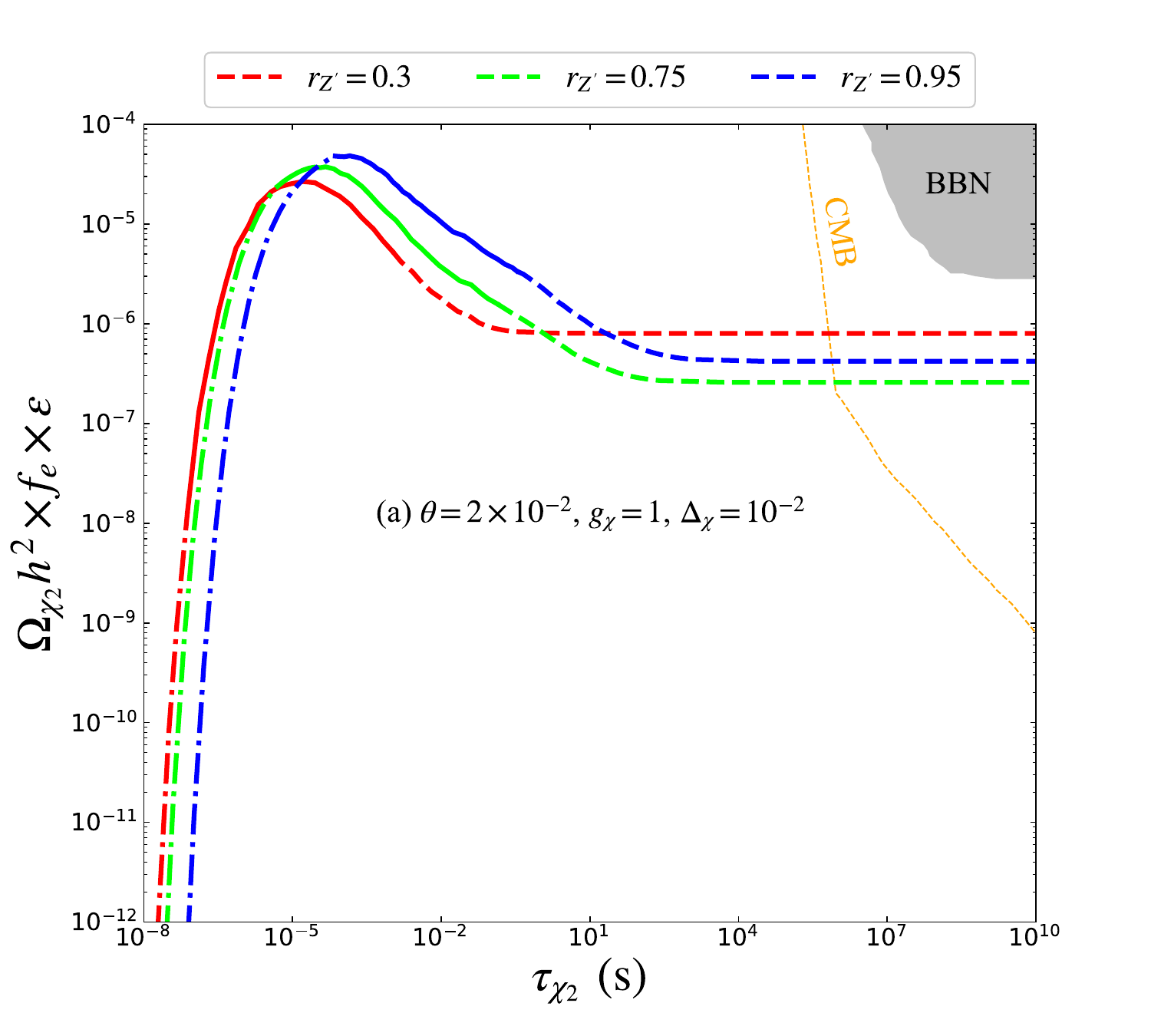}
		\includegraphics[width=0.45\linewidth]{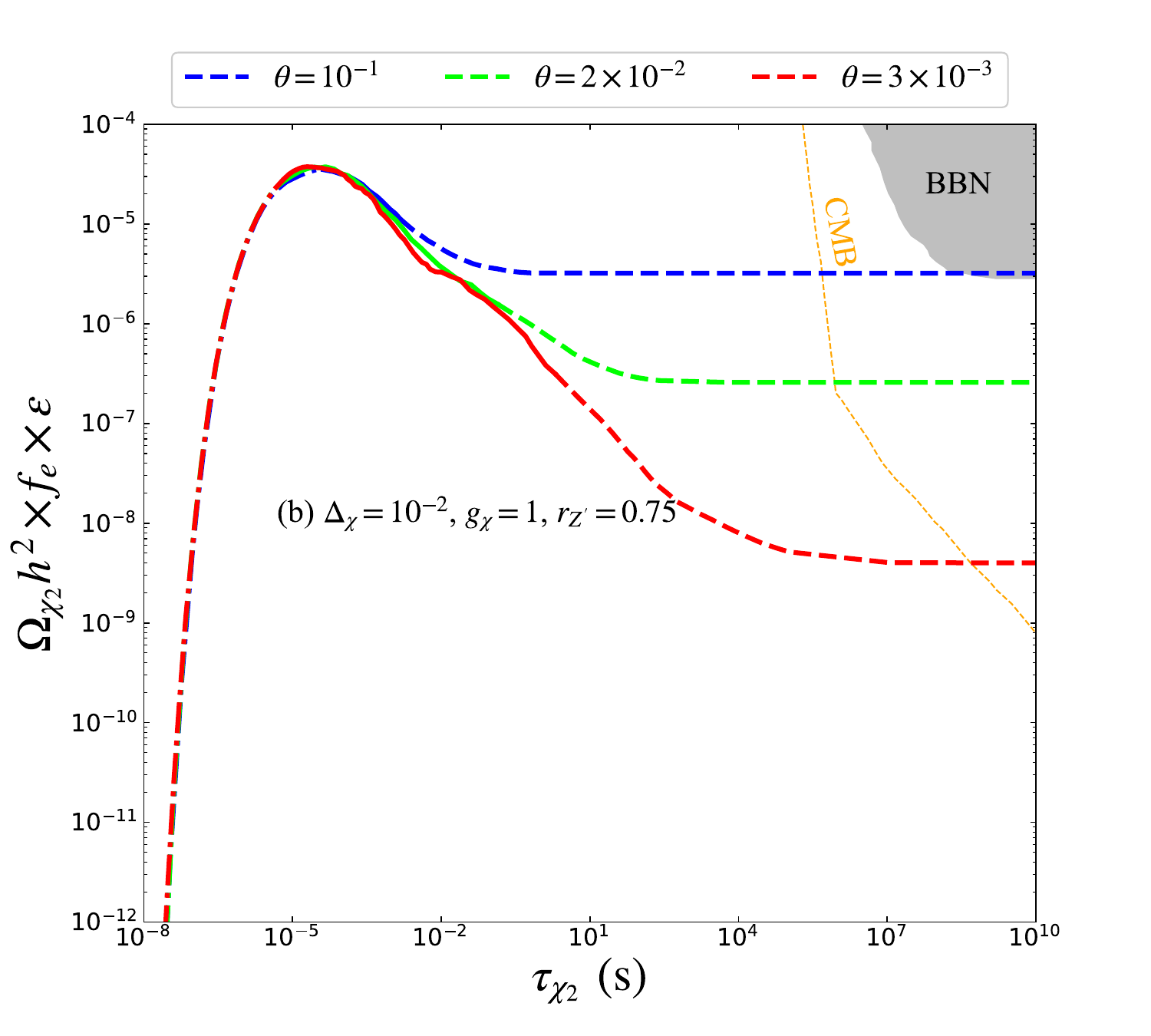}
		\includegraphics[width=0.45\linewidth]{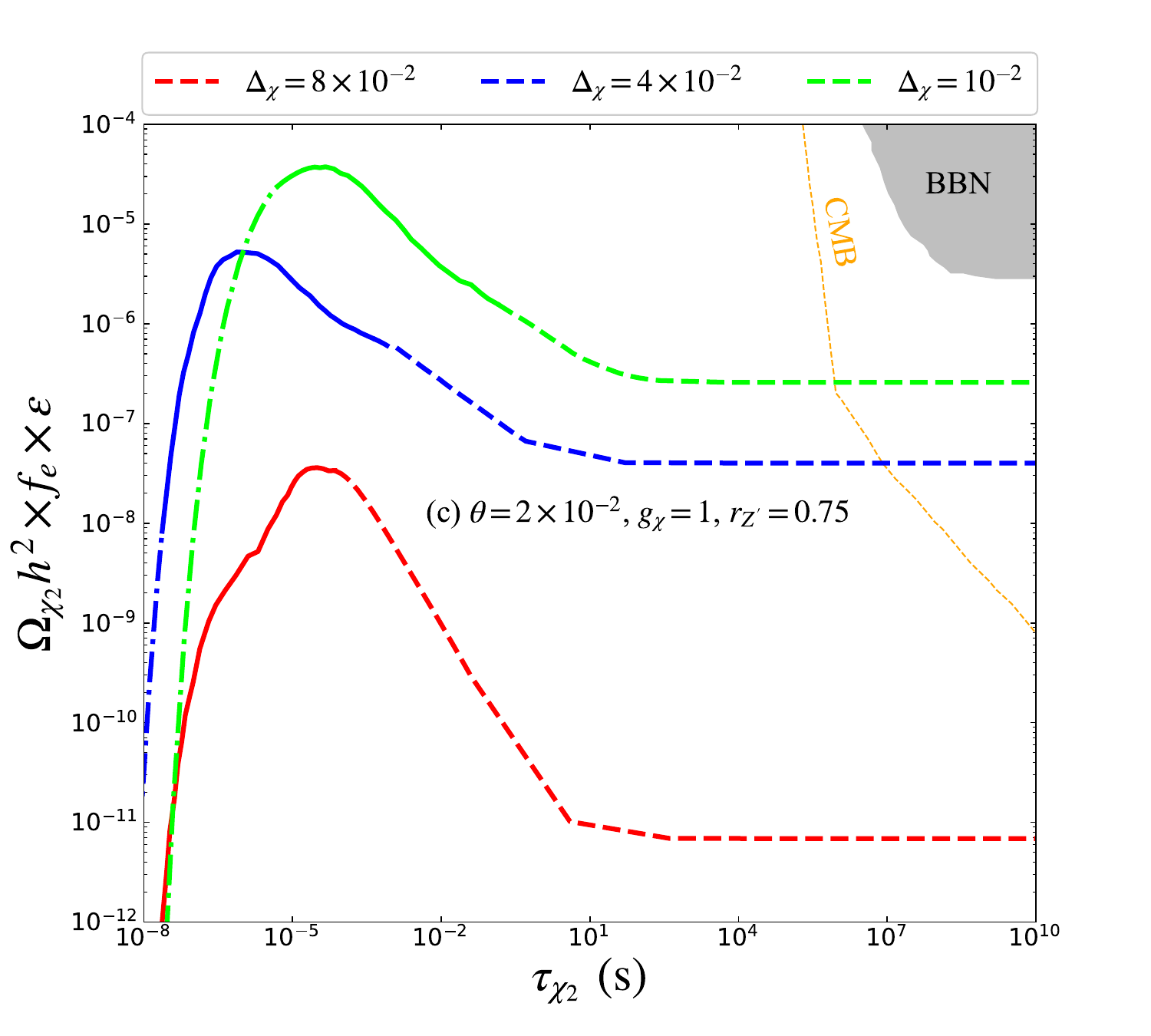}
		\includegraphics[width=0.45\linewidth]{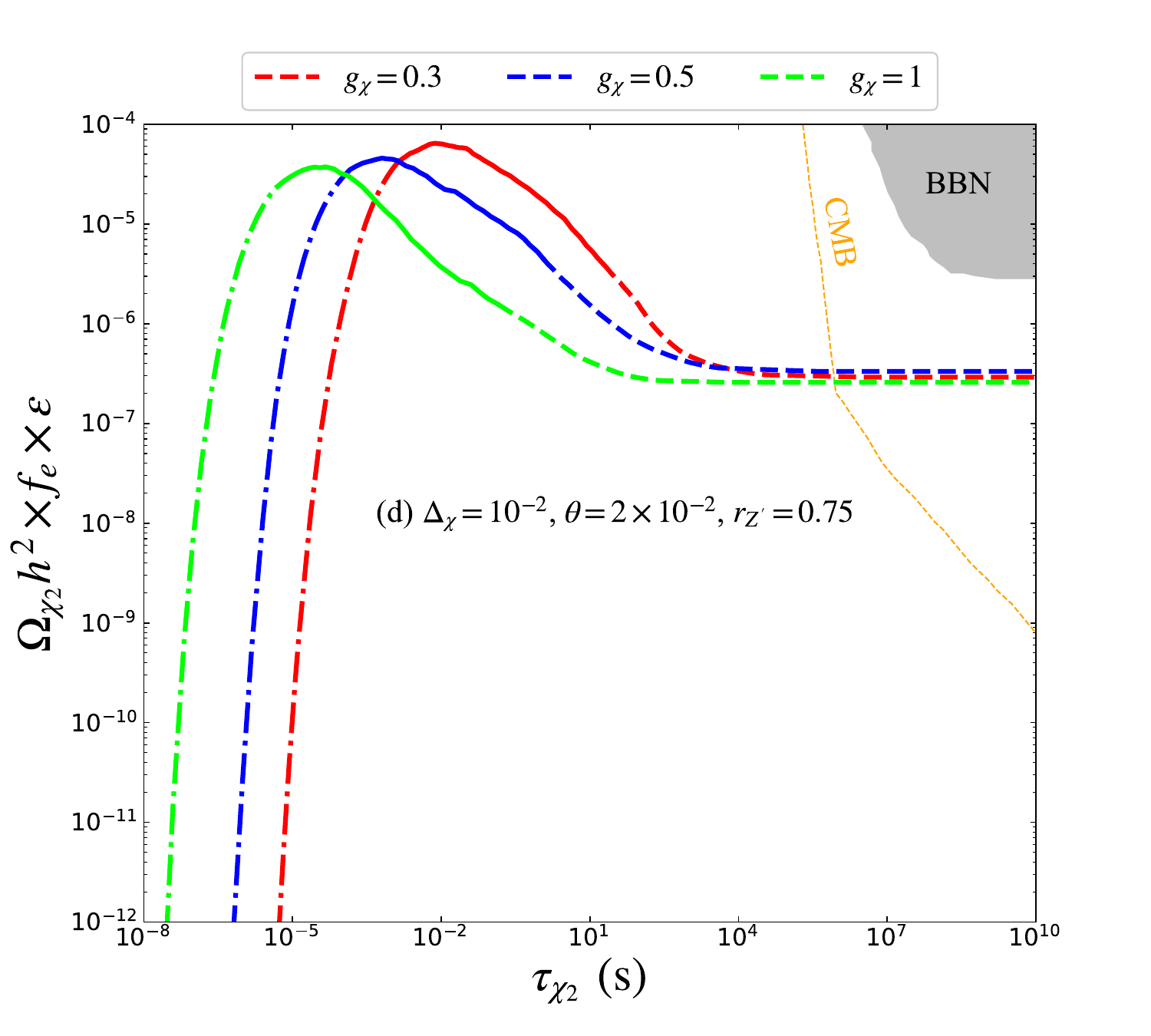}
	\end{center}
	\caption{Cosmological constraints on long-lived $\chi_2$ in the secluded scenario. The parameters appearing in the coordinate axes are defined identically to those in Figure~\ref{FIG:fig5}. The benchmarks present in each subfigure correspond one-to-one with those shown in Figure~\ref{FIG:fig7}. The gray shadow and orange dashed line represent the region excluded by the current BBN constraints and the future CMB sensitivity, respectively. 
	}
	\label{FIG:fig10}
\end{figure}

In the secluded scenario $m_{Z^\prime}<m_{\chi_2}$, the dark fermion $\chi_2$  is always generated via off-shell $Z^\prime$ at colliders, which naturally suppresses the production cross section. So the collider signature will not be considered in this scenario. Consequently, the more promising phenomenology induced by the dark partner $\chi_2$ still arises from cosmological aspects, specifically the constraints imposed by BBN \cite{Kawasaki:2017bqm} and CMB \cite{Lucca:2019rxf} on the delayed decay of $\chi_2\to\chi_1e^+e^-$. Same as the resonance scenario, they are represented in Figure~\ref{FIG:fig10} ~by the gray region and orange dashed lines, respectively.

In this scenario, we note that the conversion becomes independent of $g^\prime$ at the minimum $m_{\chi_1}$, which exceeds 1 GeV in all benchmarks. Hence, there is no need for truncation as required in the resonance scenarios. With $g^\prime$ decreasing, the lifetime $\tau_{\chi_2}$ can become significantly long.  As illustrated in Figure~\ref{FIG:fig10},  since $\tau_{\chi_2}$ is primarily governed by $g^\prime$, the increasing $\tau_{\chi_2}$ in  any  benchmark line corresponds sequentially to the coannihilation, coscattering, and conversion phase. The main factor inducing variation in the longitudinal axis is $\Omega_{\chi_2}h^2$. In the coannihilation phase, $\Omega_{\chi_2}h^2$ is predominantly suppressed by the $\chi_2 f\to \chi_1 f$ process.  The reduction in $\Gamma_{\chi_2 f\to \chi_1 f}$ alleviates its suppression on $\Omega_{\chi_2}h^2$ as $g^\prime$ decreases ($\tau_{\chi_2}$ increases). Consequently, $\Omega_{\chi_2}h^2$ begins to increase numerically. When the benchmark enters the coscattering phase, the reduction in $m_{Z^\prime}$ enhances $\Gamma_{\chi_i \chi_2\to \chi_j \chi_1}$, leading to progressively stronger suppression of $\Omega_{\chi_2}h^2$, hence $\Omega_{\chi_2}h^2$ starts to decrease. Finally, in the conversion phase, $\Omega_{\chi_2}h^2$ remains a constant as the dominant $\Gamma_{\chi_i \chi_2\to \chi_j \chi_1}$ stabilizes and no longer varies. In the phenomenological aspect,  the maximum $\tau_{\chi_2}$ of all benchmarks can reach the range where the constraints of BBN and CMB are effective.  However, different parameter choices result in varying relative relic densities on the vertical axis.  Basically speaking, as long as $\theta\lesssim0.1$, there will be no conversion excluded by BBN.  For the conversion that has the potential to be captured by future CMB at $\tau_{\chi_2}\gtrsim\mathcal{O}(10^6)$ s, it is necessary for $\theta\gtrsim3\times10^{-3}$  and $\Delta_{\chi}<0.08$.

Furthermore, regarding the CMB constraint on neutrinos produced by the delayed decay of $\chi_2\to \chi_1 \nu\bar{\nu}$. We consider the maximum $\tau_{\chi_2}$ during the CMB epoch, i.e., the calculated result $(f_\nu\epsilon~\Omega_{\chi_2}/\Omega_{\chi_1})^2\tau_{\chi_2}\lesssim\mathcal{O}(10)$ s, which is eight orders of magnitude lower than the current limit~\cite{Hambye:2021moy}. Therefore, constraints from neutrino final states do not pose any threat to our benchmarks.

\subsection{Combination and Discussion}\label{CD-S}

Compared to the resonance scenario, the secluded scenario is more promising to be tested at future experiments. Especially in the conversion phase, the relatively large secluded annihilation cross section and long lifetime of the dark partner make this regime detectable. In Figure~\ref{FIG:fig7}, we also present the combined results of future indirect detection and CMB  that exceed the detection range of colliders. Meanwhile, the sensitive coscattering  of future direct detection experiments typically requires $g'\sim\mathcal{O}(10^{-4})$ for $m_{Z'}\sim\mathcal{O}(10)$ GeV, which may also be captured by upcoming colliders.

In panel (a) of Figure~\ref{FIG:fig7}, we consider $r_{Z^\prime}$ to be greater than 0.2 to ensure the effectiveness of conversion. The future CMB is sensitive to conversion with 10 GeV $\lesssim m_{Z^\prime}\lesssim40$ GeV and $g^\prime\lesssim\mathcal{O}(10^{-8})$. The indirect detection experiments can capture conversion of $g'$ up to $\mathcal{O}(10^{-5})$, however, the sensitive region concentrates on $m_{Z^\prime}$ being a dozen GeV with $r_{Z'}\gtrsim0.75$ of the benchmarks. It should be mentioned that in the secluded scenario, the relic density  in principle does not depend on the coupling $g'$, which means it can be arbitrarily small. However, for a too small $g'<10^{-11}$, the lifetime of new gauge boson $\tau_{Z'}$ becomes too large when $m_{Z'}>1$ GeV, which may alert BBN. To avoid such an issue, we simply require $g'>10^{-10}$ in this paper.

In panel (b) of Figure~\ref{FIG:fig7}, an excessively large $\theta$ will result in $\chi_1$ being entirely produced through the coannihilation mechanism and the current BBN constraint imposes stringent restrictions on $\theta\gtrsim0.1$ as shown in Figure \ref{FIG:fig10}.  Therefore, we consider $\theta$ to be less than 0.1. Correspondingly, conversion  will be detected by CMB when $m_{Z^\prime}\lesssim90$ GeV and $g^\prime\lesssim2\times10^{-8}$.  The sensitive region of indirect detection experiments will expand upward by three orders of magnitude of $g^\prime$. 

In case (c) of Figure~\ref{FIG:fig7}, we also assume that $\Delta_\chi$ is within the range of $[10^{-3},10^{-1}]$.  For varying $\Delta_\chi$, the future CMB can only capture conversion in the vicinity of 10 GeV. It is clear that when $\Delta_{\chi}$ decreases, the promising region of CMB quickly increases up to $\mathcal{O}(10^{-6})$. The capability for indirect detection is stronger, and it is desirable to detect conversion within the range of  1.8  GeV to 18 GeV. Sub-GeV DM is  not favored by the current constraints from indirect detection, and the corresponding $\Delta_\chi$ should be less than $\mathcal{O}(0.1)$ of the benchmarks. 

In panel (d) of Figure~\ref{FIG:fig7}  where $g_\chi\lesssim1$, the future CMB shows sensitivity of $m_{Z^\prime}$ below 15~GeV and  $g^\prime$  smaller than $10^{-7}$. We also report that the sensitive limit of $g'$ from CMB decreases as $g_\chi$ increases. Similarly, the conversion that is expected to be  probed  by indirect detection experiments could still be expanded to $g^\prime\sim\mathcal{O}(10^{-5})$.

As a whole for the secluded scenario,  the majority of coannihilation is excluded by the current collider constraints, and certain coscattering within the range
of $\mathcal{O}(10^{-5})\lesssim g^\prime\lesssim\mathcal{O}(10^{-3})$  could be tested by both future colliders and direct detection experiments. Meanwhile, portions of conversion in the range of $\mathcal{O}(10^{-8})\lesssim g^\prime\lesssim\mathcal{O}(10^{-5})$ and $m_\chi\lesssim \mathcal{O}(10)$ GeV may be captured through indirect detection experiments. Furthermore, those conversions below $\mathcal{O}(10^{-8})$ hold promise for dual verification via both CMB  and indirect detection experiments. Such a distinct feature makes the secluded scenario of this model quite different from the canonical secluded DM. Because the canonical GeV-scale secluded DM is tightly constrained by indirect detection, and does not predict an observable CMB signature.

\section{Conclusion} \label{SEC:CL}

Besides the pair annihilation of DM, alternative mechanisms, such as coscattering, conversion, and coannihilation, could also significantly modify the evolution of DM relic density. In order to study these new mechanisms within the framework of $U(1)_{B-L}$ symmetry, we introduce two Dirac fermions $\tilde{\chi}_1$ and $\tilde{\chi}_2$ with nearly degenerate masses.  Among them, only the non-dark matter fermion  $\tilde{\chi}_2$ possesses a non-zero $U(1)_{B-L}$ charge, meanwhile the stability of DM is ensured by a $Z_2$ symmetry. The mass term $\delta m \bar{\tilde{\chi}}_1\tilde{\chi}_2$ induces mixing between the dark fermions, which is crucial for activating the conversion mechanism. Based on  the magnitude of the mass of mediator $Z^\prime$ and dark fermions, we categorize our study into the resonance and secluded scenarios.

In the resonance scenario, the gauge coupling $g^\prime$ in the coscattering phase is typically greater than $\mathcal{O}(10^{-2})$, which is easily excluded by the current $Z^\prime$ constraints. Therefore, we focus on the coannihilation and conversion with $g^\prime\lesssim\mathcal{O}(10^{-3})$.  For the conversion at the range of $\mathcal{O}(10^{-5})\lesssim g^\prime\lesssim\mathcal{O}(10^{-3})$, which is most likely to be probed by future colliders, the mass can vary from GeV to TeV scale, and requiring $\theta\lesssim5\times10^{-3}$ and $g_\chi\sim\mathcal{O}(0.1)$. On the other hand, the magnitude of $\Delta_\chi$ should not be excessively large to ensure the effectiveness of conversion.  Regarding the coannihilation dominated region  with  $g^\prime$ between $\mathcal{O}(10^{-8})$ and $\mathcal{O}(10^{-5})$, although detection of $Z'$ at the GeV scale may be achievable with future colliders, more promising signatures are anticipated from forthcoming CMB cosmological observables derived from the long lived $\chi_2$. The CMB could capture the dark partner up to TeV, but it is necessary to satisfy $\theta\lesssim5\times10^{-4}$, $\Delta_\chi\lesssim10^{-2}$ and $g_\chi\lesssim1$. Due to the suppression of the small mixing angle $\theta$, future direct and indirect detection experiments for DM will face significant challenges in capturing any mechanisms. Unless $\theta$ far exceeds $5\times10^{-3}$, only coannihilation has hope at this moment.

In the secluded scenario with $m_{Z'}\lesssim m_{\chi_{1,2}}$, the situation is quite different from the resonance scenario. The coannihilation with the relatively large $g^\prime$  is naturally excluded by $Z^\prime$ constraints, while a relatively small $g'$ is required by the coscattering and conversion that we are interested in. For coscattering with $\mathcal{O}(10^{-5})\lesssim g^\prime\lesssim\mathcal{O}(10^{-3})$, future  collider searches for $Z^\prime$ will be sensitive to $\theta \gtrsim 3\times10^{-3}$ and $g_\chi\sim\mathcal{O}(0.1)$, with corresponding new gauge boson mass not exceeding the TeV scale. When the DM mass is around dozens of GeV, the future direct detection experiments hold greater promise to probe the region with $\Delta_\chi\gtrsim4\times10^{-2}$  and $g_\chi\lesssim 0.5$. Although the conversion below $\mathcal{O}(100)$ GeV with $g^\prime\lesssim\mathcal{O}(10^{-5})$ is beyond the reach of future colliders, the advantages in indirect detection are even more pronounced. With minimal specific requirements for parameters, namely $\Delta_\chi\lesssim\mathcal{O}(0.1)$, the conversion can easily satisfy current constraints.  Moreover, for the lower regions of $g^\prime\lesssim\mathcal{O}(10^{-8})$,  CMB will also participate in the verification of conversion,   which is sensitive to $\mathcal{O}(10^{-3})\lesssim\theta\lesssim\mathcal{O}(0.1)$ and $\mathcal{O}(10^{-3})\lesssim\Delta_{\chi} <\mathcal{O}(10^{-2})$.

In summary, although both scenarios can generate DM through the coscattering, conversion, and coannihilation mechanisms,  there are significant differences in their phenomenological constraints. In the resonance scenario, coscattering is notably less favored, whereas in the secluded scenario, it changes into coannihilation. Furthermore, the magnitude of $\theta$ in the secluded scenario is two orders of magnitude larger than that in the resonance scenario, which makes it easier for the secluded case to be tested by future direct and indirect detection experiments. Finally, CMB can test coannihilation in the resonance scheme, but only conversion holds promise in the secluded scenario.

\section*{Acknowledgments}
We would like to thank Prof. Lei Wu for his insightful discussions in this study. This work is supported by the National Natural Science Foundation of China under Grant No. 12505112, Natural Science Foundation of Shandong Province under Grant No. ZR2024QA138, and State Key Laboratory of Dark Matter Physics, University of Jinan Disciplinary Cross-Convergence Construction Project 2024 (XKJC-202404).

%%%%%%%%%%%%%%%%%%%%%%%%%%%%%

\end{document}